\documentclass[twocolumn]{aastex631}

\usepackage{bm}
\usepackage{color}
\usepackage{natbib}
\usepackage{amsmath}

\usepackage{multirow}

\newcommand{\LD}[1]{{\color{black} #1}}
\accepted{Jul.~14, 2022}
\shortauthors{Hu et al.}

\begin{document}

\title{A Comprehensive Analysis of the Gravitational Wave Events with the Hilbert-Huang Transform: From Compact Binary Coalescence to Supernova}
\author[0000-0001-8551-2002]{Chin-Ping Hu}
\affiliation{Department of Physics, National Changhua University of Education, Changhua, 50007, Taiwan}
\author[0000-0003-4083-9567]{Lupin Chun-Che Lin}
\affiliation{Department of Physics, National Cheng-Kung University, Taiwan}
\author[0000-0002-1473-9880]{Kuo-Chuan Pan}
\affiliation{Institute of Astronomy, National Tsing Hua University, Hsinchu 30013, Taiwan}
\author[0000-0001-8229-2024]{Kwan-Lok Li}
\affiliation{Department of Physics, National Cheng Kung University, 70101 Tainan, Taiwan}
\author[0000-0001-5674-5309]{Chien-Chang Yen}
\affiliation{Department of Mathematics, Fu Jen Catholic University, 24205 New Taipei City, Taiwan}
\author[0000-0002-5105-344X]{Albert K. H. Kong}
\affiliation{Institute of Astronomy, National Tsing Hua University, Hsinchu 30013, Taiwan}
\author[0000-0003-1753-1660]{C. Y. Hui}
\affiliation{
Department of Astronomy and Space Science, Chungnam National University, Daejeon, Korea (ROK)}

\correspondingauthor{C.-P. Hu}
\email{cphu0821@gm.ncue.edu.tw}

\begin{abstract}
We analyze the gravitational wave signals with a model-independent time-frequency analysis, which is improved from the Hilbert-Huang transform (HHT) and optimized for characterizing the frequency variability on the time-frequency map. Except for the regular HHT algorithm, i.e., obtaining intrinsic mode functions with ensemble empirical mode decomposition and yielding the instantaneous frequencies, we propose an alternative algorithm that operates the ensemble mean on the time-frequency map. We systematically analyze the known gravitational wave events of the compact binary coalescence observed in LIGO O1 and O2, and in the simulated gravitational wave signals from core-collapse supernovae (CCSNe) with our method. The time-frequency maps of the binary black hole coalescence cases show much better details compared to those wavelet spectra. Moreover, the oscillation in the instantaneous frequency caused by mode-mixing could be reduced with our algorithm. For the CCSNe data, the oscillation from the proto-neutron star and the radiation from the standing accretion shock instability can be precisely determined with the HHT in great detail. More importantly, the initial stage of different modes of oscillations can be clearly separated. These results provide new hints for further establishment of the detecting algorithm, and new probes to investigate the underlying physical mechanisms. 
\end{abstract}


\section{Introduction \label{introduction}}

The successful detection of the gravitational wave signal from the Laser Interferometer Gravitational-Wave Observatory (LIGO) opens a new window in both astronomy and physics \citep{AbbottAA2016}.
The advanced Virgo (aVirgo) interferometer also joined the engineer run with two aLIGO detectors since May of 2017.
At the end of 2018 for the first and second observation runs (O1 and O2), aLIGO and the aVirgo has confirmed 11 gravitational wave (GW) transients, including 10 merging events from two massive stellar-mass black holes (BHs) and one from the coalescence of two neutron stars (NSs) \citep{AbbottAA2019}. Until the end of 3rd observation run in March of 2020, the aVirgo and aLIGO network have confirmed 90 GW events \citep{GWTC3-2021}, and 50 of which have been reported during the first half of the third observing run (O3) \citep{GWTC2-2021}. These GW events can be detected by multi-resolution techniques including wavelet Q-transform \citep{Chatterji2004}.

It has been suggested that the novel time-series analysis method, the Hilbert-Huang transform (HHT), could precisely describe the time-frequency properties of the GW signal owing to the high resolution of the HHT \citep{Camp2007}. The GW signal can be extracted from the simulated data with a BH merging signal embedded in a noisy background \citep{StroeerCC2009}. Similar work has been extended to investigate the simulated GW signal from the coalescence of binary NSs \citep{KaneyamaOT2016}.

The first and strongest event GW 150914 has been analyzed with the HHT \citep{SakaiOM2017}. The signal is not able to be decomposed into a single component through empirical mode decomposition (EMD). The ensemble EMD (EEMD) has been proposed to overcome the so-called `mode-mixing'' or ``mode-splitting'' problem \citep{Wu2004, Wu2009} but it is only valid for a signal with moderate frequency variability because EEMD is equivalent to a dyadic filter. This prevents getting a precise instantaneous frequency although the signal from different components with highest Hilbert amplitude could marginally connected with each other. The entire time-frequency map could be consistent with the theoretical prediction although some frequency modulations can be clearly seen. A systematic investigation of known GW events using the HHT is carried out by \citet{AkhshiAB2021}. A physical time delay could be derived from the reconstructed waveform.

In this paper, we comprehensively apply the Morlet wavelet analysis and the HHT on the LIGO data of the known GW events to test if the HHT is eligible to detect the trigger. Moreover, we propose an alternative method to obtain the Hilbert spectrum by stacking the time-frequency map instead of taking the ensemble mean of components in the data sets with artificial noises. Besides the coalescence events in the LIGO data sets, we also tested the algorithm on the simulated GW signal from core-collapse supernovae (CCSNe). Section \ref{sec:data_and_algorithm} describes the reduction and processing of the LIGO data, and the mathematical algorithm used in this research. We present the analysis result of two cases, GW 150914 and GW170817, in Section \ref{sec:analysis_result}. The time-frequency maps of other events are shown in Appendix \ref{sec:appendix_LIGO}. We describe the models and parameters used in generating the CCSN signal, and show the corresponding analytical result of a typical CCSN case in Section \ref{sec:CCSNe}. Cases with other rotational parameters and viewing geometries are presented in Appendix \ref{sec:appendix_CCSNe}.

\section{Data Reduction and Time-Frequency Analysis Algorithm}
\label{sec:data_and_algorithm}

\subsection{LIGO Data Reduction}
\label{sec:Data_reduction} 
In order to examine the capability of our numerical method, we check the performance of our algorithm on the known GW events in GWTC-1 (A gravitational-wave transient catalog of compact binary mergers observed by LIGO and Virgo during the first and second observing runs) within 2015--2017.
We use the 32\,s data of 4KHz released from the LIGO Open Science Center\footnote{\url{https://www.gw-openscience.org/catalog/GWTC-1-confident/html/}} \citep{VallisneriKW2015} obtained from Hanford and Livingston sites to perform our analysis.
We notice that a strong glitch and a GW candidate event (i.e., GW 170817) occurred almost simultaneously at the Livingston site, and therefore we downloaded the `clean' data after the noise subtracting following the glitch model described in \citet{AbbottAA2017_GW170817}.

Table~\ref{events} lists data sets of known GW events considered in this article, and all the signal processing was proceeded with \LD{the python based GW analysis library named as GWPY v2.1.0\footnote{\url{https://gwpy.github.io/docs/stable/}}} \citep{MacleodUC2020}.
To deal with the GW data, we followed the standard whitening process to suppress the extra colored noise at low frequencies and spectral lines led by instrumental or background effects \citep[see, e.g.,][]{UsmanNH2016, AbbottAA2019}. This operation might affect the nonlinearily of the GW signal. \LD{We note that whitening process is parameter dependent and the method is not unique \citep[see, e.g.,][]{TsukadaCH2018, HuangZL2021}.} Further developments of the de-noising techniques would be important though this is beyond the scope of this research. We therefore whitened the time series to enhance the higher-frequency content with a FFT integration length of 4\,s and 2\,s-overlap between FFTs for the given window of Hann smoothing.
We also apply a bandpass filter to reserve the data within 30--450\,Hz and then use notch filters to attenuate line noises because we want to suppress the effect from thermal noises of mirrors, quantum noise of laser or any line noise arisen from detector resonances to dominate the spectral behavior.

\begin{table*}[ht]
\caption{{\footnotesize The data files of LIGO/Virgo events investigated in this article.}}\label{events}
\centering
\begin{tabular}{lcc} 
\hline 
\multirow{2}{*}{GW150914} & Hanford & H-H1\_GWOSC\_4KHZ\_R1-1126259447-32.hdf5
\\
& Livingston & L-L1\_GWOSC\_4KHZ\_R1-1126259447-32.hdf5
\\
\hline 
\multirow{2}{*}{GW151012} & Hanford & H-H1\_GWOSC\_4KHZ\_R1-1128678885-32.hdf5
\\
& Livingston & L-L1\_GWOSC\_4KHZ\_R1-1128678885-32.hdf5
\\
\hline 
\multirow{2}{*}{GW151226} & Hanford & H-H1\_GWOSC\_4KHZ\_R1-1135136335-32.hdf5
\\
& Livingston & L-L1\_GWOSC\_4KHZ\_R1-1135136335-32.hdf5
\\
\hline 
\multirow{2}{*}{GW170104} & Hanford & H-H1\_GWOSC\_4KHZ\_R1-1167559921-32.hdf5
\\
& Livingston & L-L1\_GWOSC\_4KHZ\_R1-1167559921-32.hdf5
\\
\hline 
\multirow{2}{*}{GW170608} & Hanford & H-H1\_GWOSC\_4KHZ\_R1-1180922479-32.hdf5
\\
& Livingston & L-L1\_GWOSC\_4KHZ\_R1-1180922479-32.hdf5
\\
\hline 
\multirow{2}{*}{GW170729} & Hanford & H-H1\_GWOSC\_4KHZ\_R1-1185389792-32.hdf5
\\
& Livingston & L-L1\_GWOSC\_4KHZ\_R1-1185389792-32.hdf5
\\
\hline 
\multirow{2}{*}{GW170809} & Hanford & H-H1\_GWOSC\_4KHZ\_R1-1186302504-32.hdf5
\\
& Livingston & L-L1\_GWOSC\_4KHZ\_R1-1186302504-32.hdf5
\\
\hline 
\multirow{2}{*}{GW170814} & Hanford & H-H1\_GWOSC\_4KHZ\_R1-1186741846-32.hdf5
\\
& Livingston & L-L1\_GWOSC\_4KHZ\_R1-1186741846-32.hdf5
\\
\hline 
\multirow{2}{*}{GW170817} & Hanford & H-H1\_GWOSC\_4KHZ\_R1-1187008867-32.hdf5
\\
& Livingston & 
L-L1\_LOSC\_CLN\_4\_V1-1187007040-2048.hdf5
\\
\hline 
\multirow{2}{*}{GW170818} & Hanford & H-H1\_GWOSC\_4KHZ\_R1-1187058312-32.hdf5
\\
& Livingston & L-L1\_GWOSC\_4KHZ\_R1-1187058312-32.hdf5
\\
\hline 
\multirow{2}{*}{GW170823} & Hanford & H-H1\_GWOSC\_4KHZ\_R1-1187529241-32.hdf5
\\
& Livingston & L-L1\_GWOSC\_4KHZ\_R1-1187529241-32.hdf5
\\
\hline 
\end{tabular}
\end{table*} 

To confirm the practicability of our methods, we also check the instantaneous frequency obtained from the simulated gravitational waveform. We generate the waveform using the effective-one-body \citep[EOB;][]{TaracchiniBP2014,Purrer2016,BoheST2017} numerical-relativity (NR) model.  
The simulated GW time series can be obtained by employing apps \citep[LAL 6.19.1;][]{lalsuite} provided by the LIGO Algorithm Library\footnote{\url{https://lscsoft.docs.ligo.org/lalsuite/}}.
The basic physical parameters of GW150914 applied in the simulated waveform refer to those determined in GWTC-1 \citep{AbbottAA2019}.

\subsection{Hilbert-Huang Transform}

The Hilbert-Huang transform (HHT) is a novel time-frequency analysis algorithm that can trace frequency variability in a great detail \citep{Huang1998}. \LD{This adaptive method needs no \emph{priori} basis, and has been widely used in many fields of science and engineering \citep[see, e.g.,][]{HuangSL1999, HuangA2005, HuangS2014}.} The core idea of the HHT is the Hilbert transform of a time series $x(t)$: 
\begin{equation}\label{eq:Hilbert-transform}
    y(t)=\frac{1}{\pi}\mathcal{P}\int_{-\infty}^{\infty}\frac{x(t')}{t-t'}dt',
\end{equation}
where $\mathcal{P}$ indicates the Cauchy principal value and $y(t)$ is the Hilbert transform of $x(t)$. With this definition, we can have an analytical signal
\begin{equation}\label{eq:z(t)}
    z(t)=x(t)+iy(t)=a(t)e^{i\theta(t)}
\end{equation}
where $a(t)=\sqrt{x^2(t)+y^2(t)}$ is the Hilbert amplitude and $\theta(t)=\tan^{-1}\left[y(t)/x(t)\right]$ is the instantaneous phase function. The time derivative of $\theta(t)$
\begin{equation}\label{eq:definition_IF}
    \omega(t)=\frac{d\theta(t)}{dt}
\end{equation}
can be defined as the instantaneous frequency. As the development of the HHT, a few more methods were also proposed to obtain the frequency, such as generalized zero-crossing, and direct quadrature  \citep{Huang2009}. In our study, we mainly utilize the normalized Hilbert transform, which applying the Hilbert transform after normalizing the intrinsic mode functions (IMFs) to the range between $-1$ and $1$. This is designed to overcome the limitation imposed by the Bedrosian theorem  \citep{Bedrosian1962,XuY2006}. In some cases the intra-wave modulation, i.e., frequency modulation within one cycle that caused by possible non-linear effect of an oscillation system, is not physically important, we use the generalized zero-crossing, which is a robust method to calculate the inter-wave modulation (i.e., the variation of a cycle length) frequency  \citep{SekharS2004}. Both methods yield a fully consistent result.

\begin{figure*}
    \centering
    \includegraphics[width=0.7\textwidth]{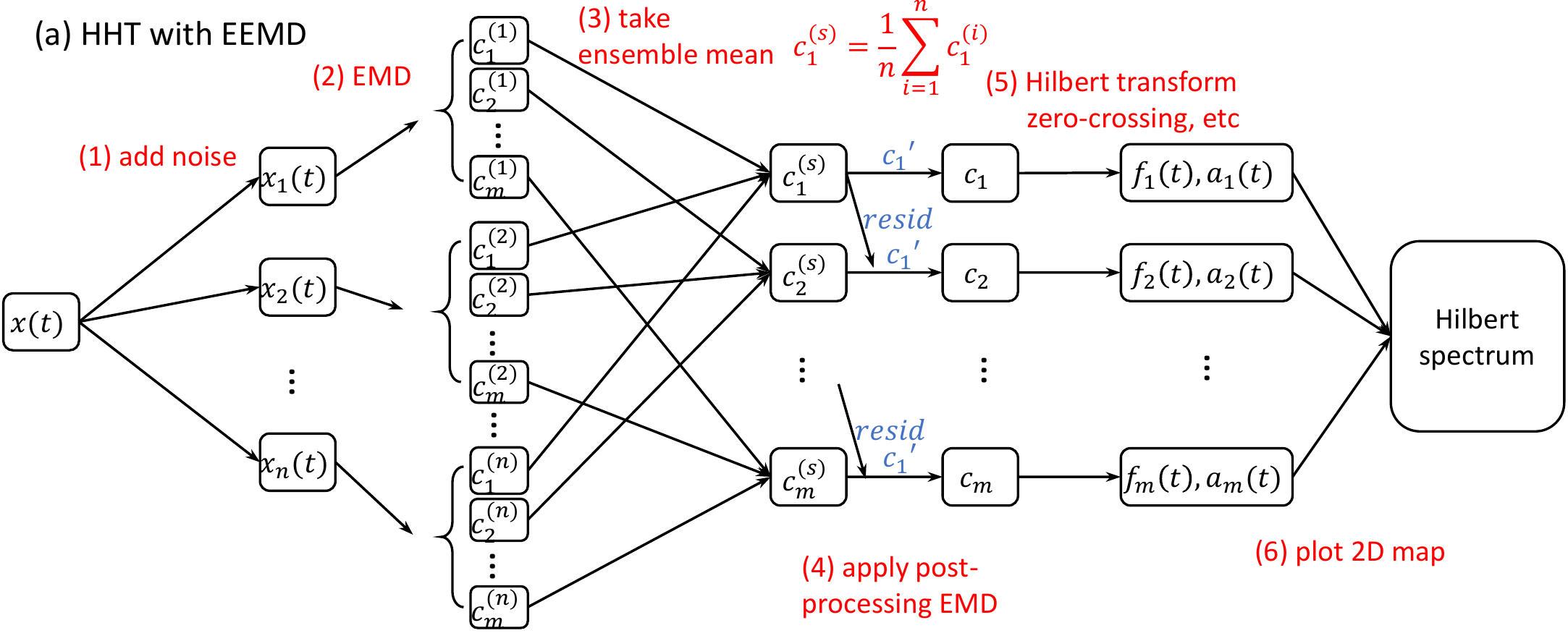}\\[0.2cm]
    \includegraphics[width=0.7\textwidth]{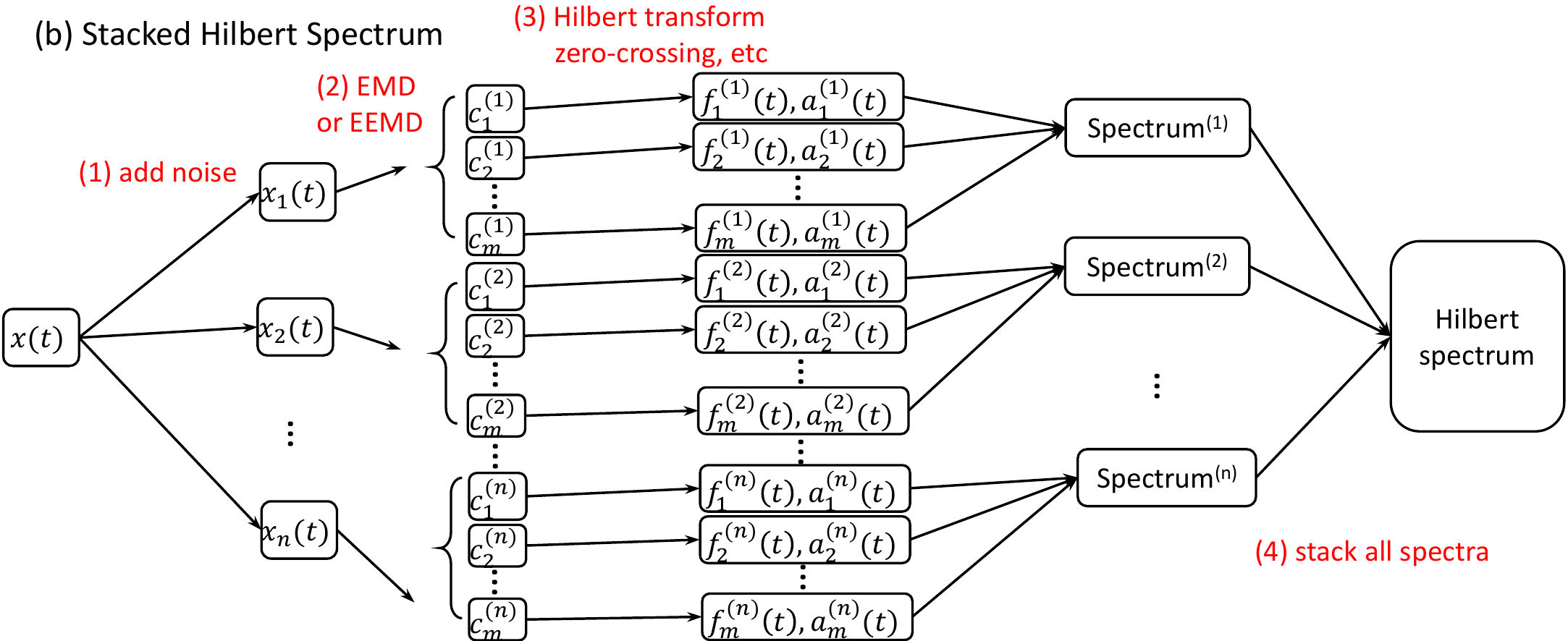}
    \caption{(a) The procedure of EEMD in this paper. We applied the post-processing EMD from the first IMF. (b) Stacked Hilbert analysis algorithm that is optimized for visualization of the time-frequency map. }
    \label{fig:hht_with_eemd_procedure}
\end{figure*}

However, the instantaneous frequency is meaningful only if the input time series is an IMF, which must be symmetry with respect to zero, and the number of extrema and zero-crossing must be the same or at least differs by one. This criterion is difficult to be achieved in real data. A pre-processing is needed. Therefore, the HHT consists of two steps: (1) decompose the time series data into finite IMFs and (2) obtain instantaneous frequencies for each IMFs  \citep{Wu2009}. 

The first method proposed for the decomposition is the empirical mode decomposition (EMD). EMD is an iterative process called ``sifting''  that extracts IMFs by removing the local mean values obtained from taking the average of the upper and lower envelope of a time series \citep{Huang1998}. It can separate time series into IMFs of different time scales, and make IMFs symmetric with respect to zero. 

In principle, there are infinite way to decompose the signal. For example, different sifting parameters, e.g., stoppage criteria, spline method, etc. A confidence limit for Hilbert spectrum can be obtained from an ensemble of valid IMFs resulting from different sifting parameters. The ideal case is that the raw time series $x(t)$ can be decomposed as
\begin{equation}
    x(t)=\sum_{j=1}^n c_j(t)+r(t)=\sum_{j=1}^n a_j(t)e^{i\int\omega_j(t)dt}+r(t)
\end{equation}
where $c_j(t)$ is the $j$-th IMF, and $r(t)$ is the non-oscillating residual. 

A signal with a narrow modulation timescale is expected to be decomposed in a single IMF. However, it could be decomposed into different IMFs using EMD if there are noises or intermittent signals. This ``mode-mixing'' problem can be solved by introducing white noise with finite amplitude into the data and taking the ensemble mean of corresponding obtained from all the simulated sets. This is the Ensemble EMD (EEMD) \citep{Wu2009}. Based on the study of white noise, the EEMD is a dyadic filter that collates signals of similar scales together. The remaining problem is that the summation of IMFs is no longer an IMF. Therefore, a post-processing EMD is needed \citep{Wu2009}. One can apply the post-processing EMD from arbitrary components of interest. Here we use it since the first IMF, the highest frequency mode of the signal. The entire procedure is shown as Figure \ref{fig:hht_with_eemd_procedure}(a). 

When applying EEMD, two more parameters are needed to be considered: the amplitude of the input noise $\sigma_n$ and the number of ensemble trials $N$. \LD{Unlike EMD, the completeness of the signal is not guaranteed in EEMD due to the input noise. Therefore, a large N is needed to minimize the incompleteness of the output signal, which costs high computing resources.}  To reduce the computing time, the EEMD has been improved by adding a pair of positive and negative noises. \LD{This operation is also named as complementary EEMD \citep[CEEMD][]{YehSH2010}}.  With this operation, a good completeness can be achieved compared to the original version of EEMD \citep[see, e.g., Figure 6 in][]{YehSH2010}. \LD{With our parameter setting, the deviation between the summation of IMFs and the original time series is $\sim10^{-15}$ times of the deviation of the original time series. This algorithm has been included in the Matlab fast EEMD package as the default EEMD algorithm, which is} developed by the Research Center for Adaptive Data Analysis at National Central University \citep{WangYY2014}. \LD{ We have successfully explored the time-frequency properties of X-ray binaries and active galactic nuclei using this package \citep[see, e.g.,][]{Hu2014,SuCH2015,HuMS2019}. A pywrapper of this package\footnote{\url{https://github.com/HHTpy/HHTpywrapper}} is also available \citep{Su2017_HHT}.} We also tried to perform the following analysis using the complete ensemble empirical mode decomposition
with adaptive noise\footnote{\url{https://github.com/macolominas/CEEMDAN}} and found that the result is fully consistent with that obtained from the fast EEMD algorithm \citep{TorresCS2011}.  

To decide the best value of $\sigma_n$, we tried different level of and choose the best $\sigma_n$ to minimize the orthogonality index $\textrm{OI}$, which is defined as
\begin{equation}\label{eq:orthogonality}
    \textrm{OI}\equiv\frac{\sum_t\sum_{i\neq j} c_i(t)c_j(t)}{\sum_t x^2(t)}
\end{equation}
The OI can be defined specifically for any IMF of interest or between all IMFs. Because the GW signal usually spread in a wide frequency range, we minimize the overall OI presented in equation \ref{eq:orthogonality}. 

However, the GW signal of the coalescence of two compact objects usually spans a wide frequency range, which is difficult to be decomposed into a single IMF with EEMD. This makes it difficult to trace the evolution of the signal in detail, especially near the boundary when the signal migrates from one IMF to another. For visualization purpose, we establish a stacked Hilbert spectrum algorithm to trace the time-frequency property of GW signals. 

The idea of our algorithm is demonstrated in Figure \ref{fig:hht_with_eemd_procedure}(b). Similar to EEMD, we first add finite noises into the signal and create $N$ simulated time series. Then, we apply EMD or EEMD on the noise-inserted time series to obtain IMFs. After this, we did not take the ensemble mean of IMFs. Instead, we calculate the instantaneous frequency and amplitudes of individual IMFs in each simulated data set. Finally, we stack all the Hilbert spectra on the time-frequency map. The basic idea is that the variability of the instantaneous frequency introduced by the mode-splitting effect would be different in each set of added white noise. As a result, the true frequency evolution would be enhanced on the stacked map, and the spurious variability would be smeared out. The detailed mathematical proof and statistical study of this method will be presented in another paper (Yen et al.~in prep).

\section{Analysis of the LIGO O1/O2 Data}\label{sec:analysis_result}
\subsection{GW150914}
\begin{figure}
\includegraphics[width=0.4\textwidth]{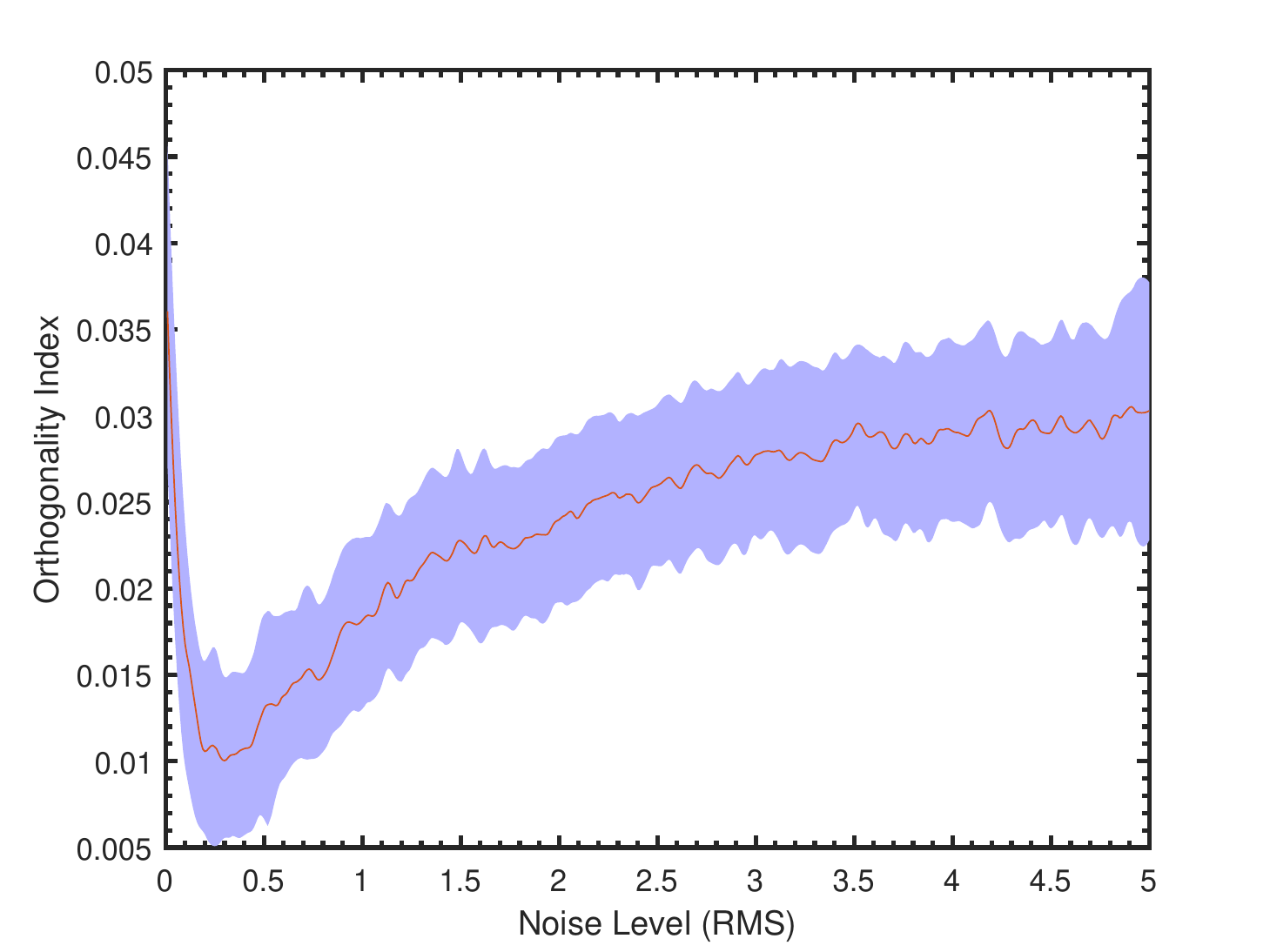}
\caption{ The orthogonality index versus EEMD's noise level in the Hanford data of GW150914 data. Blue filled area denotes the 1-sigma interval obtained from the simulated data sets. \label{fig:gw150914_oi}}
\end{figure}

We obtain the LIGO data and whiten them using the data reduction pipeline described in Section \ref{sec:Data_reduction}. \LD{We first input white noises with $\sigma=1.0$ times the deviation of the data and perform $10^4$ times of such simulations (step 1 in Figure \ref{fig:hht_with_eemd_procedure}). This is helpful to obtain confidence intervals because the whitened data can be treated as the Gaussian white noise except for a short time interval that contains GW signal. Then, we try both EMD and EEMD on the simulated time series (step 2 in Figure \ref{fig:hht_with_eemd_procedure}). Further input of noise in EEMD is needed, and we observe the variability of the OI with different noise levels between 0 and 5 times the deviation of the data (see Figure \ref{fig:gw150914_oi}). We found that the OI could be minimized with $\sigma\approx0.3$, and therefore we use it for the following analysis.} We then stack the Hilbert spectra on the 2-dimensional map to smear the boundary effect when the GW signal was split into two nearby IMFs. 

\begin{figure*}
\begin{minipage}{0.49\linewidth}
\includegraphics[width=1.01\textwidth]{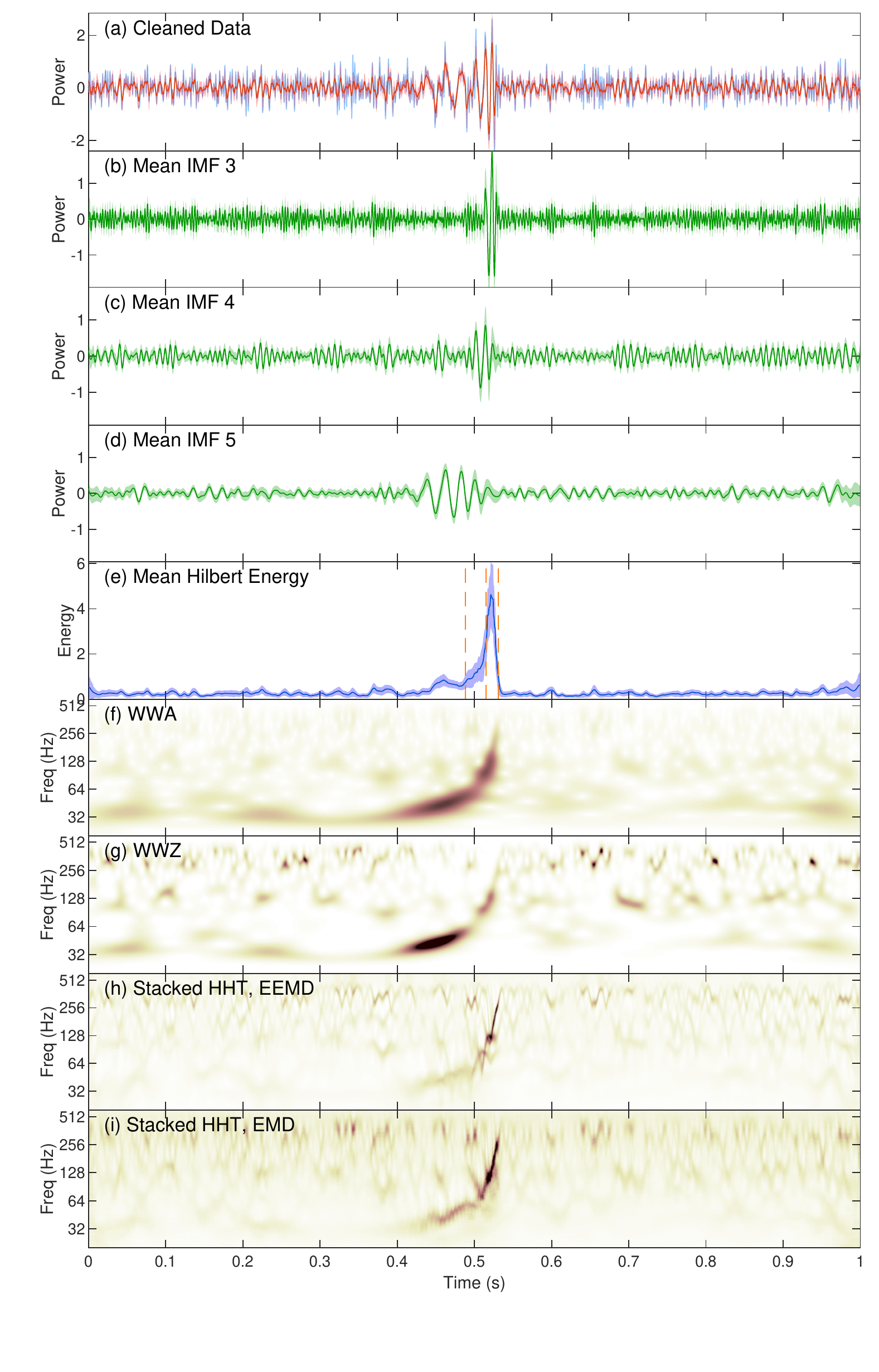}
\end{minipage}
\hspace{0.0cm}
\begin{minipage}{0.49\linewidth}
\includegraphics[width=1.01\textwidth]{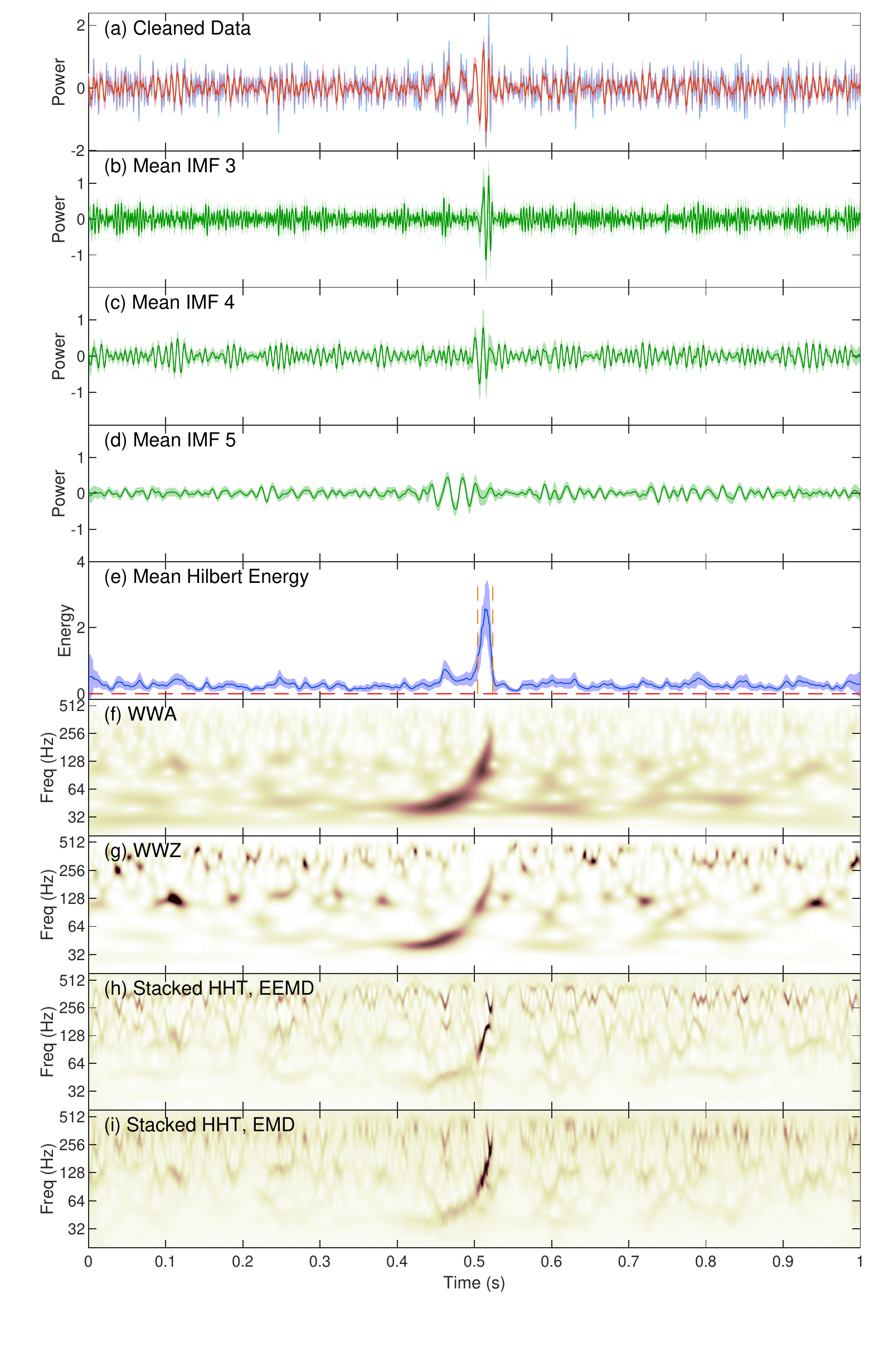}
\end{minipage}
\caption{The analysis results of GW150914 observed with LIGO Hanford (left) and Livingstone (right). (a) The whitened strain data (blue) overlaied with the EEMD band-pass filtered signal (red). (b)--(d) The EEMD decomposed IMF3--IMF5 with different input noise levels. The green shaded area denotes the 1-$\sigma$ interval among the 10000 simulations (e) The summation of the Hilbert amplitude of IMF3--IMF5. The vertical orange lines denote the time of significant change in the Hilbert amplitude (see main text). The horizontal dashed line denotes the zero level of the Hilbert amplitude. (f) WWA spectrum (g) WWZ spectrum (h) Stacked Hilbert spectrum with $10^4$ EEMD decomposed IMFs with randomly distributed noise level (i) Stacked Hilbert spectrum with $10^4$ EMD decomposed IMFs. \label{fig:gw150914_all}}
\end{figure*}

Figure \ref{fig:gw150914_all} shows the decomposition result and the Hilbert spectra of GW 150914 observed with LIGO Hanford (left) and Livingstone (right). Panel (a) shows the whitened data and the summation of IMF3--5. Because the GW signal spans from $\sim50$ Hz to $\sim300$\,Hz, it matches the characteristic time scales of IMFs 3 ($\sim 340$\,Hz)--IMF 5 ($\sim80$\,Hz) based on the fact that the EEMD is a dyadic filter and the sampling rate is 4096 Hz. In this case, the summation of IMFs 3--5 can greatly reproduce the waveform. This operation is equivalent to applying EEMD band-pass filter to the data. We plotted the decomposed IMFs 3--5 in Figure \ref{fig:gw150914_all} b--d, respectively. It clearly shows that the \LD{GW} signal firstly appears in IMF5, then migrates toward the high-frequency components, IMF4 and IMF3. 

We performed the normalized Hilbert transform to calculate the Hilbert energy, which is defined as the square of the amplitude, and instantaneous frequency of each IMFs, and stacked those obtained from IMFs 3--5 to minimize the contribution from other frequency ranges. This operation is done for all $10^4$ simulated data sets (see Figure \ref{fig:hht_with_eemd_procedure}) and thus we obtained the mean value and deviation of the band-pass filtered Hilbert energy (Figure \ref{fig:gw150914_all}e). This instantaneous Hilbert energy can be used to detect the GW signal. We use the Bayesian block technique to detect occurrence times of significant change in the Hilbert energy \citep{ScargleNJ2013}. GW150914 is the strongest event in the LIGO O1/O2 catalog where the boundaries can be easily detected if we set the null hypothesis probability $p_0$ to be $10^{-80}$.

The stacked Hilbert spectra with EEMD and EMD algorithms with $10^4$ times of simulation are presented in Figure \ref{fig:gw150914_all}h and \ref{fig:gw150914_all}i, respectively. We also performed the weighted wavelet analysis of the amplitude spectrum (WWA, Figure \ref{fig:gw150914_all}f) and the Z-transform spectrum (WWZ, Figure \ref{fig:gw150914_all}g) for comparison. \LD{The Morlet wavelet,} 
\begin{equation}
    f(z)=e^{-c\omega^2(t-\tau)^2}\left[e^{i\omega (t-\tau)-e^{-1/4c}} \right]
\end{equation}
\LD{is used in this analysis, where $\omega$ is the scale factor and $\tau$ is the time shift. The parameter $c$ determines how rapidly th wavelet decays, or the size of the Gaussian envelope.} We set it as $c=0.0125$ \citep{Foster1996}. These four algorithms yielded consistent results in the frequency evolution. However, the wavelet analysis has a trade-off between the frequency and the temporal resolutions by setting the width of the Gaussian window. The time-frequency property would be smeared out. In comparison, the Hilbert spectrum shows a strikingly good resolution on the time-frequency map and indicates the precise time of the incoming \LD{GW} signal. The mode-mixing may introduce a few nonphysical modulations when the signal crosses two IMFs. Except for this, HHT would be an ideal tool to study the detailed time-frequency structure of the \LD{GW} signals induced by two BHs coalescence. 

\begin{figure}
\includegraphics[width=0.95\linewidth]{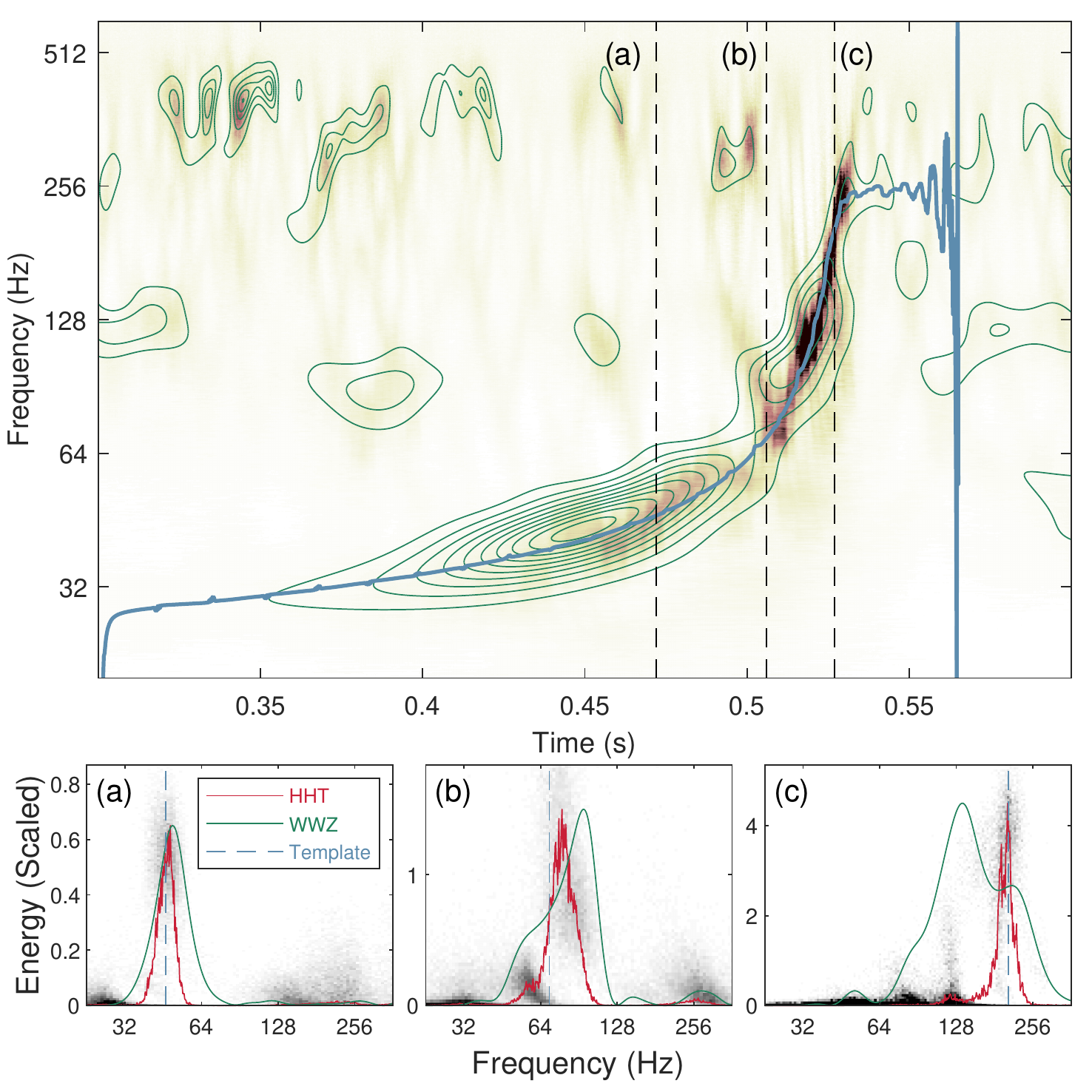}
\caption{Comparison between the stacked HHT spectrum (color map), WWZ spectrum (contour) and the instantaneous frequency of the template (blue line) of GW 150914 observed with LIGO Hanford detector. Three slices are labeled by the dashed line for low-frequency (a), intermediate frequency (b), and high-frequency (c) regimes. The amplitude versus frequency plot (or spectral density) are plotted in the lower panels. Density maps of the instantaneous amplitude and frequency of the specific time of all simulation are shown as black dots, where the red profile is the averaged profile for the instantaneous frequency. The green curve denotes the result from the WWZ, where the blue dashed line indicates the instantaneous frequency obtained from the template. \LD{The HHT and WWZ profiles in panels (a)--(c) are scaled for display purpose.} \label{wwz_vs_HHT_vs_template}}
\end{figure}

To compare the analysis result between different algorithms and the model template, we \LD{simulated the GW waveform using the python-based open software to explore astrophysical sources of GWs, \texttt{PyCBC}v2.0.1\footnote{\url{https://github.com/gwastro/pycbc}} \citep{UsmanNH2016}, with the approximant given by the effective-one-body \citep[EOB;][]{TaracchiniBP2014,Purrer2016,BoheST2017} NR model.} The instantaneous frequency of the template can be directly calculated via Hilbert transform without decomposing it into IMFs because the template fits the IMF criterion during the merging phase. The resulting instantaneous frequency of the template, the WWZ spectrum, and the stacked Hilbert spectrum are plotted in Figure \ref{wwz_vs_HHT_vs_template}. The time-frequency map of the HHT can clearly describe the instantaneous frequency in better detail compared to that of the wavelet spectrum. To demonstrate this, we choose three slices in (a) low-frequency, (b) intermediate-frequency, and (c) high-frequency regimes. The \LD{slices of the HHT and WWZ spectra}, as well as the instantaneous frequency obtained from the template, are shown in panel (a)--(c) of Figure \ref{wwz_vs_HHT_vs_template}. The gray color map denotes the density of the Hilbert energy-frequency pairs of each trial. The HHT result at epoch (b) is most ambiguous because the \LD{GW} signal switches from one IMF to another at this time. However, the performance of the HHT remains comparable to that of the WWZ. 

GW150914 is one of the strongest signals recorded in GWTC-1. We analyzed all the BH-BH coalescence events in GWTC-1 and present the results in the appendix (See Figure \ref{fig:gw151012_all} -- \ref{fig:gw170823_all}). These signals have been analyzed in a combination with the HHT and de-noise algorithm using a network array \citep{AkhshiAB2021}. The instantaneous frequency can be detected but large variability in the instantaneous frequency is occasionally seen \citep[see, e.g., Figure 4 in][]{AkhshiAB2021}. These signals could be minimized by our stacking method. 

\begin{figure*}
\begin{minipage}{0.49\linewidth}
\includegraphics[width=1.01\textwidth]{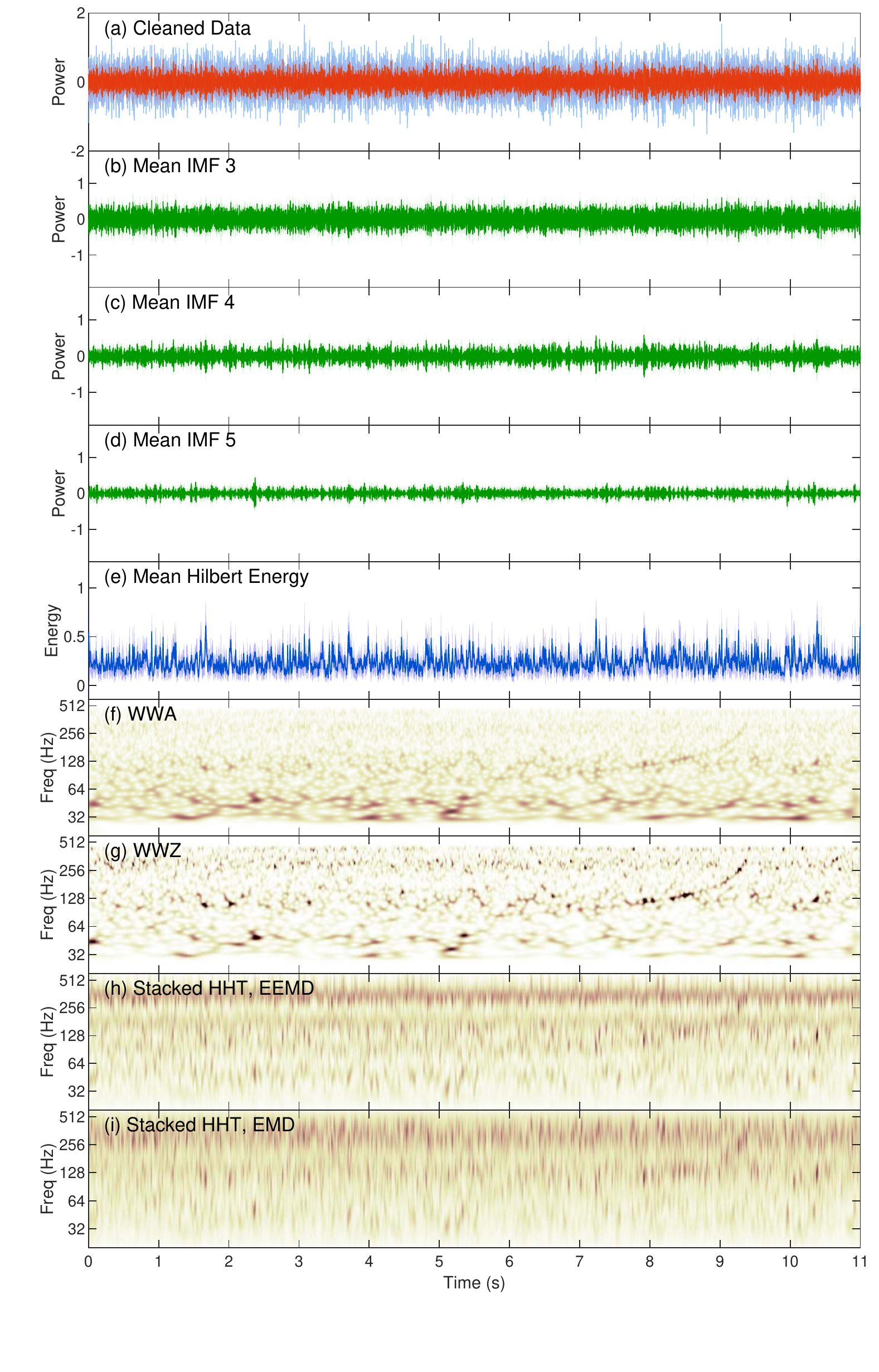}
\end{minipage}
\hspace{0.0cm}
\begin{minipage}{0.49\linewidth}
\includegraphics[width=1.01\textwidth]{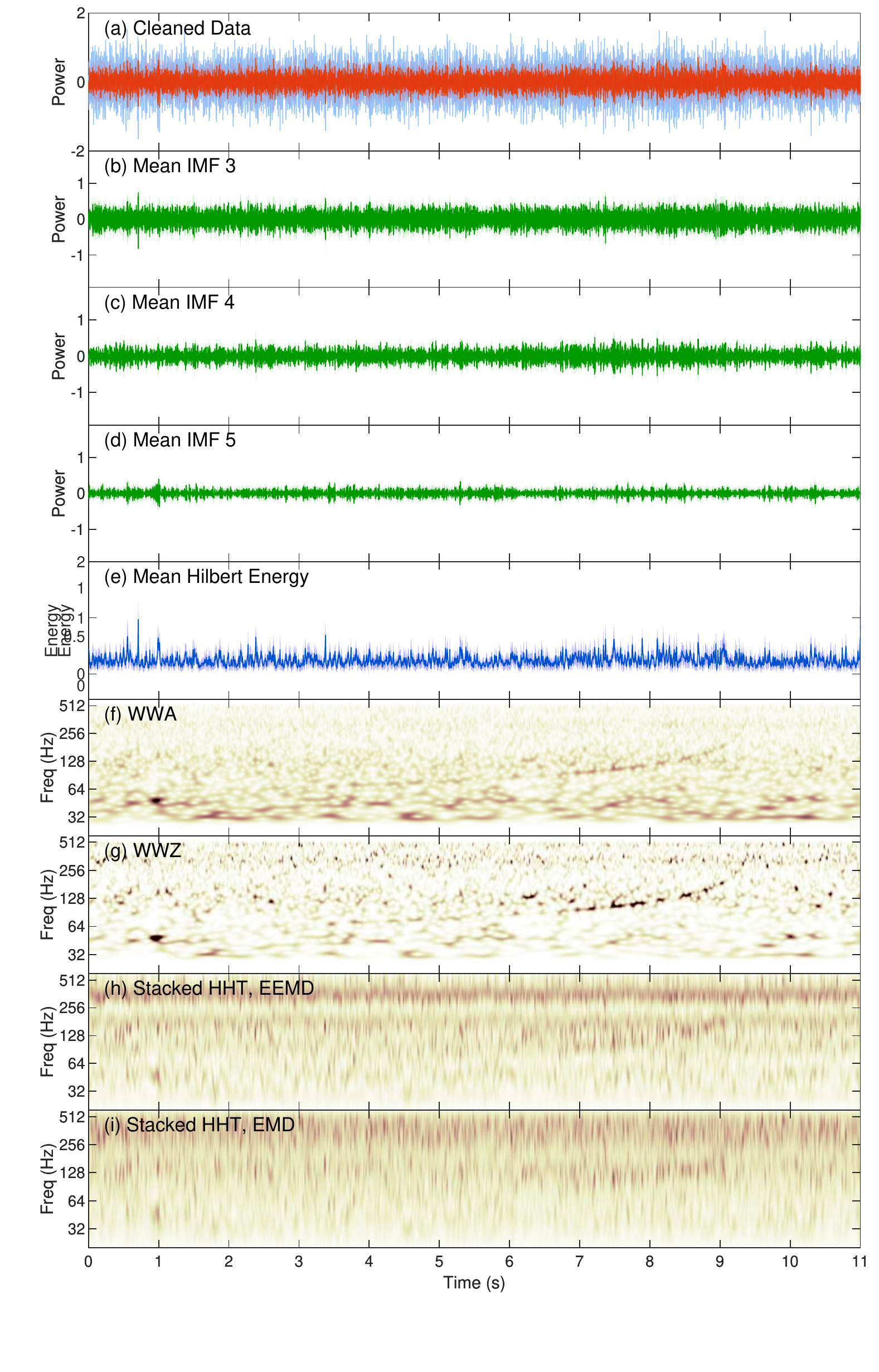}
\end{minipage}
\caption{The analysis results of GW170817 observed with LIGO Hanford (left) and Livingstone (right). Panel definitions are the same as those in Figure \ref{fig:gw150914_all}. 
\label{fig:gw170817_all}}
\end{figure*}

\subsection{GW170817}
GW170817 is the first confirmed NS-NS coalescence event \citep{AbbottAA2017_GW170817}. 
A $\gamma$-ray burst and multi-wavelength afterglows is a benchmark of the multi-messenger astronomy era \citep{AbbottAA_gamma_ray2017}. 
Compared to other BH-BH coalescence obtained in GWTC-1 \citep{AbbottAA2019}, the signal of GW170817 has a much longer coherence time but the amplitude is much weaker. 
We tested both the WWZ and the HHT algorithms on this event. 
The signal can be seen on the wavelet map with $c=0.00125$, ten times smaller than the $c$ value for BH-BH coalescence events. Compared to the signals of BH-BH events, the amplitude of the NS-NS signal is much lower, and the signal keeps its coherence for a long while. The nearly coherent signal before the merging makes it possible to be detected with this setting, but the time-frequency property near the merging phase is smeared in time due to the large Gaussian window. In contrast, the signal is difficult to be resolved in the HHT time-frequency map. 
Because the HHT is sensitive to both the instantaneous variability of the frequency and the noise, the signal-to-noise ratio of this event is insufficient for the HHT to identify its intrinsic time-frequency property. Therefore, the HHT could be powerful for future NS-NS coalescence events with improved \LD{GW} observatories in the next generation that could have a high signal-to-noise ratio.

\section{Core-Collapse Supernovae}\label{sec:CCSNe}
Detection of GW signals from a nearby CCSN will be the next milestone of gravitational wave astrophysics \citep{AbbottAA2016c, AbbottAA2020}.
It would be crucial for multi-messenger astronomy because 
the possible co-detection of gravitational waves and neutrinos would allow astronomers to probe the activity of the center of a massive star during the final evolutionary stage. Thanks to the growth of computing power, our knowledge of the explosion engine of a CCSN improved dramatically. Owing to the asymmetric and complex fluid motion and the oscillation/evolution of the proto-neutron star (PNS), complex gravitational signals with multiple components are expected to be investigated from three-dimensional high-fidelity self-consistent CCSN simulations \citep{JankaMS2016, BurrowsV2021}. CCSN waveforms depend on several physical quantities, such as the core angular momentum and the composition of the collapsar, the nuclear equation of state, and the dynamics of fluid instabilities. The future multi-messenger observations will place meaningful constraints on the supernova engines and the microphysics.

Nevertheless, recent theoretical works have reached consensuses on a few main structures of the GW signal on the time-frequency domain. 
These provide a strong basis to improve searching algorithms. On the other hand, the HHT is likely the most precise tool to investigate the detailed time-frequency properties of the CCSNe. Here we adopted the gravitational waveforms from the self-consistent three-dimensional CCSN simulations with neutrino transport in \citet{PanLC2021}. 

\subsection{Waveforms and Simulations}

The gravitational waveforms of CCSNe simulations used in this project are adopted from the three-dimensional self-consistent CCSN simulations with the Isotropic Diffusion Source Approximation \citep{LiebendoerferWF2009} for the neutrino transport, using the FLASH code \citep{FryxellOR2000} and are presented in \citet{PanLC2021}.
The data are derived from a $40 M_\odot$ progenitor at zero-age main sequence with solar metallicity from \citet{WoosleyH2007} with three different initial rotational speeds ($\Omega_0= 0, 0.5$ and 1~rad~s$^{-1}$) at the onset of core-collapse.
A post-Newtonian treatment with the effective general relativity potential \citep{MarekDJ2006} and the Lattimer$-$Swesty (LS220) equation of state are considered \citep{LattimerS1991}. 

The GW strains are evaluated by the first time derivative of the mass quadrupole moment with an assumption of a fixed distance $d= 10$~kpc to an observer.  
The second time derivative of the mass quadrupole moment is calculated by the finite difference in post-processing, using the formula described in \citet{ScheideggerFW2008} and \citet{MurphyOB2009}. We simulated two viewing angles, where the observers are in the infinity in the direction of the equatorial plane or the rotational axis. 

\subsection{Analysis of Simulated CCSNe Waveforms}
\begin{figure}
    \centering
    \includegraphics[width=0.49\textwidth]{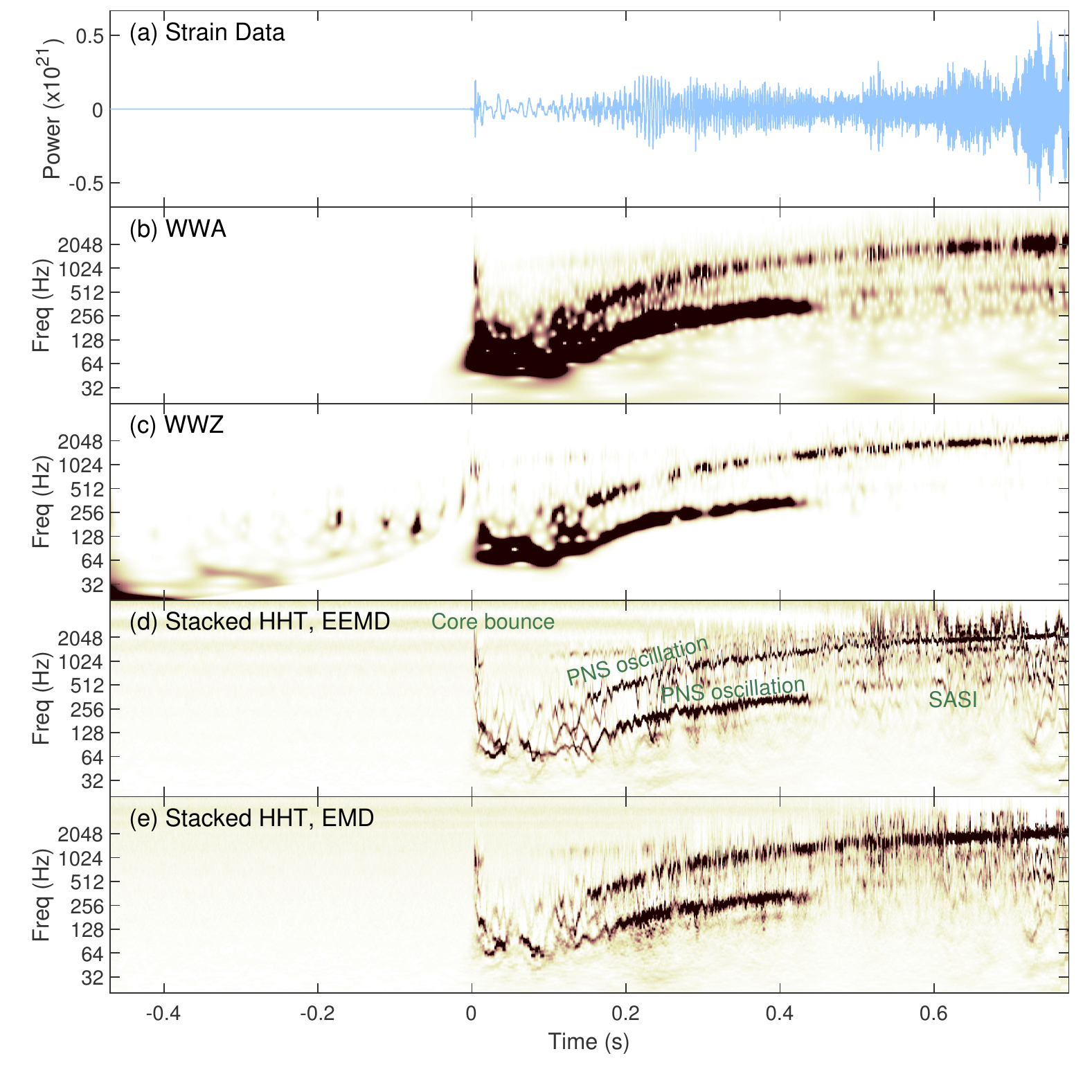}
    \caption{(a) The cross-component of the gravitational waveform of a CCSN seen from the equatorial plane with an initial stellar mass of $40$~$M_{\odot}$ and without rotation. (b)--(c) The WWA and WWZ map of the gravitational signal. (d)--(e) Stacked Hilbert spectra based on EEMD (d) and EMD (e). }
    \label{fig:ccsn_nor_CrossEquator}
\end{figure}

Here we take the gravitational waveform of the non-rotating model with an observer viewed from the equatorial direction as an example for our HHT analysis. The gravitational waveforms of other models with different parameters are shown in the appendix (Figure \ref{fig:ccsn_nor_CrossPole} -- \ref{fig:ccsn_fr_PlusPole}). 
Figure \ref{fig:ccsn_nor_CrossEquator} (a) shows the simulated strains, which followed by the wavelet spectra (b) -- (c) and the stacked HHT spectra (d) -- (e). 
The waveform starts from the core collapse and reaches the core bounce at $t=0$~s. Since this model is non-rotating and the proto-neutron star convection happens after the core bounce, the gravitational strain shows a loud bounce signal at time zero, and then it is followed by a low-frequency peak PNS oscillation, which could be caused by the g-mode oscillation although the nature remains controversial \citep{MullerJM2013, PanLC2021}.  Multiple low-frequency oscillation signals dominate the first $\sim0.15$ s and they can be well resolved with the HHT. In comparison, the limitation of the wavelet spectrum prevents us to identify them.  As the frequency of the oscillation increases with time, another high-frequency mode oscillation signal appears with similar strength as the peak PNS mode. The low-frequency PNS oscillation vanishes at $t\sim0.5$ s. Then, interestingly, it followed by the quadruple radiation of the standing accretion shock instability \citep[SASI][]{BlondinMD2003} signal, \LD{which was suggested to be seen in the GW data} \cite{KurodaKT2016, PanLC2021}. \LD{The detailed waveform of the SASI signal is model dependent, but its time-frequency property has been shown in a few analysis either with the spectrogram or the HHT \citep[see, e.g.,][]{AndresenMM2017, MezzacappaML2020, TakedaHK2021}}. In our analysis, the SASI signal can be seen on both the wavelet and the HHT spectra. The wavelet and the HHT results for other simulated CCSN events are shown in the appendix \ref{sec:appendix_CCSNe}. The Hilbert spectra of all the data sets show much better resolution than the wavelet spectra. 

\section{Discussion}

A \LD{Morlet wavelet consists of a complex exponential carrier multiplied by a Gaussian window}. It provides a window size for both the low- and high-frequency regime that contains the same number of waves. The time-frequency map is distorted, and the signal is smeared if the frequency of a signal changes drastically within a few cycles (see. e.g., Figure \ref{fig:gw150914_all} (f) -- (g) and Figure \ref{fig:ccsn_nor_CrossEquator} (b) -- (c)). A narrower width of the Gaussian envelope can result in a better time resolution, although the frequency resolution would be sacrificed. 

On the other hand, the Hilbert analysis calculates the instantaneous frequency of a signal. This helps trace the signal's detailed evolution of the time-frequency property. Moreover, the complex multi-component, low-frequency signals of CCSNe can be well resolved with the HHT. For example, the SASI signal in Figure \ref{fig:ccsn_nor_CrossEquator} and a few multiple low-frequency signals in Figures \ref{fig:ccsn_nor_PlusPole} -- \ref{fig:ccsn_fr_PlusPole} can only be clearly resolved in the Hilbert spectra. 

Nevertheless, the HHT can decompose a fast-changing signal into different IMFs. Fake frequency modulation in the instantaneous frequency is seen when the signal crosses two components. This can be reduced with the stacking algorithm suggested in this study, although they cannot be entirely removed. Moreover, the signal cannot be seen for low signal-to-noise ratio events. For example, the NS-NS coalescence event GW 170817 can be clearly detected with the wavelet analysis with a large Gaussian width. This is because the signal keeps its coherence for a few more cycles compared to the BH-BH coalescence events, but the signal-to-noise ratio within a cycle is much lower. 

Before stacking the time-frequency map, the evolution of Hilbert energy has gave hints of the GW signal. A detailed time delay among instruments can be further characterized by cross-correlating the EMD filtered profiles \citep{AkhshiAB2021}. In addition, the special pattern of the GW signal across IMFs (see, e.g., Figure \ref{fig:gw150914_all}b -- d) would form a remarkable feature in the two-dimensional series where one dimension is time and another is IMF index. This would be helpful for detecting the GW signal using convolution neural network. 

The GW signal from a CCSN is believed to contain multiple components, where they could across a large range in the both the frequency and the energy. The high resolutions of the HHT in both the time and frequency domains are helpful in distinguishing them from each other, especially the low-frequency transient signals within $\sim0.4$ s after the core bounce. 

\section{Summary}
The HHT is a novel technique for time-frequency analysis although the mode-splitting problem is severe when working with signals with huge frequency variability like GW signals. We propose a method that could minimize the unreal frequency modulation by stacking the time-frequency map instead of taking the ensemble mean of the IMFs. This helps visualize the evolution of the instantaneous frequency. We carried out an analysis for LIGO O1 and O2 events and found that most of their frequency evolution can be well tracked with our algorithm. Furthermore, the EEMD filtered data is used to determine the time delay between Livingstone and Hanford sites. Except for the known GW sources, we also tested our algorithm on the simulated CCSN events. We found that multiple GW components can be obtained with the HHT. Compared to the wavelet analysis, the unprecedented resolutions in both the time and frequency domains of the HHT are an ideal tool for further investigating the time-frequency properties of a signal. This provides evidence that the HHT can be a powerful tool in detecting CCSNe events for future multi-messenger campaigns. 

\begin{acknowledgments}
We thank the anonymous reviewer for valuable comments that improved this paper. This research has made use of data, software and/or web tools obtained from the Gravitational Wave Open Science Center (\url{https://www.gw-openscience.org}), a service of LIGO Laboratory, the LIGO Scientific Collaboration and the Virgo Collaboration. LIGO is funded by the U.S. National Science Foundation. Virgo is funded by the French Centre National de Recherche Scientifique (CNRS), the Italian Istituto Nazionale della Fisica Nucleare (INFN) and the Dutch Nikhef, with contributions by Polish and Hungarian institutes. C.-P.H.~acknowledges support from the Ministry of Science and Technology in Taiwan through grant MOST 109-2112-M-018-009-MY3.
L.C.C.L. acknowledges support from the Ministry of Science and Technology in Taiwan through grant Nos. MOST 110-2811-M-006-515, \LD{MOST 110-2811-M-006-012} and MOST 110-2112-M-006-006-MY3.
K.C.P. is supported by the Ministry of Science and Technology of Taiwan through grant MOST 110-2112-M-007-019.
K.L.L. is supported by the Ministry of Science and Technology of the Republic of China (Taiwan) through grant 110-2636-M-006-013, and he is a Yushan Young Fellow of the Ministry of Education of the Republic of China (Taiwan).
C.-C.Y. is supported by the Ministry of Science and Technology of Taiwan through grant MOST 
109-2115-M-030 -004 -MY2. A.K.H.K. is supported by the Ministry of Science and Technology of Taiwan under the grants 109-2628-M-007-005-RSP and 110-2628-M-007-005, and the Inter-University Research Program of the Institute for Cosmic Ray Research (ICRR), the University of Tokyo. 
CYH is supported by the National Research Foundation of Korea through grants 2016R1A5A1013277.
\end{acknowledgments}

\software{Matlab \citep{matlab2018b}, Python3 \citep{python3}, GWpy \citep{MacleodUC2020}, PyCBC \citep{UsmanNH2016}}

\appendix
\setcounter{table}{0}
\setcounter{figure}{0}
\renewcommand{\thetable}{A\arabic{table}}
\renewcommand\thefigure{A\arabic{figure}} 

\section{Resulting Spectra of LIGO Events} \label{sec:appendix_LIGO}
All the remaining LIGO O1/O2 events and corresponding time-frequency maps of BH-BH mergers are shown in this section.

\begin{figure*}
\begin{minipage}{0.49\linewidth}
\includegraphics[width=1.01\textwidth]{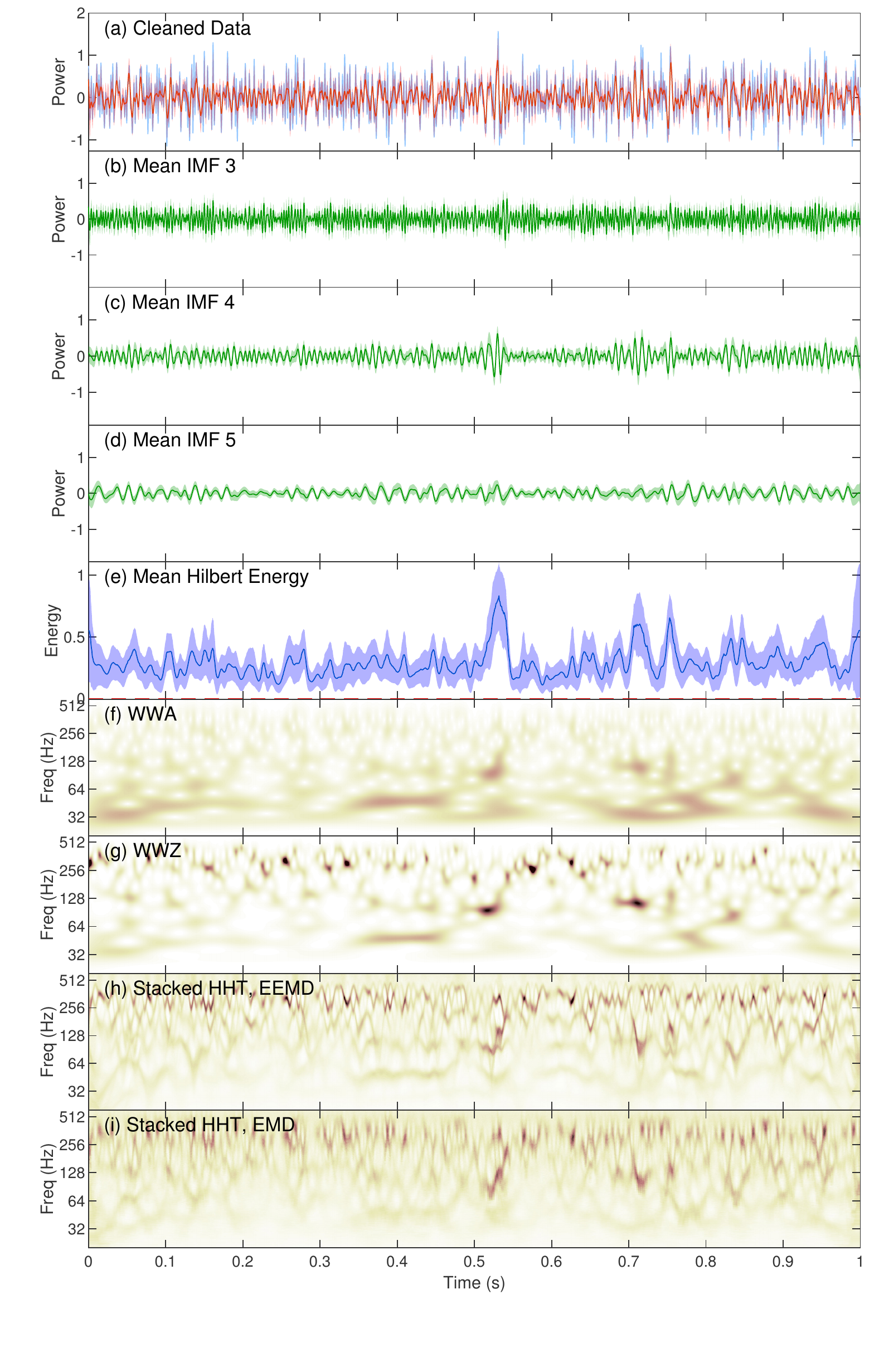}
\end{minipage}
\hspace{0.0cm}
\begin{minipage}{0.49\linewidth}
\includegraphics[width=1.01\textwidth]{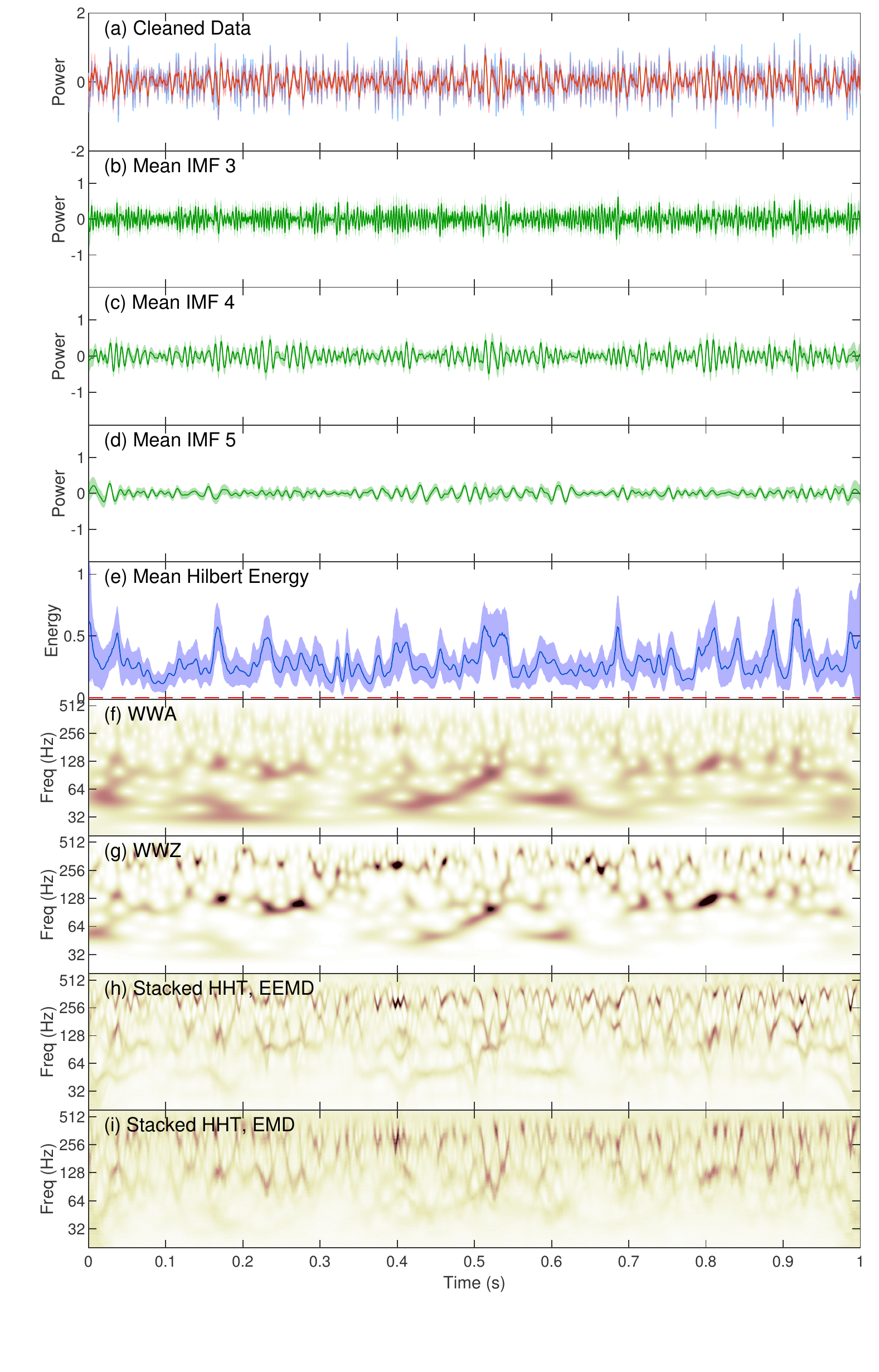}
\end{minipage}
\caption{The analysis results of GW151012 observed with LIGO Hanford (left) and Livingstone (right). Panels (a) -- (i) have the same definition as those in Figure \ref{fig:gw150914_all}. \label{fig:gw151012_all}}
\end{figure*}

\begin{figure*}
\begin{minipage}{0.49\linewidth}
\includegraphics[width=1.01\textwidth]{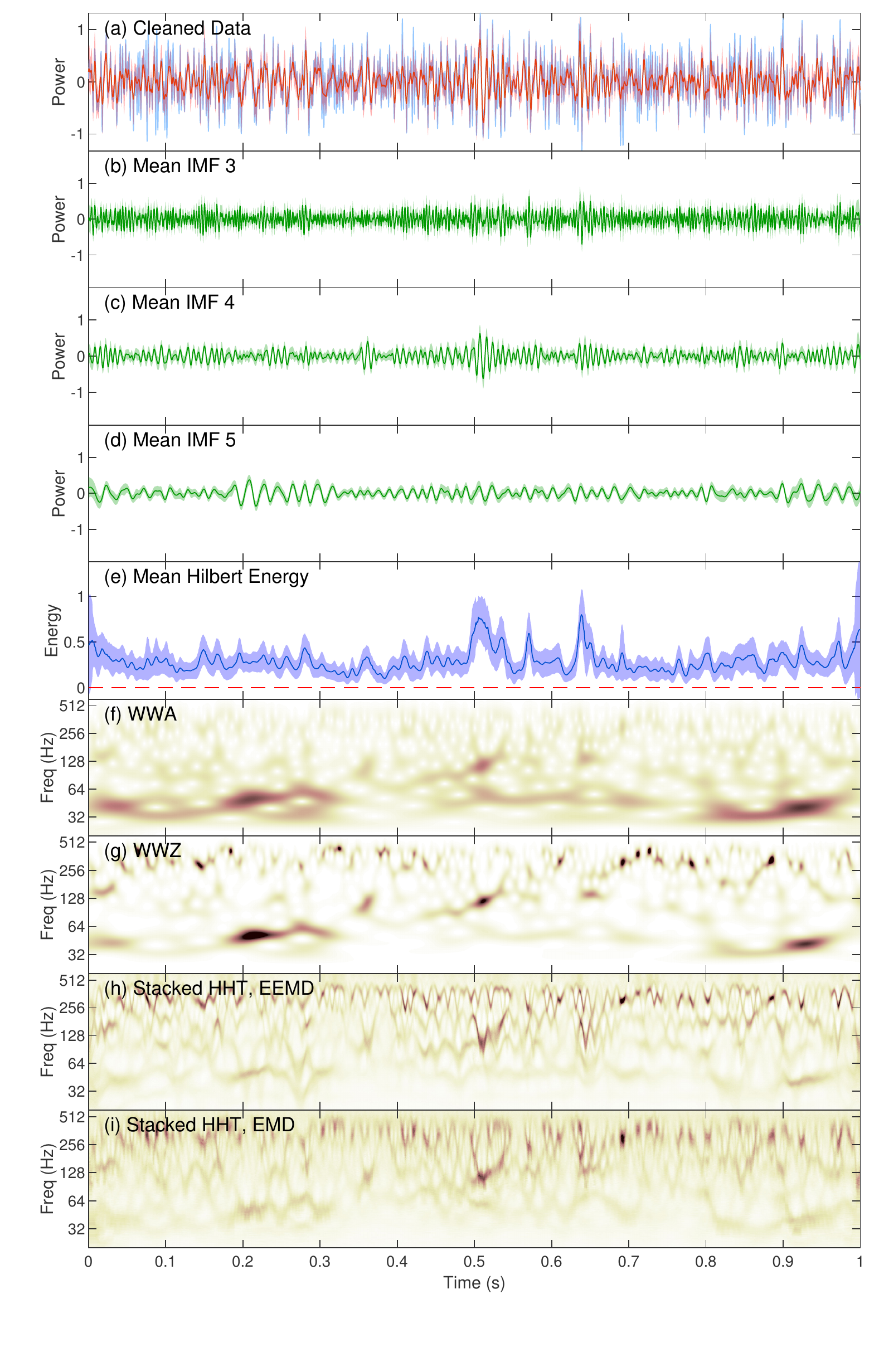}
\end{minipage}
\hspace{0.0cm}
\begin{minipage}{0.49\linewidth}
\includegraphics[width=1.01\textwidth]{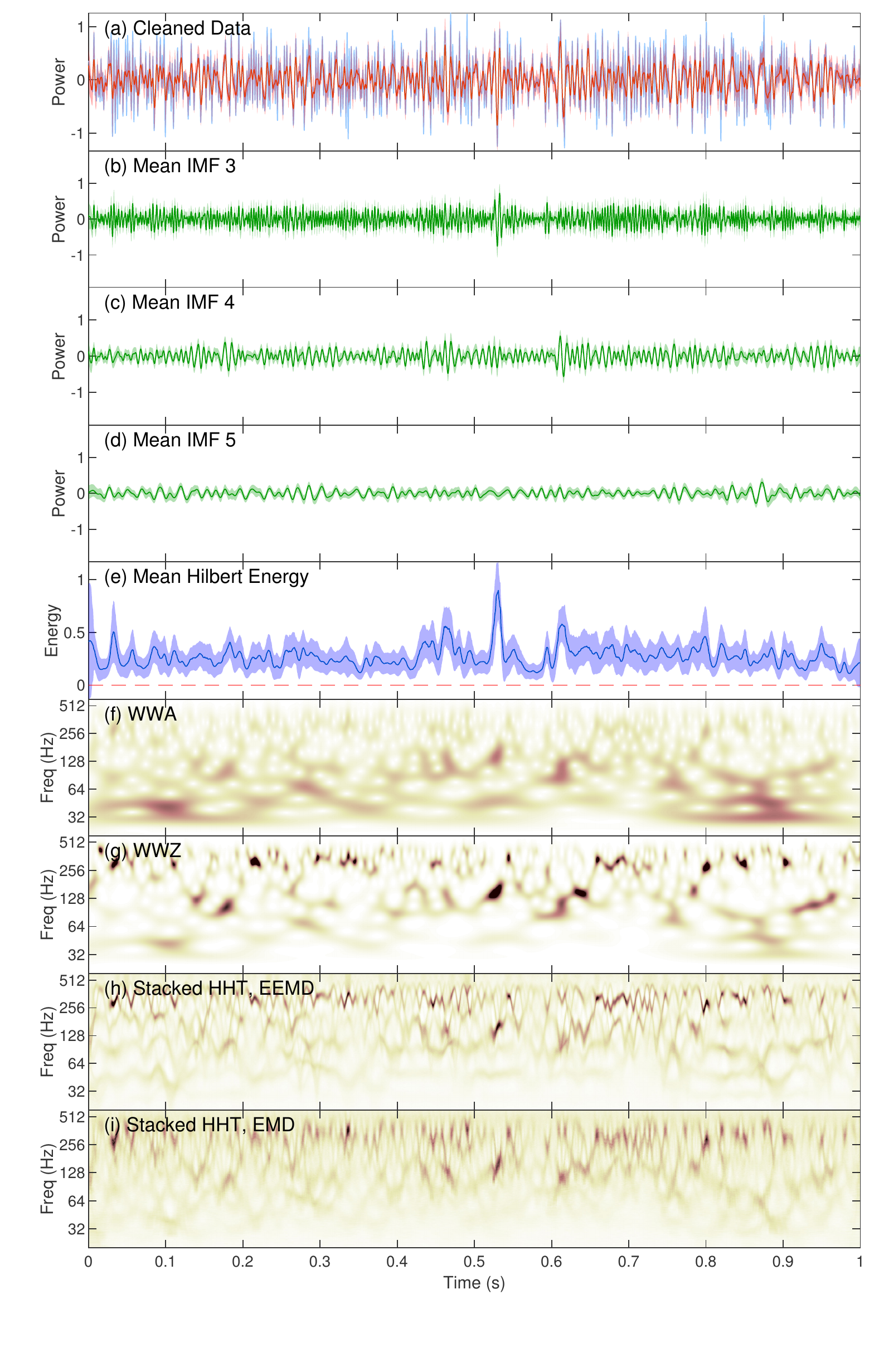}
\end{minipage}
\caption{The analysis results of GW151226 observed with LIGO Hanford (left) and Livingstone (right). Panels (a) -- (i) have the same definition as those in Figure \ref{fig:gw150914_all}. \label{fig:gw151226_all}}
\end{figure*}

\begin{figure*}
\begin{minipage}{0.49\linewidth}
\includegraphics[width=1.01\textwidth]{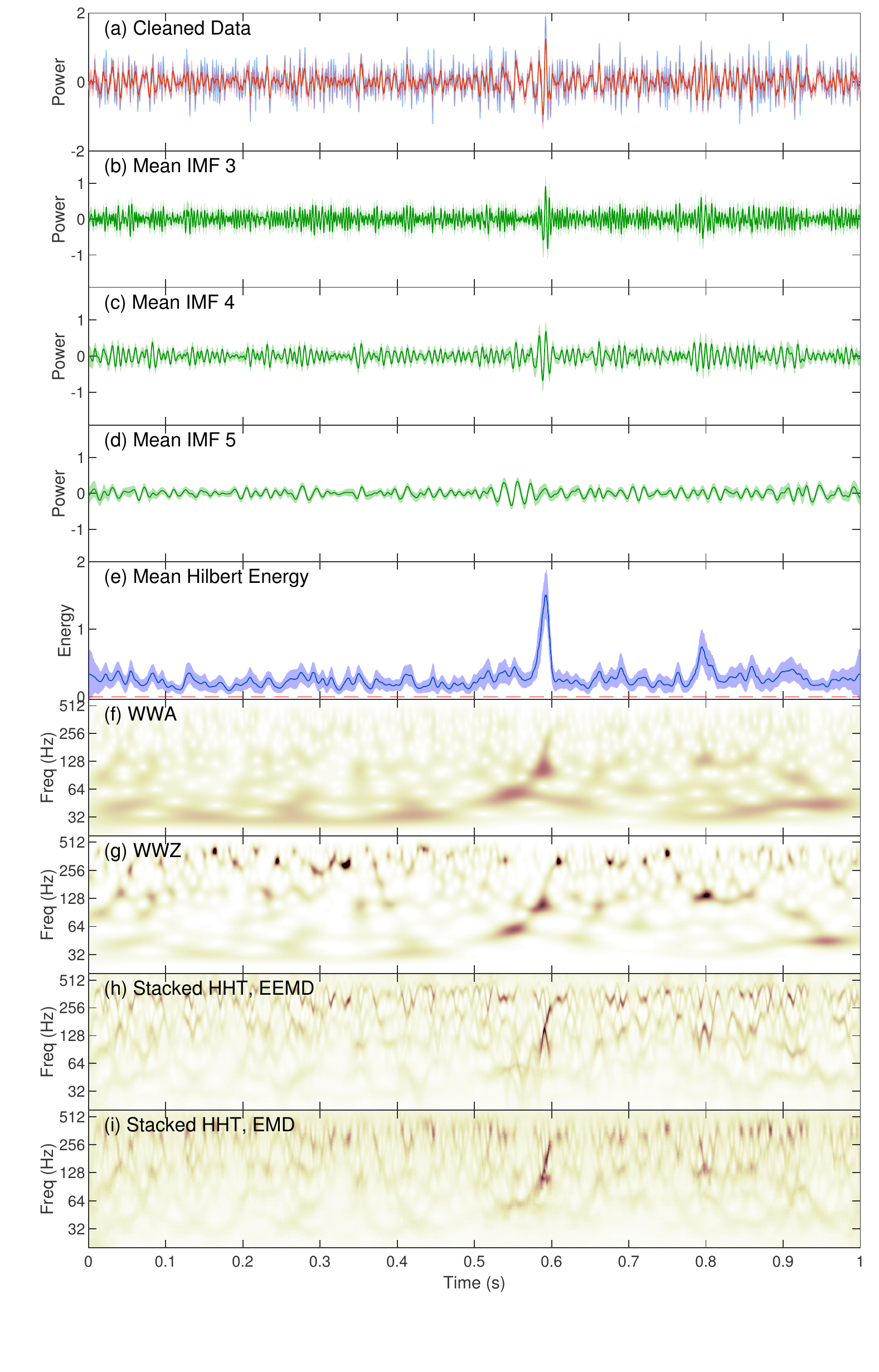}
\end{minipage}
\hspace{0.0cm}
\begin{minipage}{0.49\linewidth}
\includegraphics[width=1.01\textwidth]{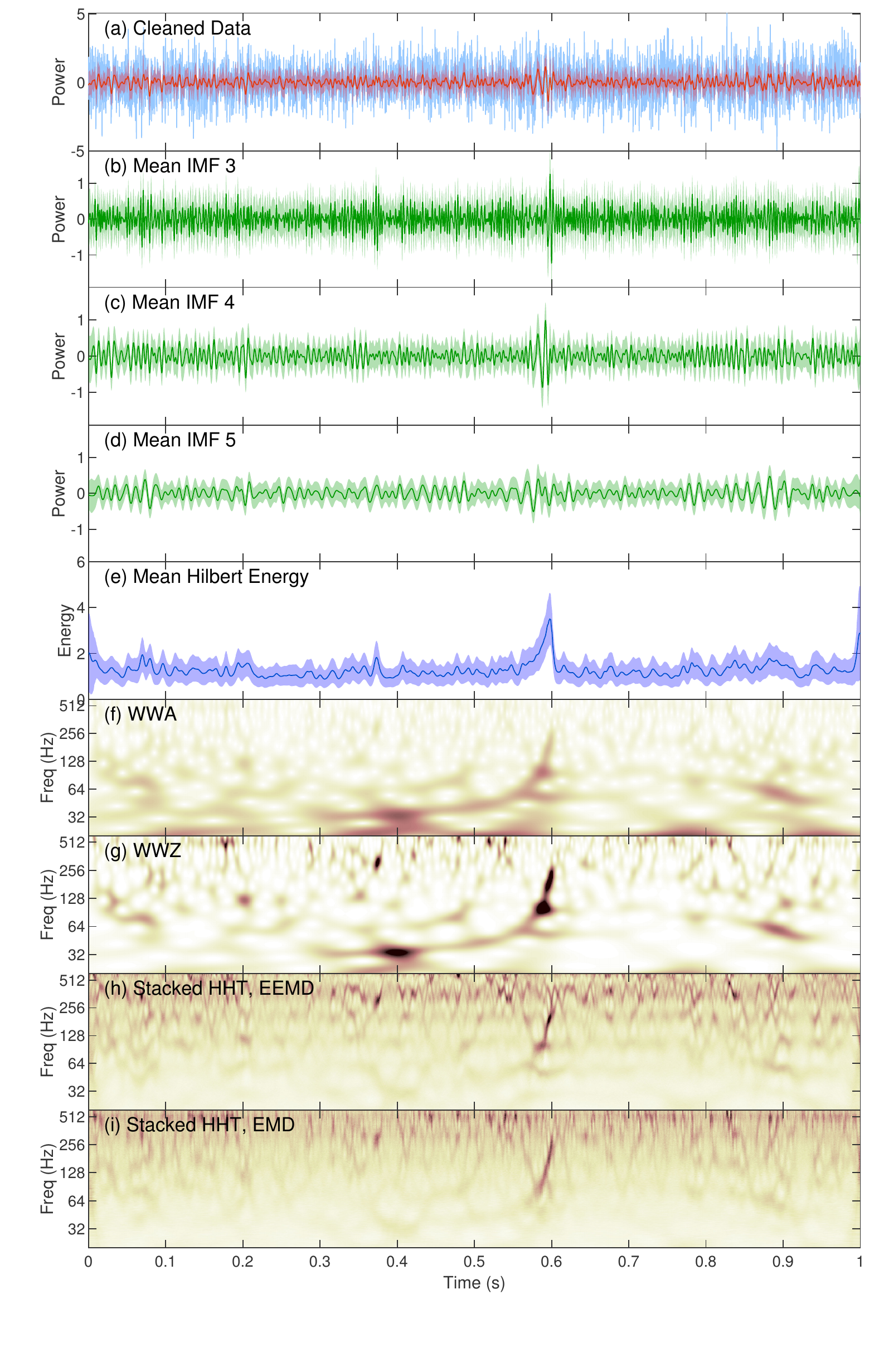}
\end{minipage}
\caption{The analysis results of GW170104 observed with LIGO Hanford (left) and Livingstone (right). Panels (a) -- (i) have the same definition as those in Figure \ref{fig:gw150914_all}. \label{fig:gw170104_all}}
\end{figure*}

\begin{figure*}
\begin{minipage}{0.49\linewidth}
\includegraphics[width=1.01\textwidth]{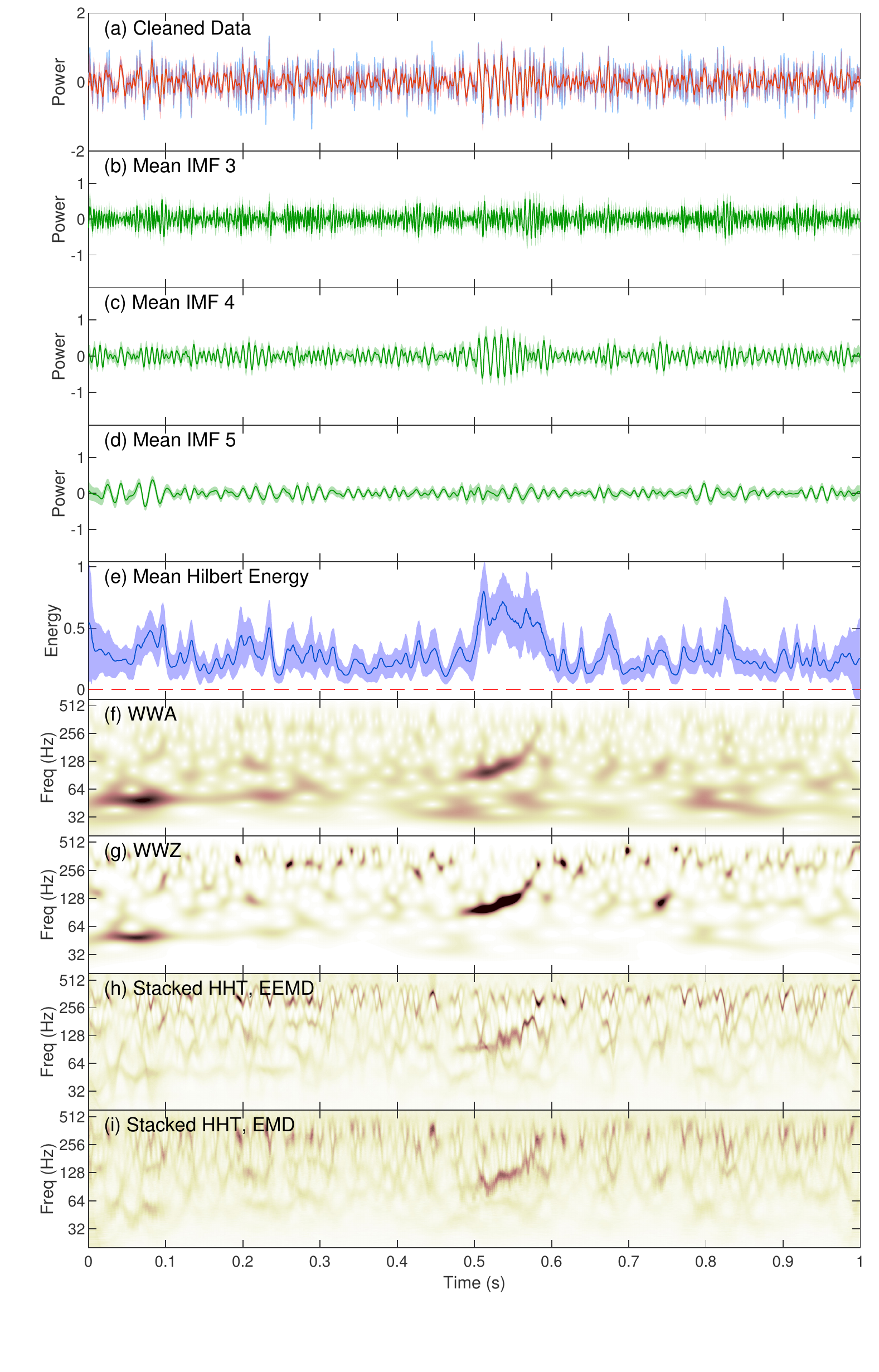}
\end{minipage}
\hspace{0.0cm}
\begin{minipage}{0.49\linewidth}
\includegraphics[width=1.01\textwidth]{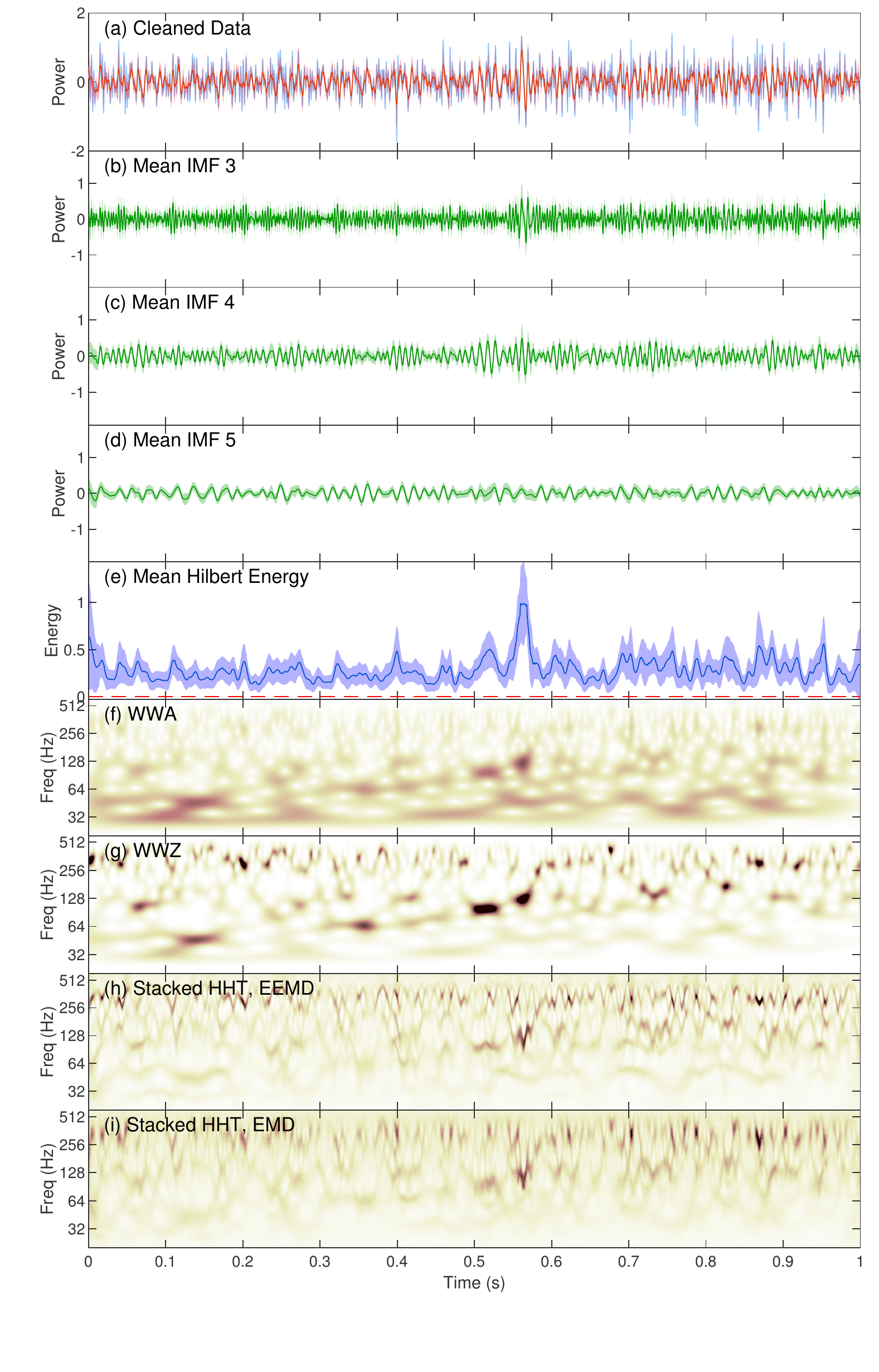}
\end{minipage}
\caption{The analysis results of GW170608 observed with LIGO Hanford (left) and Livingstone (right). Panels (a) -- (i) have the same definition as those in Figure \ref{fig:gw150914_all}. \label{fig:gw170608_all}}
\end{figure*}

\begin{figure*}
\begin{minipage}{0.49\linewidth}
\includegraphics[width=1.01\textwidth]{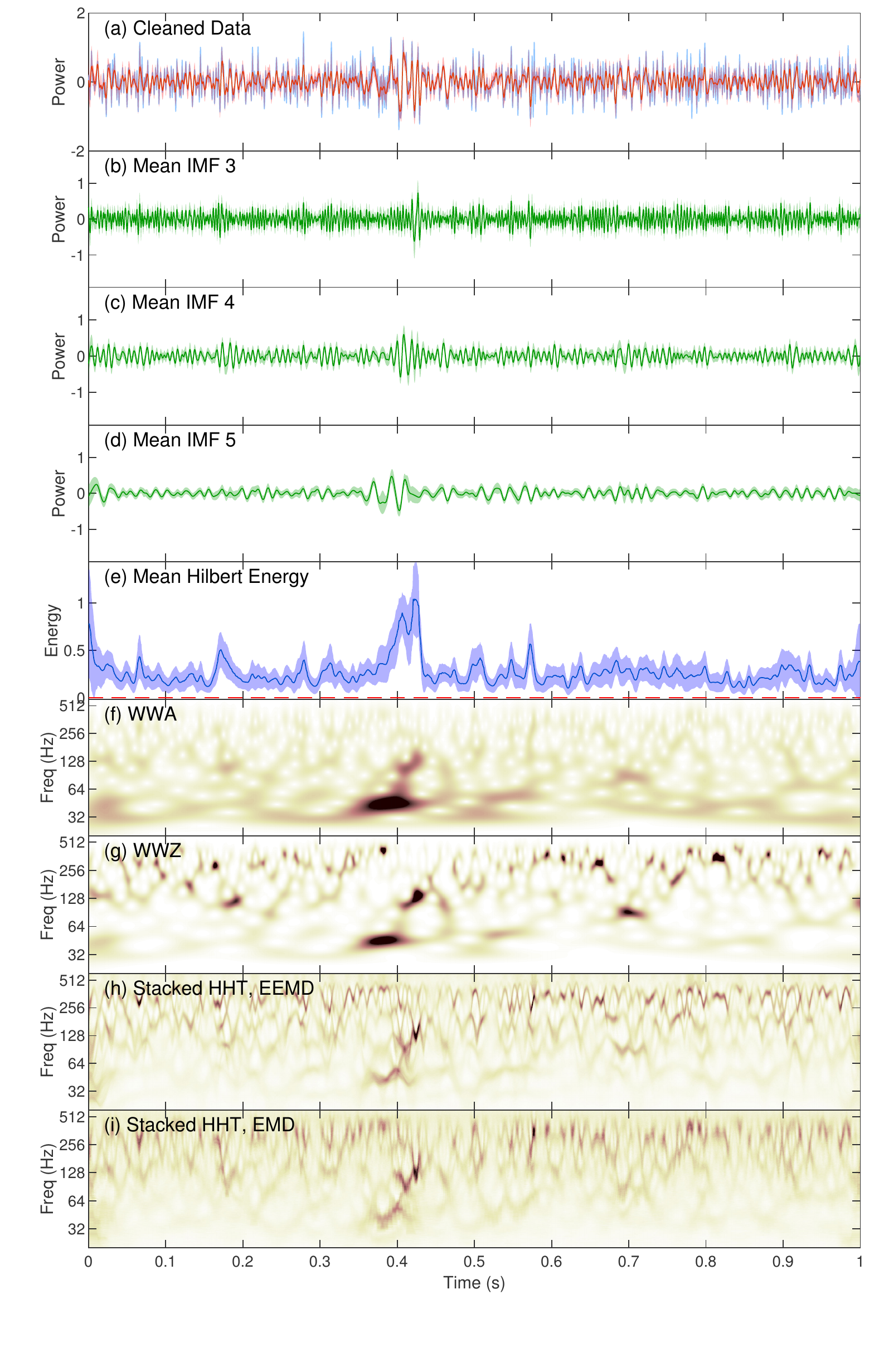}
\end{minipage}
\hspace{0.0cm}
\begin{minipage}{0.49\linewidth}
\includegraphics[width=1.01\textwidth]{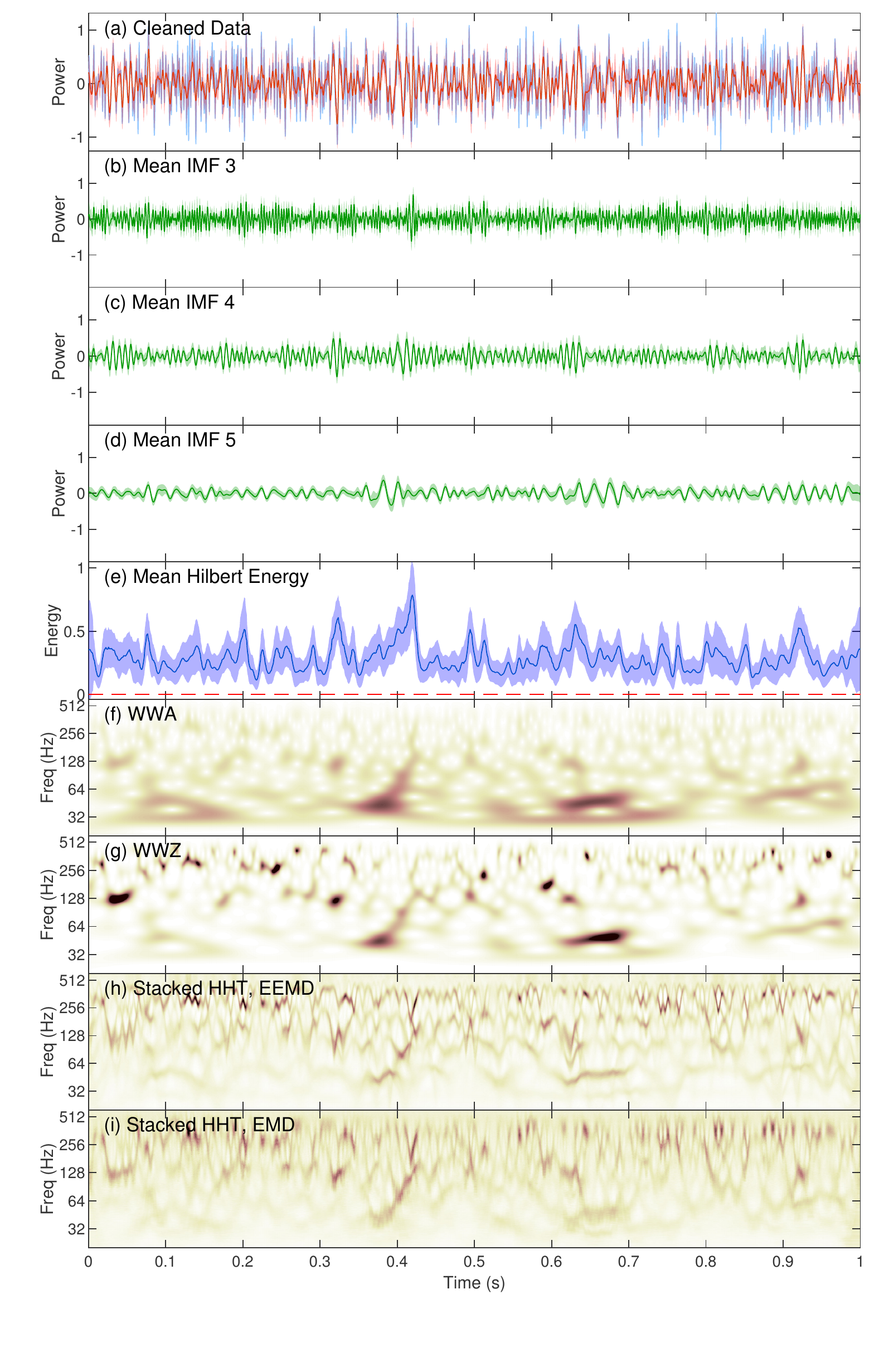}
\end{minipage}
\caption{The analysis results of GW170729 observed with LIGO Hanford (left) and Livingstone (right). Panels (a) -- (i) have the same definition as those in Figure \ref{fig:gw150914_all}. \label{fig:gw170729_all}}
\end{figure*}

\begin{figure*}
\begin{minipage}{0.49\linewidth}
\includegraphics[width=1.01\textwidth]{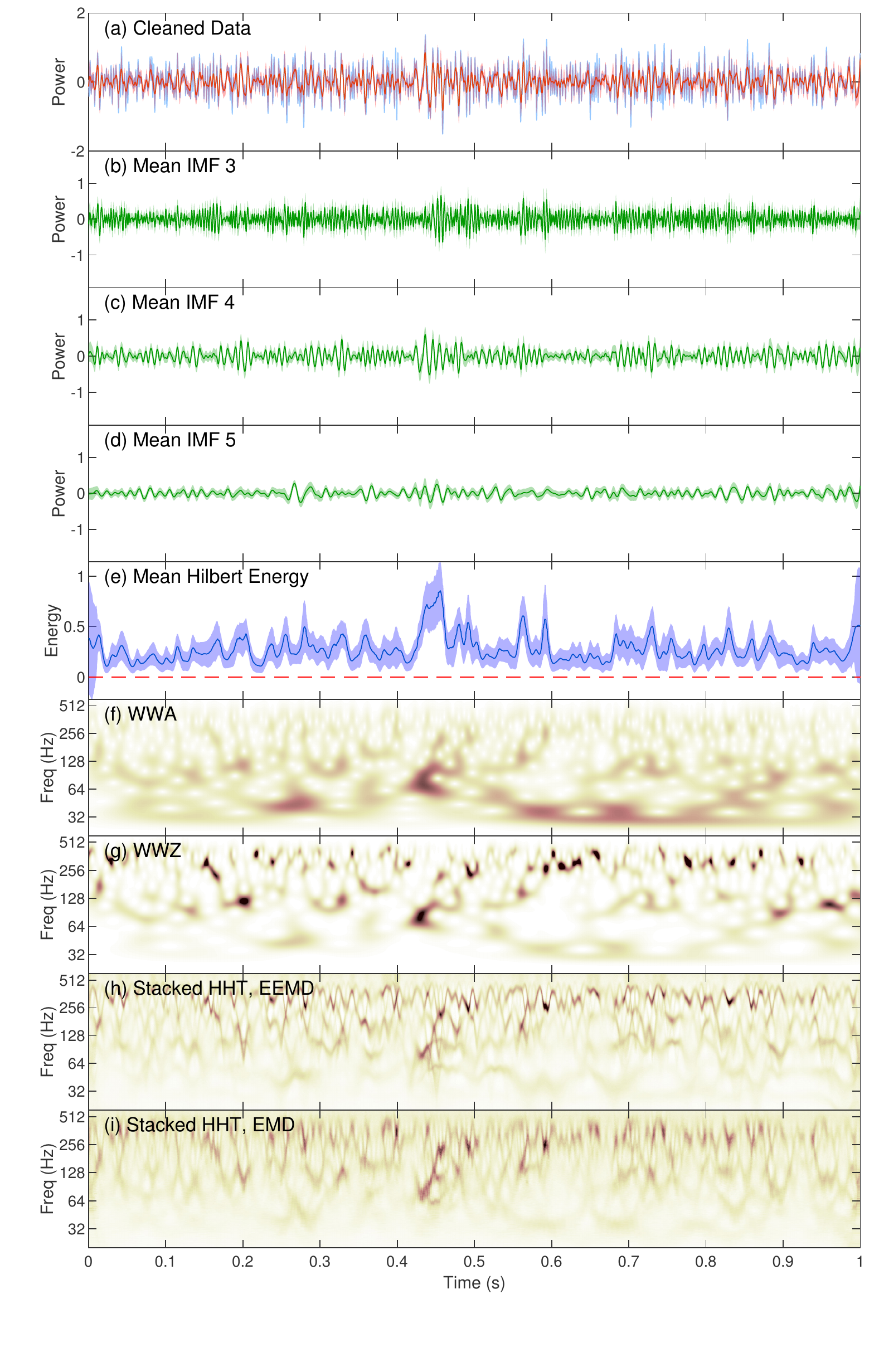}
\end{minipage}
\hspace{0.0cm}
\begin{minipage}{0.49\linewidth}
\includegraphics[width=1.01\textwidth]{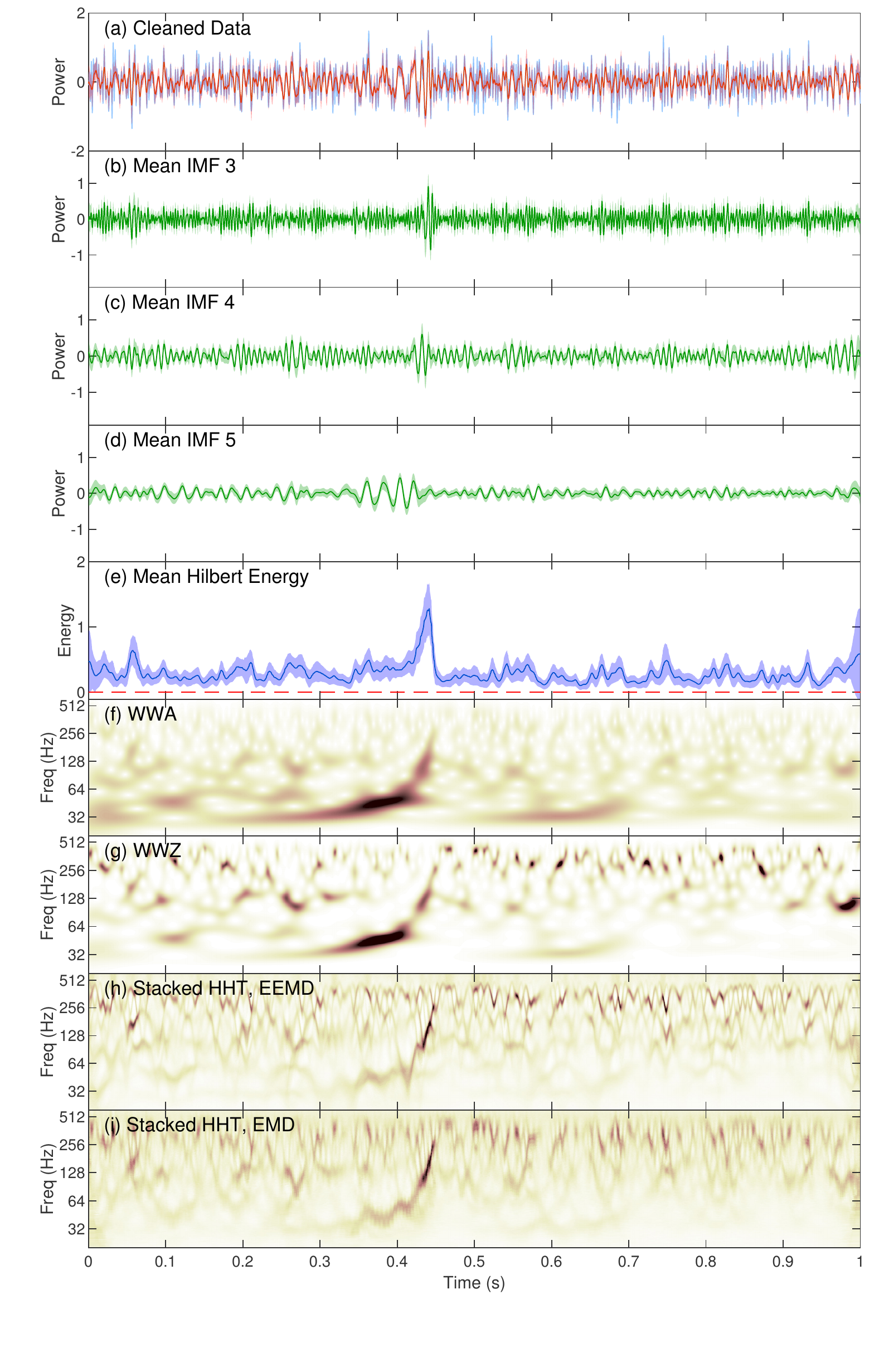}
\end{minipage}
\caption{The analysis results of GW170809 observed with LIGO Hanford (left) and Livingstone (right). Panels (a) -- (i) have the same definition as those in Figure \ref{fig:gw150914_all}. \label{fig:gw170809_all}}
\end{figure*}

\begin{figure*}
\begin{minipage}{0.49\linewidth}
\includegraphics[width=1.01\textwidth]{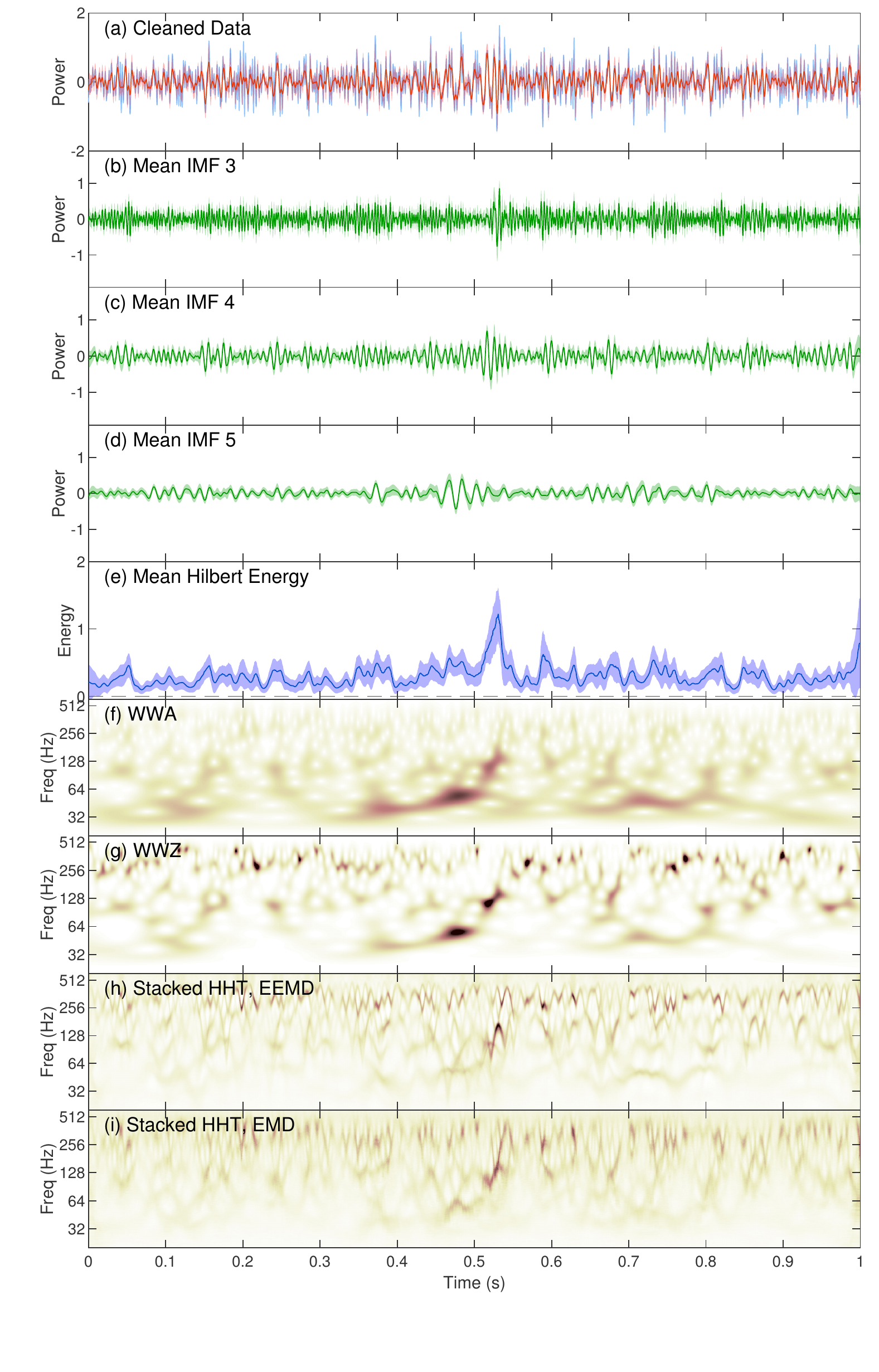}
\end{minipage}
\hspace{0.0cm}
\begin{minipage}{0.49\linewidth}
\includegraphics[width=1.01\textwidth]{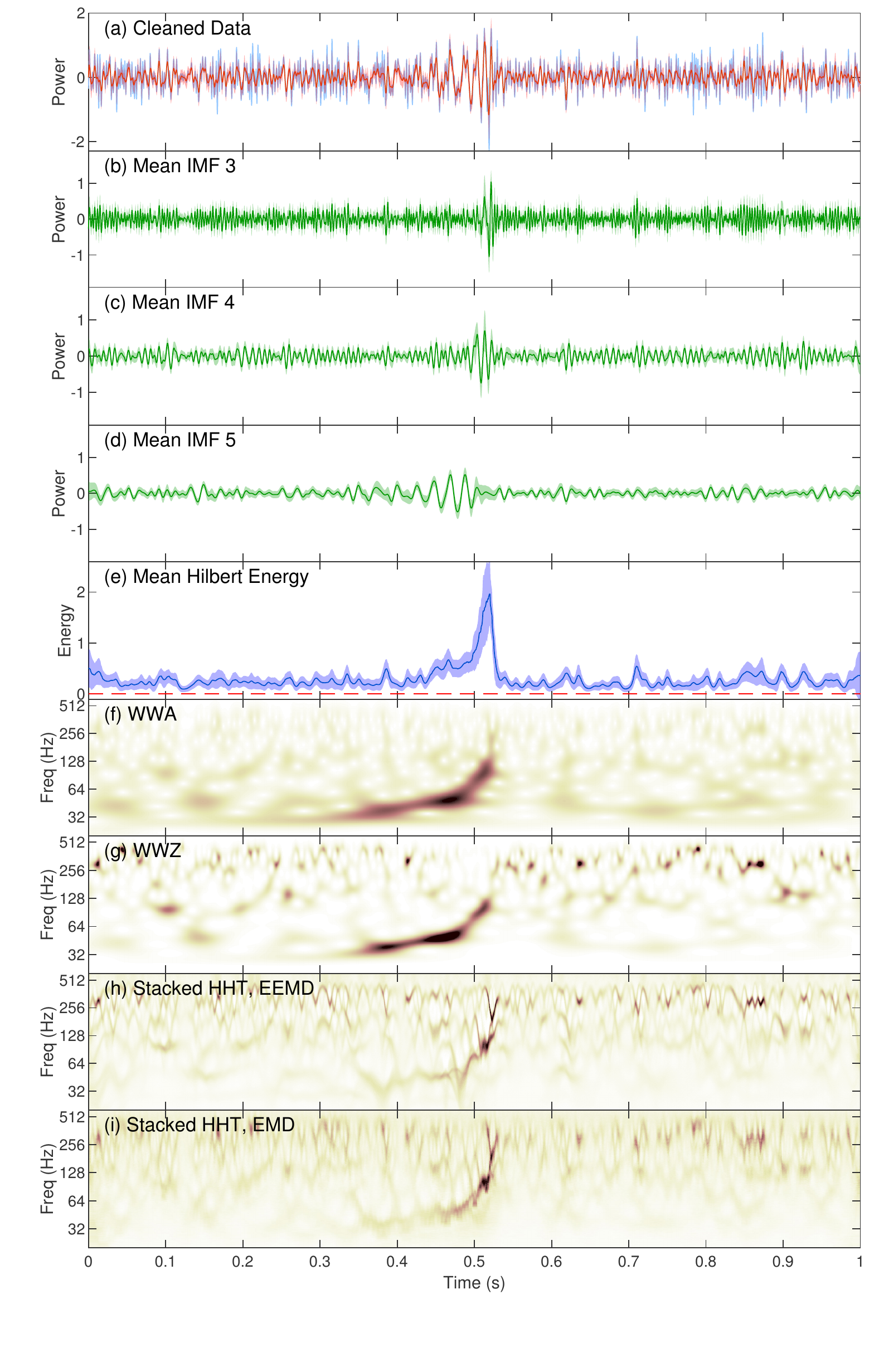}
\end{minipage}
\caption{The analysis results of GW170814 observed with LIGO Hanford (left) and Livingstone (right). Panels (a) -- (i) have the same definition as those in Figure \ref{fig:gw150914_all}. \label{fig:gw170814_all}}
\end{figure*}

\begin{figure*}
\begin{minipage}{0.49\linewidth}
\includegraphics[width=1.01\textwidth]{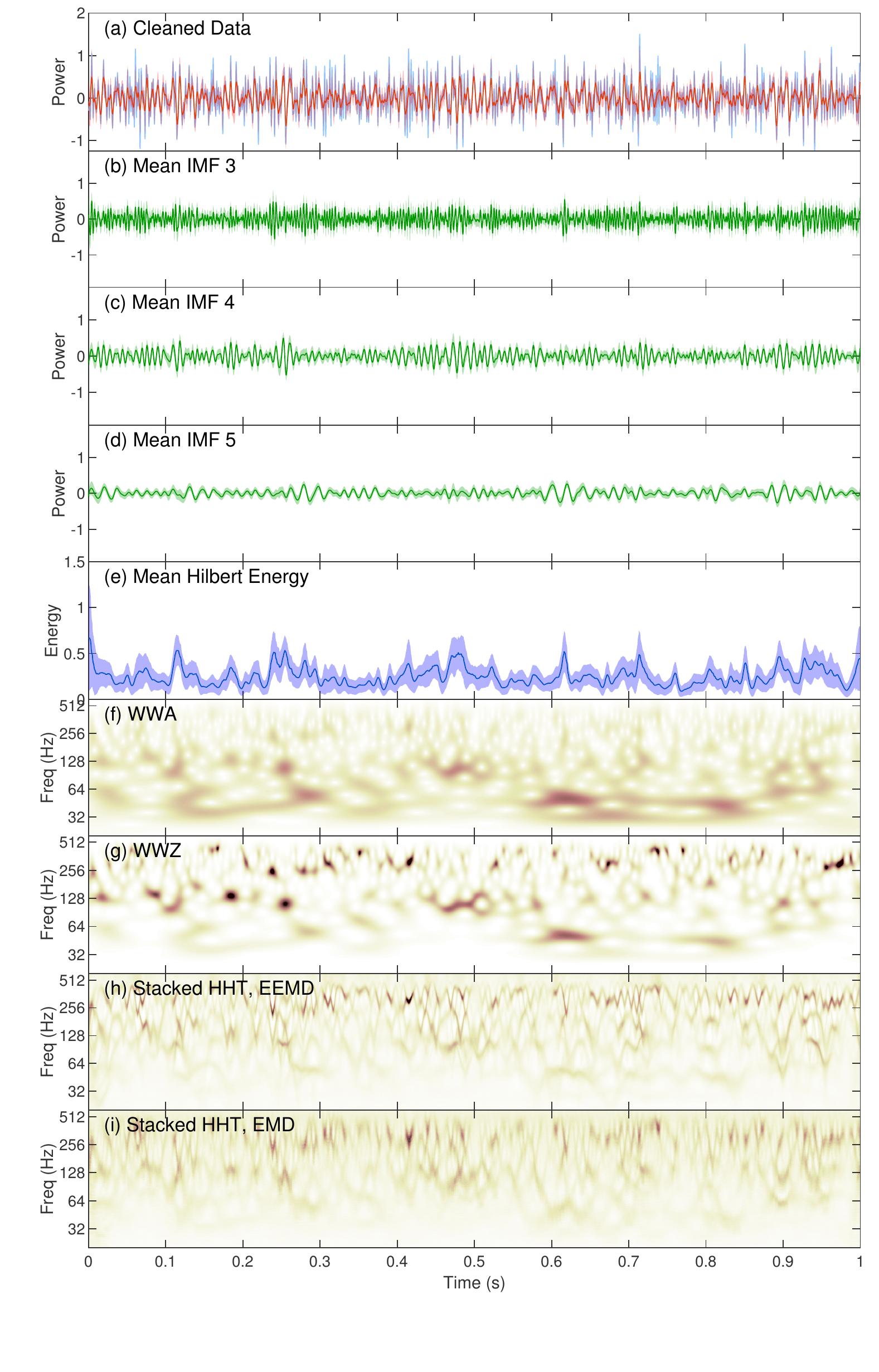}
\end{minipage}
\hspace{0.0cm}
\begin{minipage}{0.49\linewidth}
\includegraphics[width=1.01\textwidth]{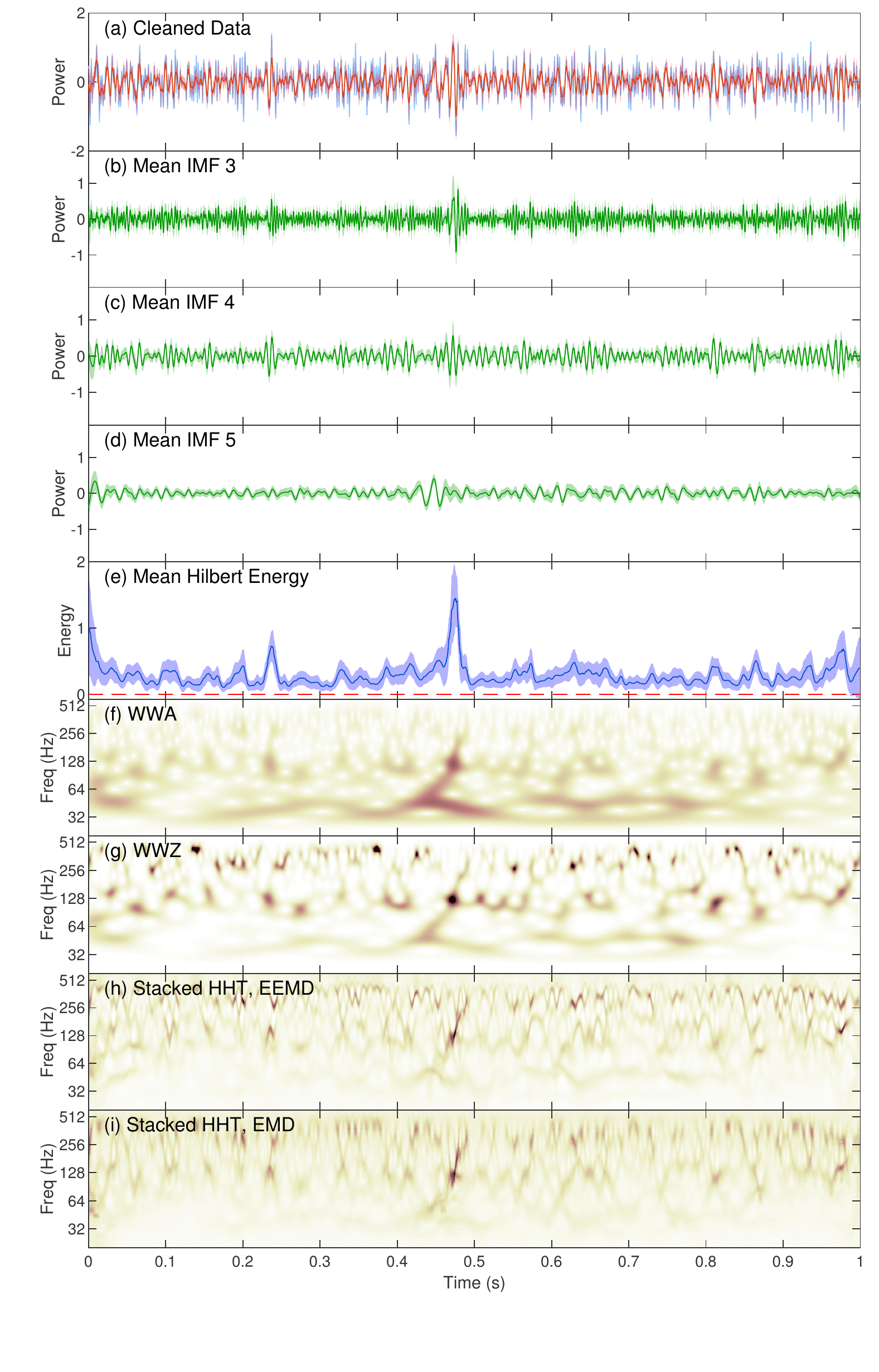}
\end{minipage}
\caption{The analysis results of GW170818 observed with LIGO Hanford (left) and Livingstone (right). Panels (a) -- (i) have the same definition as those in Figure \ref{fig:gw150914_all}. \label{fig:gw170818_all}}
\end{figure*}

\begin{figure*}
\begin{minipage}{0.49\linewidth}
\includegraphics[width=1.01\textwidth]{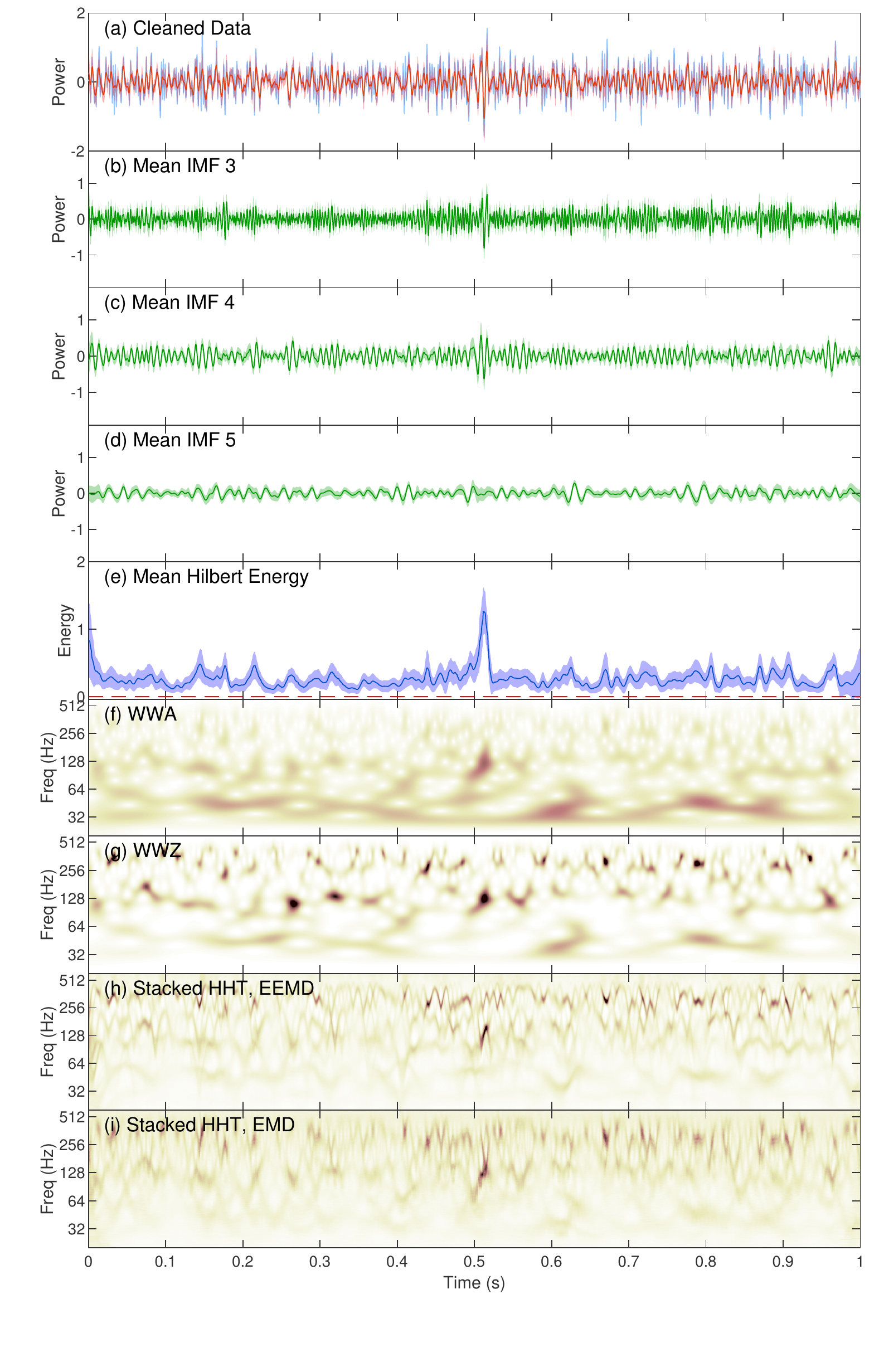}
\end{minipage}
\hspace{0.0cm}
\begin{minipage}{0.49\linewidth}
\includegraphics[width=1.01\textwidth]{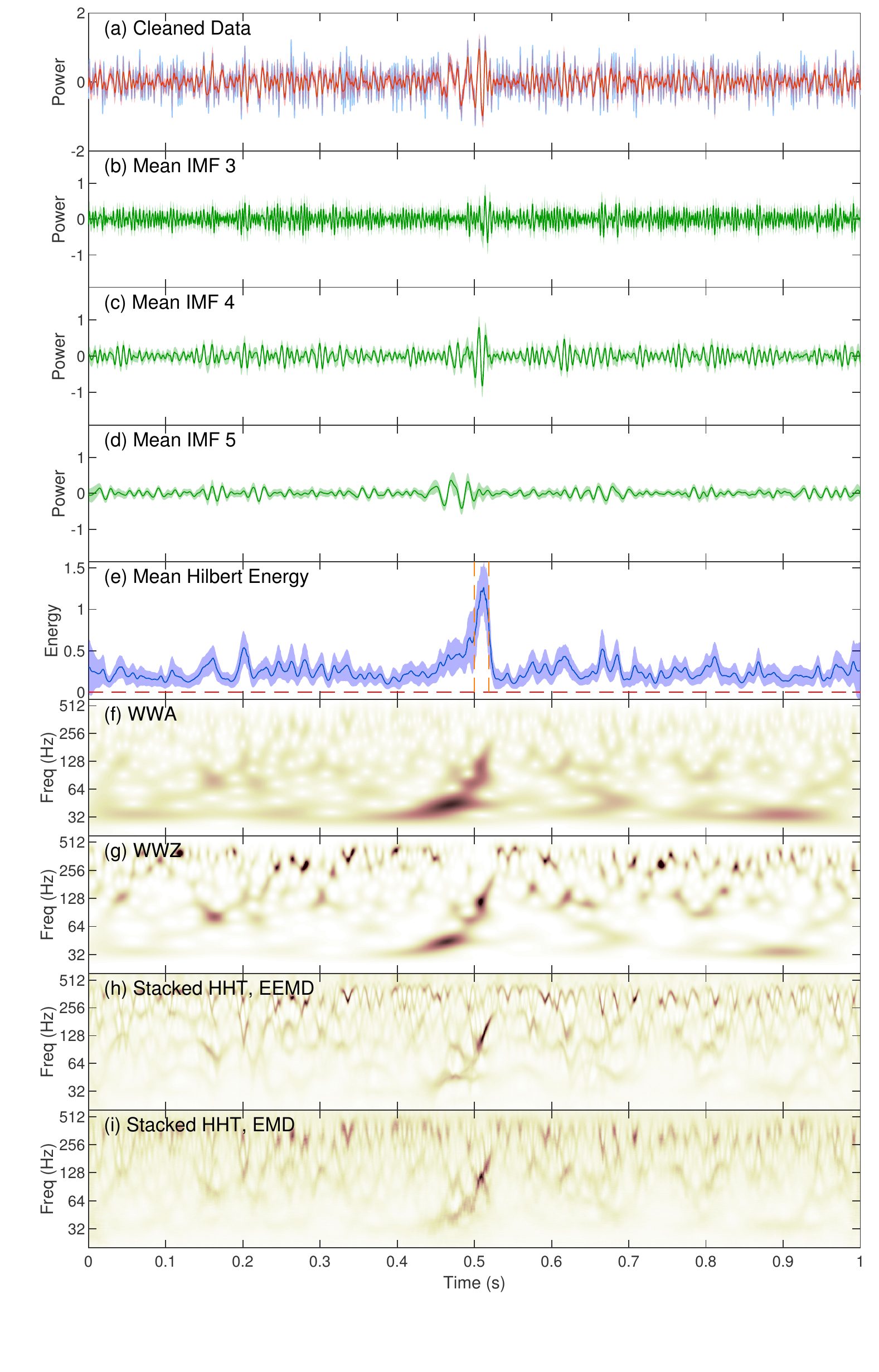}
\end{minipage}
\caption{The analysis results of GW170823 observed with LIGO Hanford (left) and Livingstone (right). Panels (a) -- (i) have the same definition as those in Figure \ref{fig:gw150914_all}. \label{fig:gw170823_all}}
\end{figure*}

\section{Resulting Spectra of CCSNe Events}\label{sec:appendix_CCSNe}
All the remaining simulated CCSN GW data, corresponding wavelet and Hilbert spectra are shown in this section. 

\begin{figure*}
\begin{minipage}{0.49\linewidth}
    \includegraphics[width=0.99\textwidth]{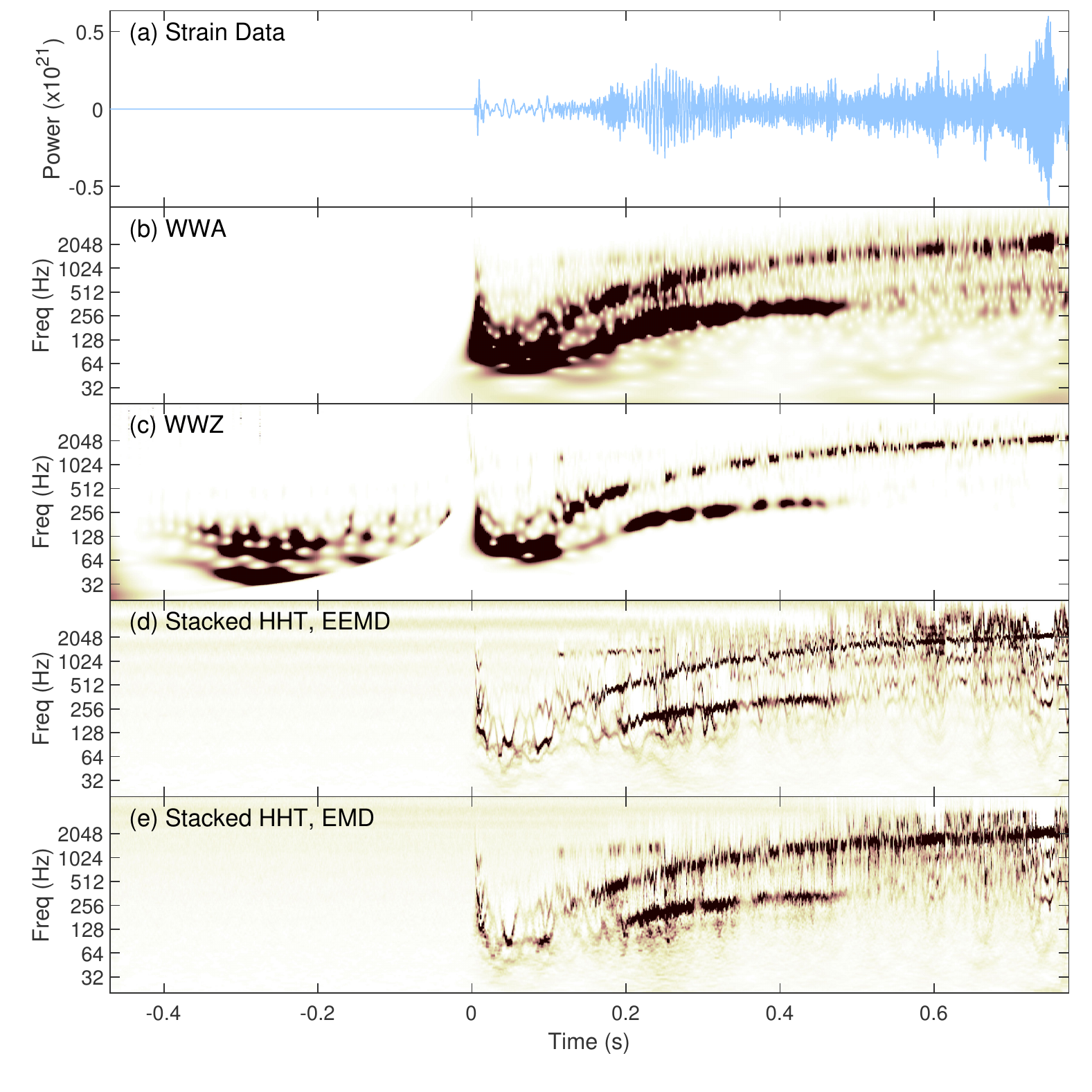}
    \caption{(a) The cross mode of the Gravitational waveform of a CCSN with an initial stellar mass of $40$~$M_{\odot}$ and without rotation viewed from the polar direction. Panels (b) -- (e) shows wavelet and Hilbert maps with the same definition as those in Figure \ref{fig:ccsn_nor_CrossEquator}. }
    \label{fig:ccsn_nor_CrossPole}
\end{minipage}
\hspace{0.0cm}
\begin{minipage}{0.49\linewidth}
    \includegraphics[width=0.99\textwidth]{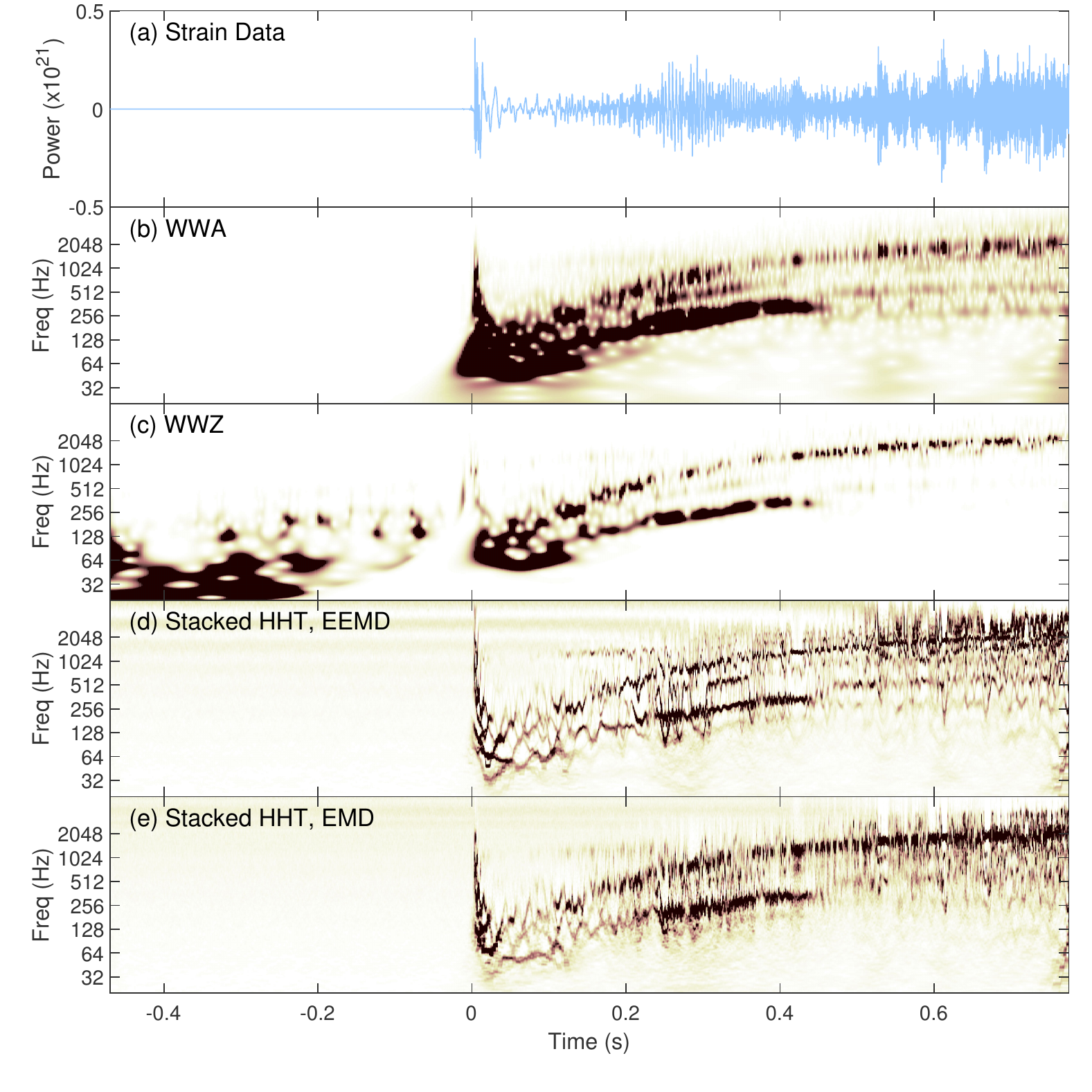}
    \caption{(a) The plus mode of the Gravitational waveform of a CCSN with an initial stellar mass of $40$~$M_{\odot}$ and without rotation viewed from the equatorial direction. Panels (b) -- (e) shows wavelet and Hilbert maps with the same definition as those in Figure \ref{fig:ccsn_nor_CrossEquator}. }
    \label{fig:ccsn_nor_PlusEquator}
\end{minipage}
\end{figure*}

\begin{figure*}
\begin{minipage}{0.49\linewidth}
    \includegraphics[width=0.99\textwidth]{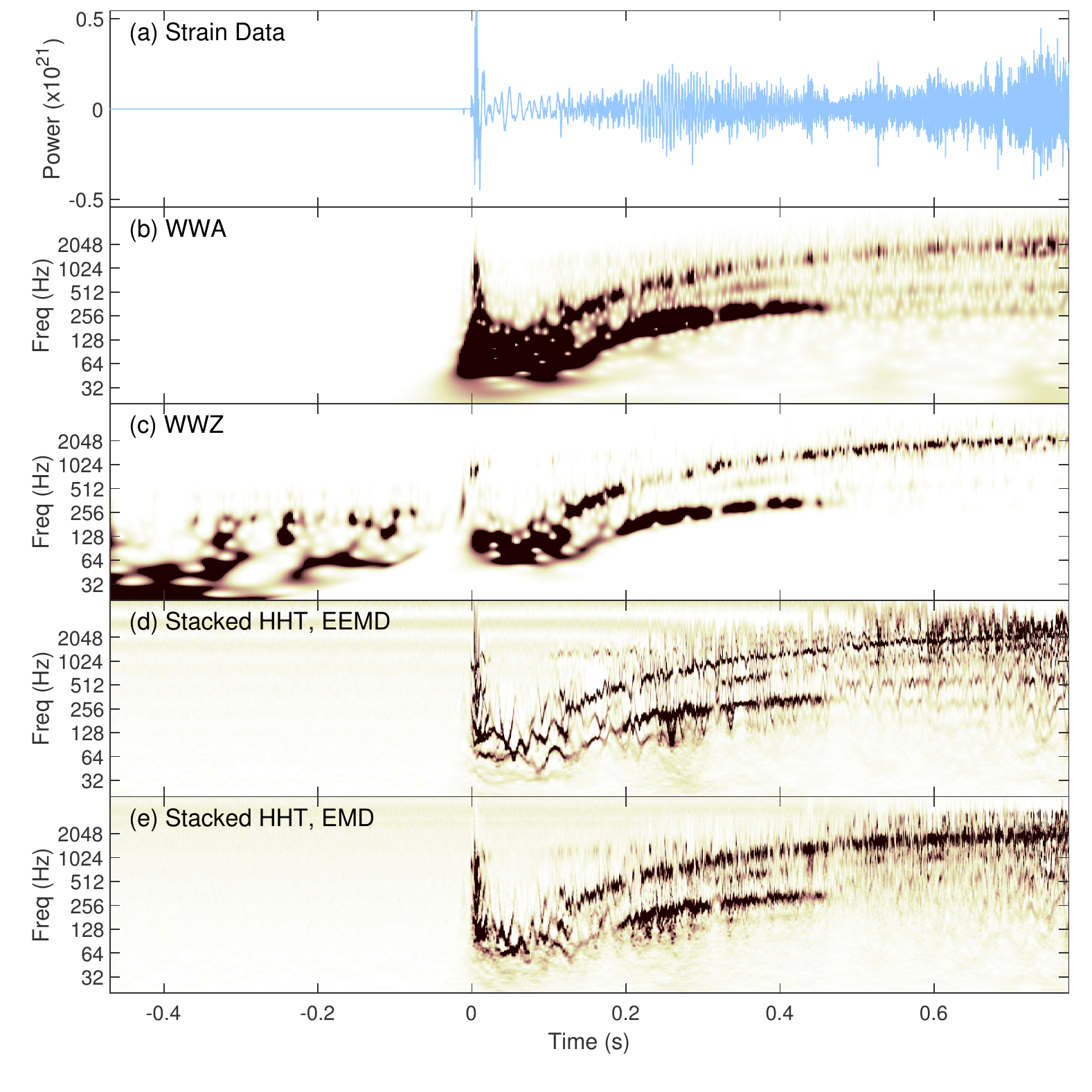}
    \caption{(a) The plus mode of the Gravitational waveform of a CCSN with an initial stellar mass of $40$~$M_{\odot}$ and without rotation viewed from the polar direction. Panels (b) -- (e) shows wavelet and Hilbert maps with the same definition as those in Figure \ref{fig:ccsn_nor_CrossEquator}. }
    \label{fig:ccsn_nor_PlusPole}
\end{minipage}
\hspace{0.0cm}
\begin{minipage}{0.49\linewidth}
    \includegraphics[width=0.99\textwidth]{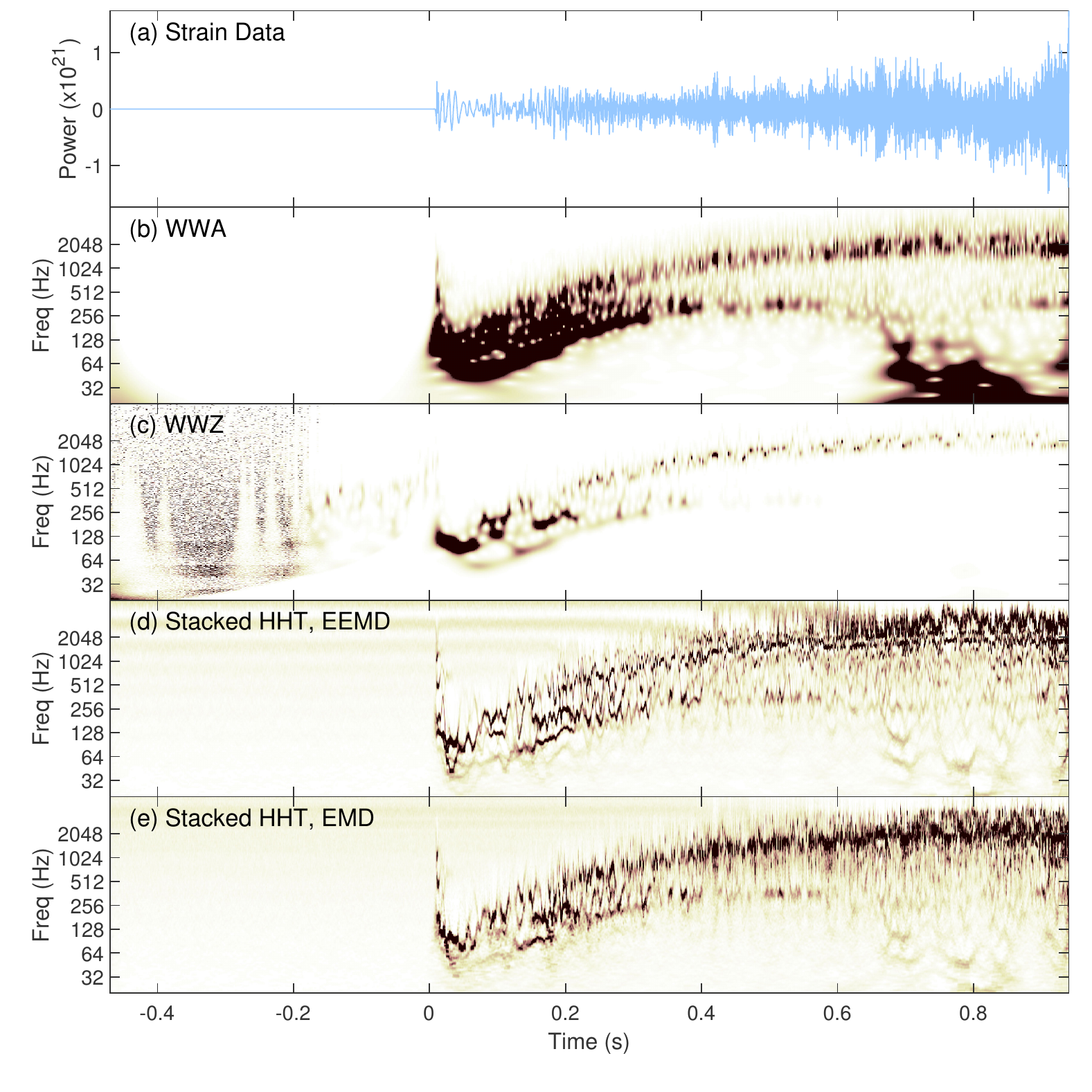}
    \caption{(a) The cross mode of the Gravitational waveform of a CCSN with an initial stellar mass of $40$~$M_{\odot}$ and slow rotation (0.5~rad~s$^{-1}$) viewed from the equatorial direction. Panels (b) -- (e) shows wavelet and Hilbert maps with the same definition as those in Figure \ref{fig:ccsn_nor_CrossEquator}. }
    \label{fig:ccsn_sr_CrossEquator}
\end{minipage}
\end{figure*}

\begin{figure*}
\begin{minipage}{0.49\linewidth}
    \includegraphics[width=0.99\textwidth]{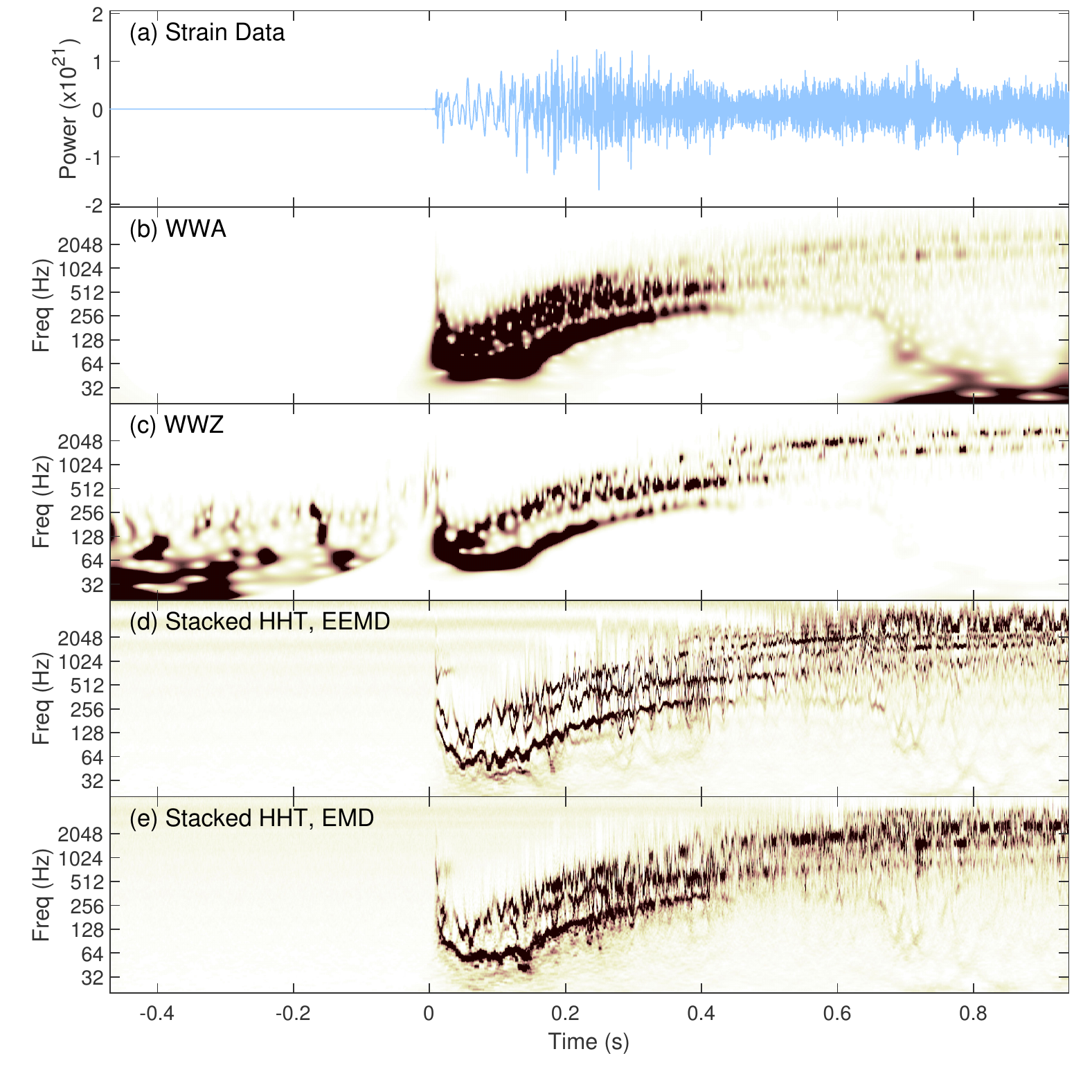}
    \caption{(a) The cross mode of the Gravitational waveform of a CCSN with an initial stellar mass of $40$~$M_{\odot}$ and and slow rotation (0.5~rad~s$^{-1}$) viewed from the polar direction. Panels (b) -- (e) shows wavelet and Hilbert maps with the same definition as those in Figure \ref{fig:ccsn_nor_CrossEquator}. }
    \label{fig:ccsn_sr_CrossPole}
\end{minipage}
\hspace{0.0cm}
\begin{minipage}{0.49\linewidth}
    \includegraphics[width=0.99\textwidth]{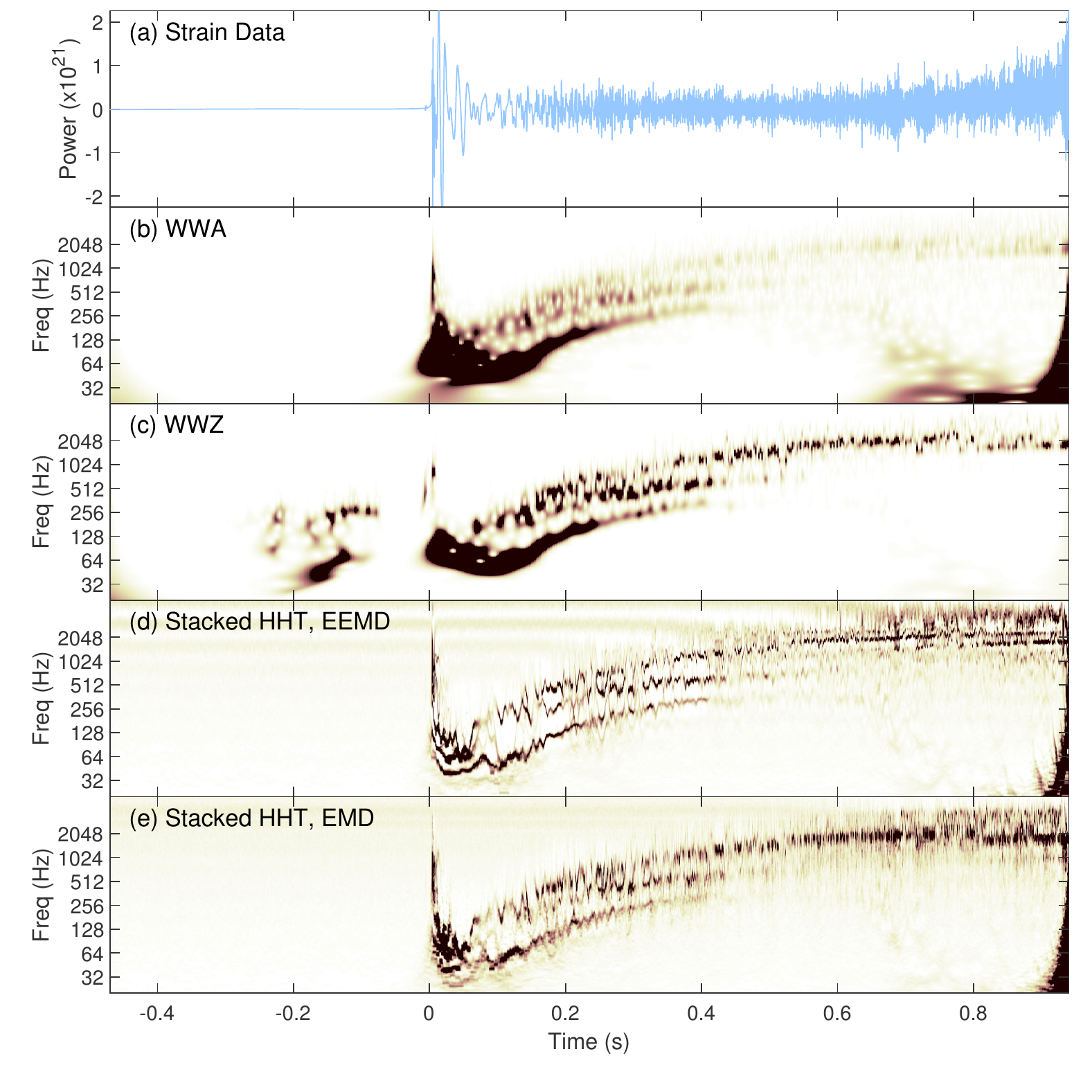}
    \caption{(a) The plus mode of the Gravitational waveform of a CCSN with an initial stellar mass of $40$~$M_{\odot}$ and slow rotation (0.5~rad~s$^{-1}$) viewed from the equatorial direction. Panels (b) -- (e) shows wavelet and Hilbert maps with the same definition as those in Figure \ref{fig:ccsn_nor_CrossEquator}. }
    \label{fig:ccsn_sr_PlusEquator}
\end{minipage}
\end{figure*}

\begin{figure*}
\begin{minipage}{0.49\linewidth}
    \includegraphics[width=0.99\textwidth]{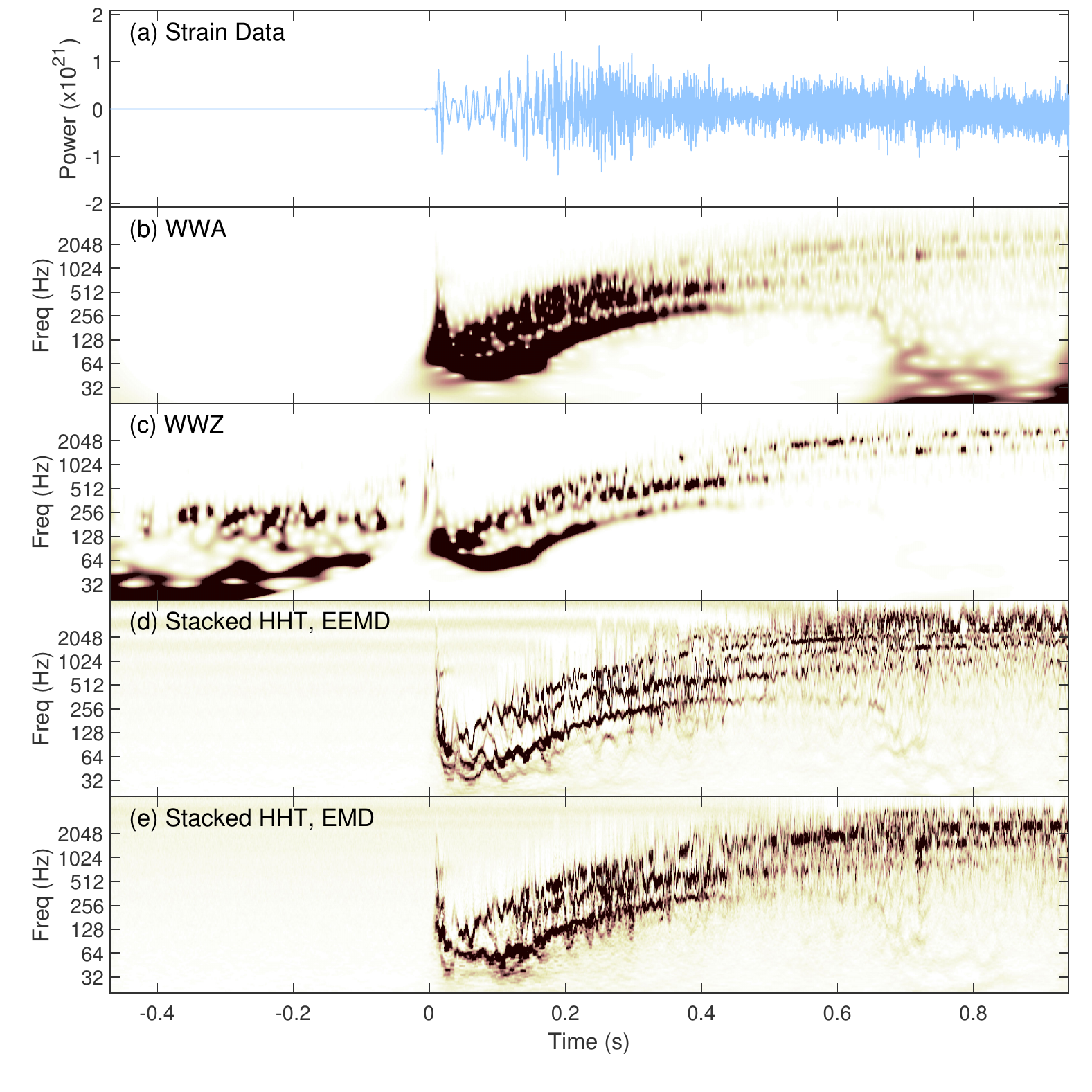}
    \caption{(a) The plus mode of the Gravitational waveform of a CCSN with an initial stellar mass of $40$~$M_{\odot}$ and and slow rotation (0.5~rad~s$^{-1}$) viewed from the polar direction. Panels (b) -- (e) shows wavelet and Hilbert maps with the same definition as those in Figure \ref{fig:ccsn_nor_CrossEquator}. }
    \label{fig:ccsn_sr_PlusPole}
\end{minipage}
\hspace{0.0cm}
\begin{minipage}{0.49\linewidth}
    \includegraphics[width=0.99\textwidth]{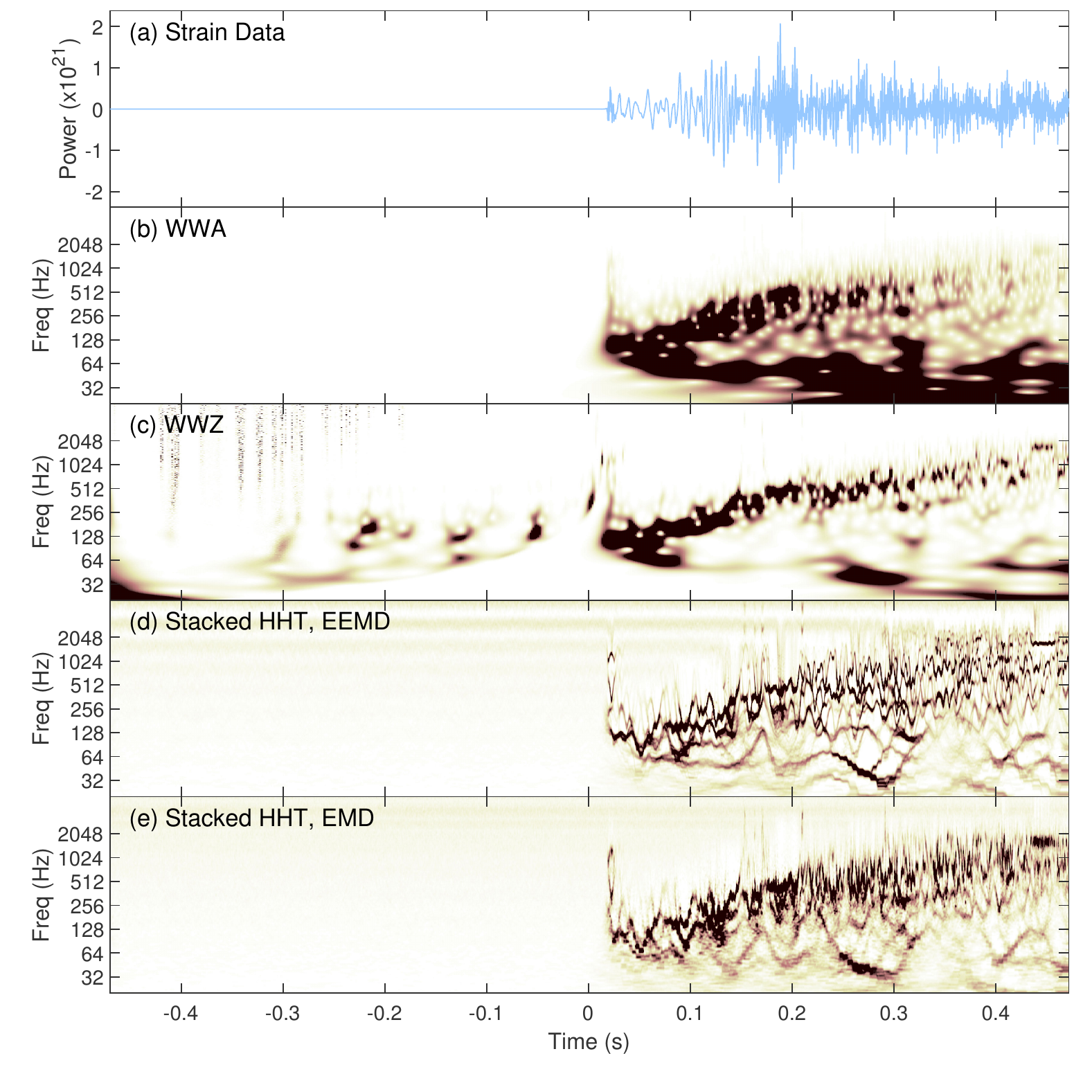}
    \caption{(a) The cross mode of the Gravitational waveform of a CCSN with an initial stellar mass of $40$~$M_{\odot}$ and fast rotation (1~rad~s$^{-1}$) viewed from the equatorial direction. Panels (b) -- (e) shows wavelet and Hilbert maps with the same definition as those in Figure \ref{fig:ccsn_nor_CrossEquator}. }
    \label{fig:ccsn_fr_CrossEquator}
\end{minipage}
\end{figure*}

\begin{figure*}
\begin{minipage}{0.49\linewidth}
    \includegraphics[width=0.99\textwidth]{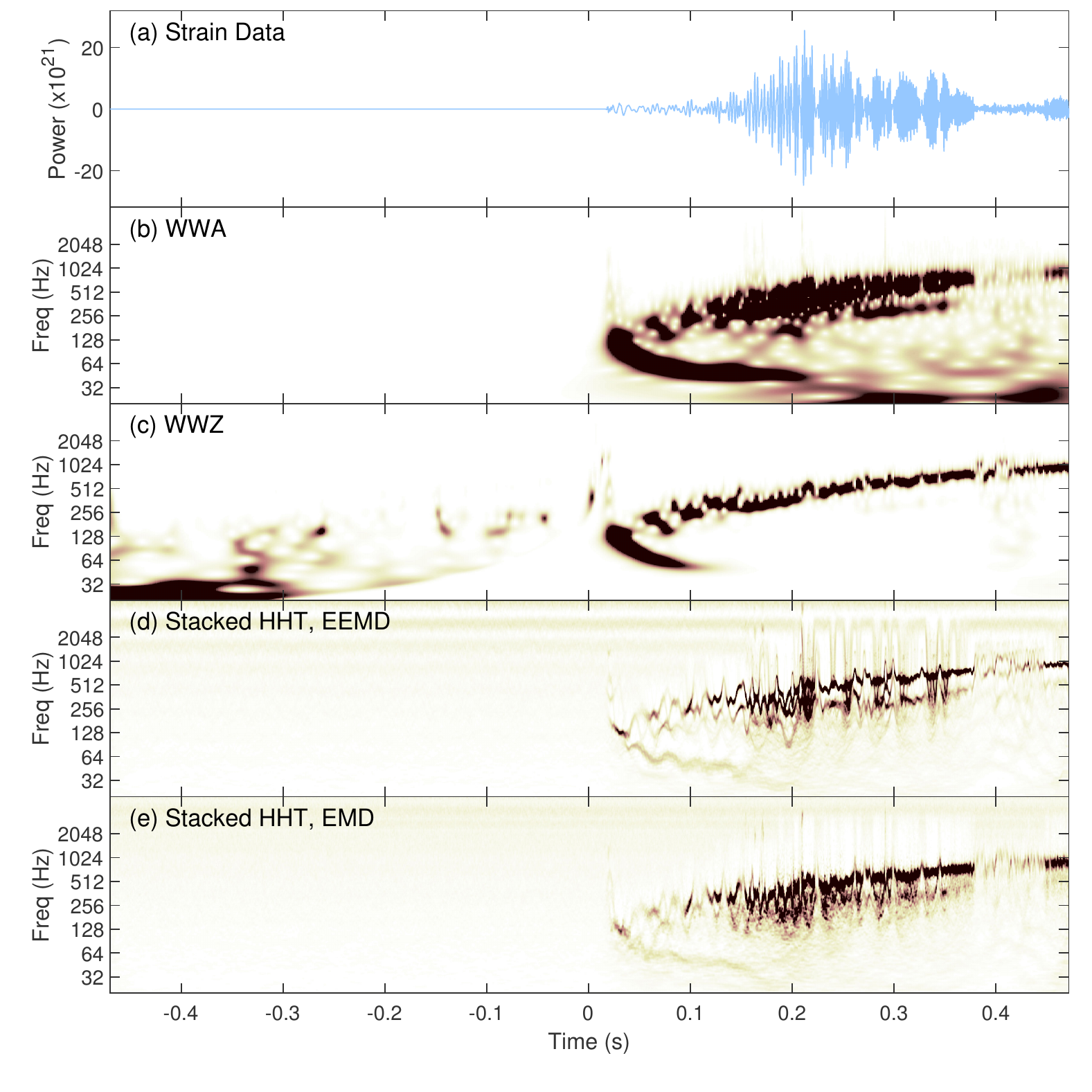}
    \caption{(a) The cross mode of the Gravitational waveform of a CCSN with an initial stellar mass of $40$~$M_{\odot}$ and and fast rotation (1~rad~s$^{-1}$) viewed from the polar direction. Panels (b) -- (e) shows wavelet and Hilbert maps with the same definition as those in Figure \ref{fig:ccsn_nor_CrossEquator}. }
    \label{fig:ccsn_fr_CrossPole}
\end{minipage}
\hspace{0.0cm}
\begin{minipage}{0.49\linewidth}
    \includegraphics[width=0.99\textwidth]{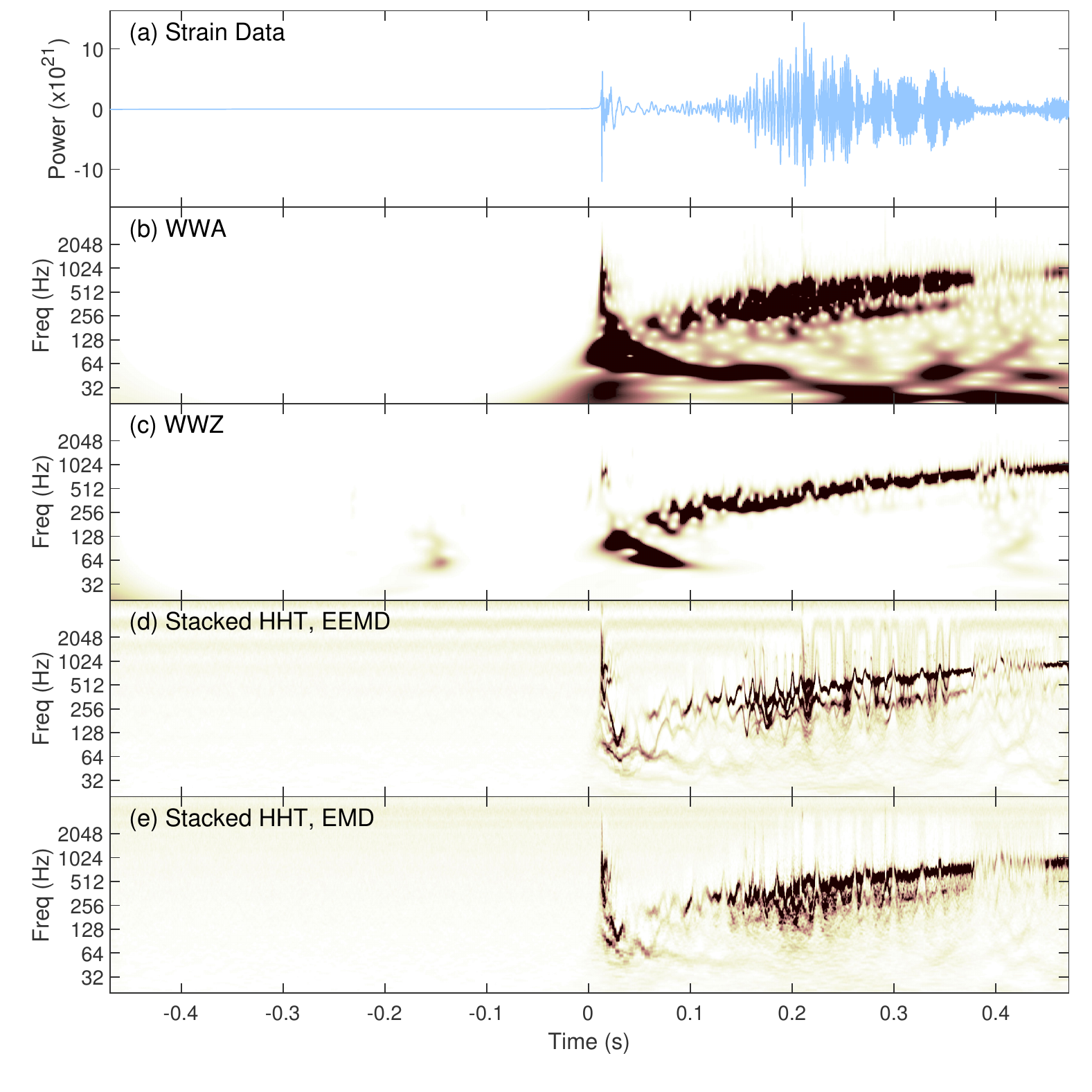}
    \caption{(a) The plus mode of the Gravitational waveform of a CCSN with an initial stellar mass of $40$~$M_{\odot}$ and fast rotation (1~rad~s$^{-1}$) viewed from the equatorial direction. Panels (b) -- (e) shows wavelet and Hilbert maps with the same definition as those in Figure \ref{fig:ccsn_nor_CrossEquator}. }
    \label{fig:ccsn_fr_PlusEquator}
\end{minipage}
\end{figure*}

\begin{figure}
\begin{minipage}{0.49\linewidth}
\centering
    \includegraphics[width=0.99\textwidth]{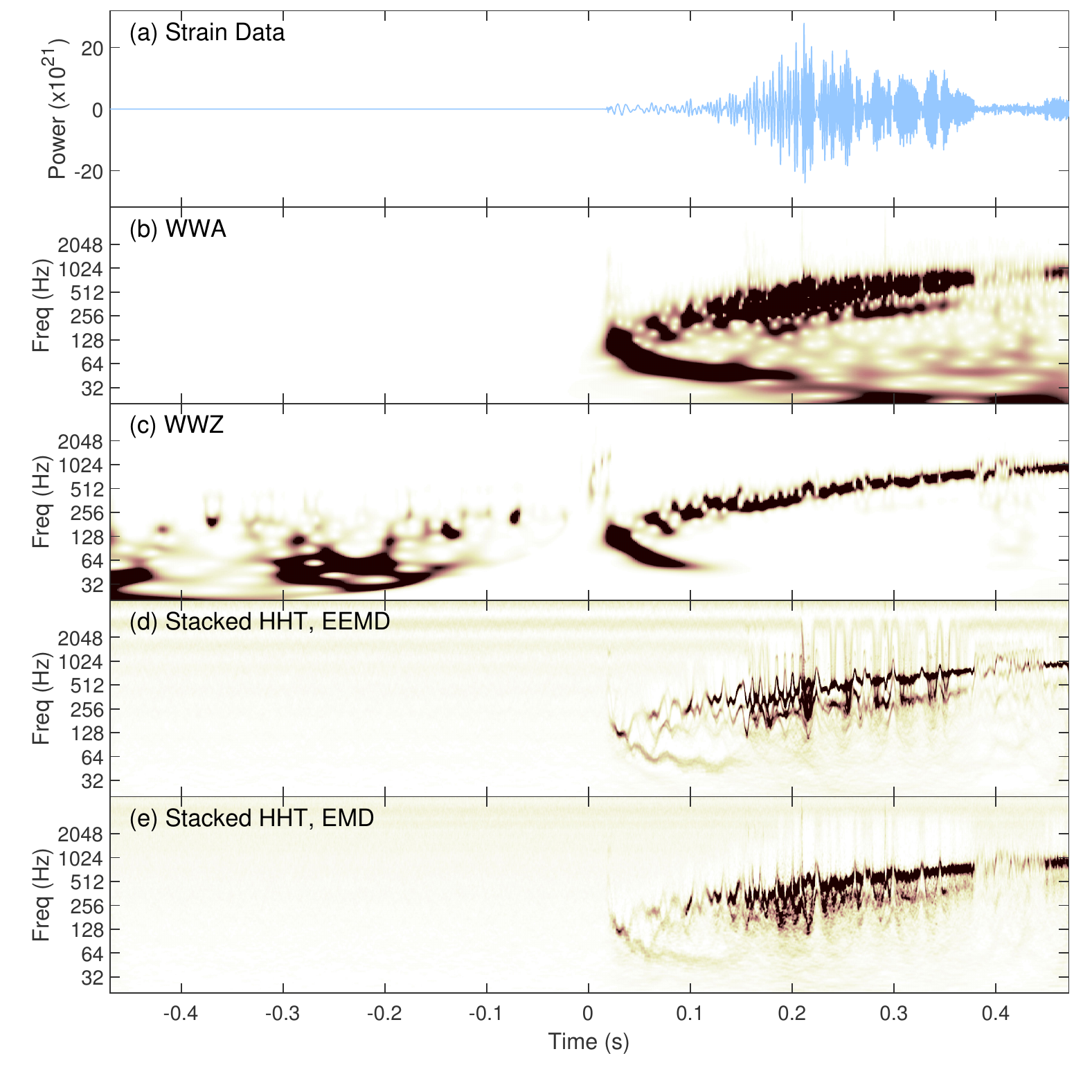}
    \caption{(a) The plus mode of the Gravitational waveform of a CCSN with an initial stellar mass of $40$~$M_{\odot}$ and and fast rotation (1~rad~s$^{-1}$) viewed from the polar direction. Panels (b) -- (e) shows wavelet and Hilbert maps with the same definition as those in Figure \ref{fig:ccsn_nor_CrossEquator}. }
    \label{fig:ccsn_fr_PlusPole}
\end{minipage}
\end{figure}


\begin{thebibliography}{}
\expandafter\ifx\csname natexlab\endcsname\relax\def\natexlab#1{#1}\fi
\providecommand{\url}[1]{\href{#1}{#1}}
\providecommand{\dodoi}[1]{doi:~\href{http://doi.org/#1}{\nolinkurl{#1}}}
\providecommand{\doeprint}[1]{\href{http://ascl.net/#1}{\nolinkurl{http://ascl.net/#1}}}
\providecommand{\doarXiv}[1]{\href{https://arxiv.org/abs/#1}{\nolinkurl{https://arxiv.org/abs/#1}}}

\bibitem[{{Abbott} {et~al.}(2016{\natexlab{a}}){Abbott}, {Abbott}, {Abbott},
  {Abernathy}, {Acernese}, {Ackley}, {Adams}, {Adams}, {Addesso}, {Adhikari},
  \& et~al.}]{AbbottAA2016}
{Abbott}, B.~P., {Abbott}, R., {Abbott}, T.~D., {et~al.} 2016{\natexlab{a}},
  Physical Review Letters, 116, 061102, \dodoi{10.1103/PhysRevLett.116.061102}

\bibitem[{{Abbott} {et~al.}(2016{\natexlab{b}}){Abbott}, {Abbott}, {Abbott},
  {Abernathy}, {Acernese}, {Ackley}, {Adams}, {Adams}, {Addesso}, {Adhikari},
  \& et~al.}]{AbbottAA2016c}
---. 2016{\natexlab{b}}, \prd, 94, 102001, \dodoi{10.1103/PhysRevD.94.102001}

\bibitem[{{Abbott} {et~al.}(2017{\natexlab{a}}){Abbott}, {Abbott}, {Abbott},
  {Acernese}, {Ackley}, {Adams}, {Adams}, {Addesso}, {Adhikari}, {Adya},
  {Affeldt}, {Afrough}, {Agarwal}, {Agathos}, {Agatsuma}, {Aggarwal}, {Aguiar},
  {Aiello}, {Ain}, {Ajith}, {Allen}, {Allen}, {Allocca}, {Altin}, {Amato},
  {Ananyeva}, {Anderson}, {Anderson}, {Angelova}, {Antier}, {Appert}, {Arai},
  {Araya}, {Areeda}, {Arnaud}, {Arun}, {Ascenzi}, {Ashton}, {Ast}, {Aston},
  {Astone}, {Atallah}, {Aufmuth}, {Aulbert}, {AultONeal}, {Austin},
  {Avila-Alvarez}, {Babak}, {Bacon}, {Bader}, {Bae}, {Bailes}, {Baker},
  {Baldaccini}, {Ballardin}, {Ballmer}, {Banagiri}, {Barayoga}, {Barclay},
  {Barish}, {Barker}, {Barkett}, {Barone}, {Barr}, {Barsotti}, {Barsuglia},
  {Barta}, {Barthelmy}, {Bartlett}, {Bartos}, {Bassiri}, {Basti}, {Batch},
  {Bawaj}, {Bayley}, {Bazzan}, {B{\'e}csy}, {Beer}, {Bejger}, {Belahcene},
  {Bell}, {Berger}, {Bergmann}, {Bernuzzi}, {Bero}, {Berry}, {Bersanetti},
  {Bertolini}, {Betzwieser}, {Bhagwat}, {Bhandare}, {Bilenko}, {Billingsley},
  {Billman}, {Birch}, {Birney}, {Birnholtz}, {Biscans}, {Biscoveanu}, {Bisht},
  {Bitossi}, {Biwer}, {Bizouard}, {Blackburn}, {Blackman}, {Blair}, {Blair},
  {Blair}, {Bloemen}, {Bock}, {Bode}, {Boer}, {Bogaert}, {Bohe}, {Bondu},
  {Bonilla}, {Bonnand}, {Boom}, {Bork}, {Boschi}, {Bose}, {Bossie},
  {Bouffanais}, {Bozzi}, {Bradaschia}, {Brady}, {Branchesi}, {Brau}, {Briant},
  {Brillet}, {Brinkmann}, {Brisson}, {Brockill}, {Broida}, {Brooks}, {Brown},
  {Brown}, {Brunett}, {Buchanan}, {Buikema}, {Bulik}, {Bulten}, {Buonanno},
  {Buskulic}, {Buy}, {Byer}, {Cabero}, {Cadonati}, {Cagnoli}, {Cahillane},
  {Calder{\'o}n Bustillo}, {Callister}, {Calloni}, {Camp}, {Canepa},
  {Canizares}, {Cannon}, {Cao}, {Cao}, {Capano}, {Capocasa}, {Carbognani},
  {Caride}, {Carney}, {Carullo}, {Casanueva Diaz}, {Casentini}, {Caudill},
  {Cavagli{\`a}}, {Cavalier}, {Cavalieri}, {Cella}, {Cepeda},
  {Cerd{\'a}-Dur{\'a}n}, {Cerretani}, {Cesarini}, {Chamberlin}, {Chan}, {Chao},
  {Charlton}, {Chase}, {Chassande-Mottin}, {Chatterjee}, {Chatziioannou},
  {Cheeseboro}, {Chen}, {Chen}, {Chen}, {Cheng}, {Chia}, {Chincarini},
  {Chiummo}, {Chmiel}, {Cho}, {Cho}, {Chow}, {Christensen}, {Chu}, {Chua},
  {Chua}, {Chung}, {Chung}, {Ciani}, {Ciolfi}, {Cirelli}, {Cirone}, {Clara},
  {Clark}, {Clearwater}, {Cleva}, {Cocchieri}, {Coccia}, {Cohadon}, {Cohen},
  {Colla}, {Collette}, {Cominsky}, {Constancio}, {Conti}, {Cooper}, {Corban},
  {Corbitt}, {Cordero-Carri{\'o}n}, {Corley}, {Cornish}, {Corsi}, {Cortese},
  {Costa}, {Coughlin}, {Coughlin}, {Coulon}, {Countryman}, {Couvares}, {Covas},
  {Cowan}, {Coward}, {Cowart}, {Coyne}, {Coyne}, {Creighton}, {Creighton},
  {Cripe}, {Crowder}, {Cullen}, {Cumming}, {Cunningham}, {Cuoco}, {Dal Canton},
  {D{\'a}lya}, {Danilishin}, {D'Antonio}, {Danzmann}, {Dasgupta}, {Da Silva
  Costa}, {Dattilo}, {Dave}, {Davier}, {Davis}, {Daw}, {Day}, {De}, {DeBra},
  {Degallaix}, {De Laurentis}, {Del{\'e}glise}, {Del Pozzo}, {Demos}, {Denker},
  {Dent}, {De Pietri}, {Dergachev}, {De Rosa}, {DeRosa}, {De Rossi}, {DeSalvo},
  {de Varona}, {Devenson}, {Dhurandhar}, {D{\'\i}az}, {Dietrich}, {Di Fiore},
  {Di Giovanni}, {Di Girolamo}, {Di Lieto}, {Di Pace}, {Di Palma}, {Di Renzo},
  {Doctor}, {Dolique}, {Donovan}, {Dooley}, {Doravari}, {Dorrington},
  {Douglas}, {Dovale {\'A}lvarez}, {Downes}, {Drago}, {Dreissigacker},
  {Driggers}, {Du}, {Ducrot}, {Dudi}, {Dupej}, {Dwyer}, {Edo}, {Edwards},
  {Effler}, {Eggenstein}, {Ehrens}, {Eichholz}, {Eikenberry}, {Eisenstein},
  {Essick}, {Estevez}, {Etienne}, {Etzel}, {Evans}, {Evans}, {Factourovich},
  {Fafone}, {Fair}, {Fairhurst}, {Fan}, {Farinon}, {Farr}, {Farr},
  {Fauchon-Jones}, {Favata}, {Fays}, {Fee}, {Fehrmann}, {Feicht}, {Fejer},
  {Fernandez-Galiana}, {Ferrante}, {Ferreira}, {Ferrini}, {Fidecaro},
  {Finstad}, {Fiori}, {Fiorucci}, {Fishbach}, {Fisher}, {Fitz-Axen},
  {Flaminio}, {Fletcher}, {Fong}, {Font}, {Forsyth}, {Forsyth}, {Fournier},
  {Frasca}, {Frasconi}, {Frei}, {Freise}, {Frey}, {Frey}, {Fries}, {Fritschel},
  {Frolov}, {Fulda}, {Fyffe}, {Gabbard}, {Gadre}, {Gaebel}, {Gair},
  {Gammaitoni}, {Ganija}, {Gaonkar}, {Garcia-Quiros}, {Garufi}, {Gateley},
  {Gaudio}, {Gaur}, {Gayathri}, {Gehrels}, {Gemme}, {Genin}, {Gennai},
  {George}, {George}, {Gergely}, {Germain}, {Ghonge}, {Ghosh}, {Ghosh},
  {Ghosh}, {Giaime}, {Giardina}, {Giazotto}, {Gill}, {Glover}, {Goetz},
  {Goetz}, {Gomes}, {Goncharov}, {Gonz{\'a}lez}, {Gonzalez Castro},
  {Gopakumar}, {Gorodetsky}, {Gossan}, {Gosselin}, {Gouaty}, {Grado}, {Graef},
  {Granata}, {Grant}, {Gras}, {Gray}, {Greco}, {Green}, {Gretarsson}, {Groot},
  {Grote}, {Grunewald}, {Gruning}, {Guidi}, {Guo}, {Gupta}, {Gupta}, {Gushwa},
  {Gustafson}, {Gustafson}, {Halim}, {Hall}, {Hall}, {Hamilton}, {Hammond},
  {Haney}, {Hanke}, {Hanks}, {Hanna}, {Hannam}, {Hannuksela}, {Hanson},
  {Hardwick}, {Harms}, {Harry}, {Harry}, {Hart}, {Haster}, {Haughian}, {Healy},
  {Heidmann}, {Heintze}, {Heitmann}, {Hello}, {Hemming}, {Hendry}, {Heng},
  {Hennig}, {Heptonstall}, {Heurs}, {Hild}, {Hinderer}, {Ho}, {Hoak}, {Hofman},
  {Holt}, {Holz}, {Hopkins}, {Horst}, {Hough}, {Houston}, {Howell}, {Hreibi},
  {Hu}, {Huerta}, {Huet}, {Hughey}, {Husa}, {Huttner}, {Huynh-Dinh}, {Indik},
  {Inta}, {Intini}, {Isa}, {Isac}, {Isi}, {Iyer}, {Izumi}, {Jacqmin}, {Jani},
  {Jaranowski}, {Jawahar}, {Jim{\'e}nez-Forteza}, {Johnson},
  {Johnson-McDaniel}, {Jones}, {Jones}, {Jonker}, {Ju}, {Junker}, {Kalaghatgi},
  {Kalogera}, {Kamai}, {Kandhasamy}, {Kang}, {Kanner}, {Kapadia}, {Karki},
  {Karvinen}, {Kasprzack}, {Kastaun}, {Katolik}, {Katsavounidis}, {Katzman},
  {Kaufer}, {Kawabe}, {K{\'e}f{\'e}lian}, {Keitel}, {Kemball}, {Kennedy},
  {Kent}, {Key}, {Khalili}, {Khan}, {Khan}, {Khan}, {Khazanov}, {Kijbunchoo},
  {Kim}, {Kim}, {Kim}, {Kim}, {Kim}, {Kim}, {Kimbrell}, {King}, {King},
  {Kinley-Hanlon}, {Kirchhoff}, {Kissel}, {Kleybolte}, {Klimenko}, {Knowles},
  {Koch}, {Koehlenbeck}, {Koley}, {Kondrashov}, {Kontos}, {Korobko}, {Korth},
  {Kowalska}, {Kozak}, {Kr{\"a}mer}, {Kringel}, {Krishnan}, {Kr{\'o}lak},
  {Kuehn}, {Kumar}, {Kumar}, {Kumar}, {Kuo}, {Kutynia}, {Kwang}, {Lackey},
  {Lai}, {Landry}, {Lang}, {Lange}, {Lantz}, {Lanza}, {Larson},
  {Lartaux-Vollard}, {Lasky}, {Laxen}, {Lazzarini}, {Lazzaro}, {Leaci},
  {Leavey}, {Lee}, {Lee}, {Lee}, {Lee}, {Lee}, {Lehmann}, {Lenon}, {Leon},
  {Leonardi}, {Leroy}, {Letendre}, {Levin}, {Li}, {Linker}, {Littenberg},
  {Liu}, {Liu}, {Lo}, {Lockerbie}, {London}, {Lord}, {Lorenzini}, {Loriette},
  {Lormand}, {Losurdo}, {Lough}, {Lousto}, {Lovelace}, {L{\"u}ck}, {Lumaca},
  {Lundgren}, {Lynch}, {Ma}, {Macas}, {Macfoy}, {Machenschalk}, {MacInnis},
  {Macleod}, {Maga{\~n}a Hernandez}, {Maga{\~n}a-Sandoval}, {Maga{\~n}a
  Zertuche}, {Magee}, {Majorana}, {Maksimovic}, {Man}, {Mandic}, {Mangano},
  {Mansell}, {Manske}, {Mantovani}, {Marchesoni}, {Marion}, {M{\'a}rka},
  {M{\'a}rka}, {Markakis}, {Markosyan}, {Markowitz}, {Maros}, {Marquina},
  {Marsh}, {Martelli}, {Martellini}, {Martin}, {Martin}, {Martynov}, {Marx},
  {Mason}, {Massera}, {Masserot}, {Massinger}, {Masso-Reid}, {Mastrogiovanni},
  {Matas}, {Matichard}, {Matone}, {Mavalvala}, {Mazumder}, {McCarthy},
  {McClelland}, {McCormick}, {McCuller}, {McGuire}, {McIntyre}, {McIver},
  {McManus}, {McNeill}, {McRae}, {McWilliams}, {Meacher}, {Meadors}, {Mehmet},
  {Meidam}, {Mejuto-Villa}, {Melatos}, {Mendell}, {Mercer}, {Merilh},
  {Merzougui}, {Meshkov}, {Messenger}, {Messick}, {Metzdorff}, {Meyers},
  {Miao}, {Michel}, {Middleton}, {Mikhailov}, {Milano}, {Miller}, {Miller},
  {Miller}, {Millhouse}, {Milovich-Goff}, {Minazzoli}, {Minenkov}, {Ming},
  {Mishra}, {Mitra}, {Mitrofanov}, {Mitselmakher}, {Mittleman}, {Moffa},
  {Moggi}, {Mogushi}, {Mohan}, {Mohapatra}, {Molina}, {Montani}, {Moore},
  {Moraru}, {Moreno}, {Morisaki}, {Morriss}, {Mours}, {Mow-Lowry}, {Mueller},
  {Muir}, {Mukherjee}, {Mukherjee}, {Mukherjee}, {Mukund}, {Mullavey}, {Munch},
  {Mu{\~n}iz}, {Muratore}, {Murray}, {Nagar}, {Napier}, {Nardecchia},
  {Naticchioni}, {Nayak}, {Neilson}, {Nelemans}, {Nelson}, {Nery}, {Neunzert},
  {Nevin}, {Newport}, {Newton}, {Ng}, {Nguyen}, {Nguyen}, {Nichols}, {Nielsen},
  {Nissanke}, {Nitz}, {Noack}, {Nocera}, {Nolting}, {North}, {Nuttall},
  {Oberling}, {O'Dea}, {Ogin}, {Oh}, {Oh}, {Ohme}, {Okada}, {Oliver},
  {Oppermann}, {Oram}, {O'Reilly}, {Ormiston}, {Ortega}, {O'Shaughnessy},
  {Ossokine}, {Ottaway}, {Overmier}, {Owen}, {Pace}, {Page}, {Page}, {Pai},
  {Pai}, {Palamos}, {Palashov}, {Palomba}, {Pal-Singh}, {Pan}, {Pan}, {Pang},
  {Pang}, {Pankow}, {Pannarale}, {Pant}, {Paoletti}, {Paoli}, {Papa}, {Parida},
  {Parker}, {Pascucci}, {Pasqualetti}, {Passaquieti}, {Passuello}, {Patil},
  {Patricelli}, {Pearlstone}, {Pedraza}, {Pedurand}, {Pekowsky}, {Pele},
  {Penn}, {Perez}, {Perreca}, {Perri}, {Pfeiffer}, {Phelps}, {Piccinni},
  {Pichot}, {Piergiovanni}, {Pierro}, {Pillant}, {Pinard}, {Pinto}, {Pirello},
  {Pitkin}, {Poe}, {Poggiani}, {Popolizio}, {Porter}, {Post}, {Powell},
  {Prasad}, {Pratt}, {Pratten}, {Predoi}, {Prestegard}, {Prijatelj},
  {Principe}, {Privitera}, {Prix}, {Prodi}, {Prokhorov}, {Puncken}, {Punturo},
  {Puppo}, {P{\"u}rrer}, {Qi}, {Quetschke}, {Quintero}, {Quitzow-James},
  {Raab}, {Rabeling}, {Radkins}, {Raffai}, {Raja}, {Rajan}, {Rajbhandari},
  {Rakhmanov}, {Ramirez}, {Ramos-Buades}, {Rapagnani}, {Raymond}, {Razzano},
  {Read}, {Regimbau}, {Rei}, {Reid}, {Reitze}, {Ren}, {Reyes}, {Ricci},
  {Ricker}, {Rieger}, {Riles}, {Rizzo}, {Robertson}, {Robie}, {Robinet},
  {Rocchi}, {Rolland}, {Rollins}, {Roma}, {Romano}, {Romano}, {Romel}, {Romie},
  {Rosi{\'n}ska}, {Ross}, {Rowan}, {R{\"u}diger}, {Ruggi}, {Rutins}, {Ryan},
  {Sachdev}, {Sadecki}, {Sadeghian}, {Sakellariadou}, {Salconi}, {Saleem},
  {Salemi}, {Samajdar}, {Sammut}, {Sampson}, {Sanchez}, {Sanchez},
  {Sanchis-Gual}, {Sandberg}, {Sanders}, {Sassolas}, {Sathyaprakash},
  {Saulson}, {Sauter}, {Savage}, {Sawadsky}, {Schale}, {Scheel}, {Scheuer},
  {Schmidt}, {Schmidt}, {Schnabel}, {Schofield}, {Sch{\"o}nbeck}, {Schreiber},
  {Schuette}, {Schulte}, {Schutz}, {Schwalbe}, {Scott}, {Scott}, {Seidel},
  {Sellers}, {Sengupta}, {Sentenac}, {Sequino}, {Sergeev}, {Shaddock},
  {Shaffer}, {Shah}, {Shahriar}, {Shaner}, {Shao}, {Shapiro}, {Shawhan},
  {Sheperd}, {Shoemaker}, {Shoemaker}, {Siellez}, {Siemens}, {Sieniawska},
  {Sigg}, {Silva}, {Singer}, {Singh}, {Singhal}, {Sintes}, {Slagmolen},
  {Smith}, {Smith}, {Smith}, {Somala}, {Son}, {Sonnenberg}, {Sorazu},
  {Sorrentino}, {Souradeep}, {Spencer}, {Srivastava}, {Staats}, {Staley},
  {Steinke}, {Steinlechner}, {Steinlechner}, {Steinmeyer}, {Stevenson},
  {Stone}, {Stops}, {Strain}, {Stratta}, {Strigin}, {Strunk}, {Sturani},
  {Stuver}, {Summerscales}, {Sun}, {Sunil}, {Suresh}, {Sutton}, {Swinkels},
  {Szczepa{\'n}czyk}, {Tacca}, {Tait}, {Talbot}, {Talukder}, {Tanner},
  {T{\'a}pai}, {Taracchini}, {Tasson}, {Taylor}, {Taylor}, {Tewari}, {Theeg},
  {Thies}, {Thomas}, {Thomas}, {Thomas}, {Thorne}, {Thorne}, {Thrane},
  {Tiwari}, {Tiwari}, {Tokmakov}, {Toland}, {Tonelli}, {Tornasi},
  {Torres-Forn{\'e}}, {Torrie}, {T{\"o}yr{\"a}}, {Travasso}, {Traylor},
  {Trinastic}, {Tringali}, {Trozzo}, {Tsang}, {Tse}, {Tso}, {Tsukada}, {Tsuna},
  {Tuyenbayev}, {Ueno}, {Ugolini}, {Unnikrishnan}, {Urban}, {Usman},
  {Vahlbruch}, {Vajente}, {Valdes}, {Vallisneri}, {van Bakel}, {van Beuzekom},
  {van den Brand}, {Van Den Broeck}, {Vander-Hyde}, {van der Schaaf}, {van
  Heijningen}, {van Veggel}, {Vardaro}, {Varma}, {Vass}, {Vas{\'u}th},
  {Vecchio}, {Vedovato}, {Veitch}, {Veitch}, {Venkateswara}, {Venugopalan},
  {Verkindt}, {Vetrano}, {Vicer{\'e}}, {Viets}, {Vinciguerra}, {Vine}, {Vinet},
  {Vitale}, {Vo}, {Vocca}, {Vorvick}, {Vyatchanin}, {Wade}, {Wade}, {Wade},
  {Walet}, {Walker}, {Wallace}, {Walsh}, {Wang}, {Wang}, {Wang}, {Wang},
  {Wang}, {Ward}, {Warner}, {Was}, {Watchi}, {Weaver}, {Wei}, {Weinert},
  {Weinstein}, {Weiss}, {Wen}, {Wessel}, {We{\ss}els}, {Westerweck},
  {Westphal}, {Wette}, {Whelan}, {Whitcomb}, {Whiting}, {Whittle}, {Wilken},
  {Williams}, {Williams}, {Williamson}, {Willis}, {Willke}, {Wimmer},
  {Winkler}, {Wipf}, {Wittel}, {Woan}, {Woehler}, {Wofford}, {Wong}, {Worden},
  {Wright}, {Wu}, {Wysocki}, {Xiao}, {Yamamoto}, {Yancey}, {Yang}, {Yap},
  {Yazback}, {Yu}, {Yu}, {Yvert}, {Zadro{\.Z}ny}, {Zanolin}, {Zelenova},
  {Zendri}, {Zevin}, {Zhang}, {Zhang}, {Zhang}, {Zhang}, {Zhao}, {Zhou},
  {Zhou}, {Zhu}, {Zhu}, {Zimmerman}, {Zucker}, {Zweizig}, {LIGO Scientific
  Collaboration}, \& {Virgo Collaboration}}]{AbbottAA2017_GW170817}
---. 2017{\natexlab{a}}, \prl, 119, 161101,
  \dodoi{10.1103/PhysRevLett.119.161101}

\bibitem[{{Abbott} {et~al.}(2017{\natexlab{b}}){Abbott}, {Abbott}, {Abbott},
  {Acernese}, {Ackley}, {Adams}, {Adams}, {Addesso}, {Adhikari}, {Adya},
  {Affeldt}, {Afrough}, {Agarwal}, {Agathos}, {Agatsuma}, {Aggarwal}, {Aguiar},
  {Aiello}, {Ain}, {Ajith}, {Allen}, {Allen}, {Allocca}, {Aloy}, {Altin},
  {Amato}, {Ananyeva}, {Anderson}, {Anderson}, {Angelova}, {Antier}, {Appert},
  {Arai}, {Araya}, {Areeda}, {Arnaud}, {Arun}, {Ascenzi}, {Ashton}, {Ast},
  {Aston}, {Astone}, {Atallah}, {Aufmuth}, {Aulbert}, {AultONeal}, {Austin},
  {Avila-Alvarez}, {Babak}, {Bacon}, {Bader}, {Bae}, {Baker}, {Baldaccini},
  {Ballardin}, {Ballmer}, {Banagiri}, {Barayoga}, {Barclay}, {Barish},
  {Barker}, {Barkett}, {Barone}, {Barr}, {Barsotti}, {Barsuglia}, {Barta},
  {Bartlett}, {Bartos}, {Bassiri}, {Basti}, {Batch}, {Bawaj}, {Bayley},
  {Bazzan}, {B{\'e}csy}, {Beer}, {Bejger}, {Belahcene}, {Bell}, {Berger},
  {Bergmann}, {Bero}, {Berry}, {Bersanetti}, {Bertolini}, {Betzwieser},
  {Bhagwat}, {Bhandare}, {Bilenko}, {Billingsley}, {Billman}, {Birch},
  {Birney}, {Birnholtz}, {Biscans}, {Biscoveanu}, {Bisht}, {Bitossi}, {Biwer},
  {Bizouard}, {Blackburn}, {Blackman}, {Blair}, {Blair}, {Blair}, {Bloemen},
  {Bock}, {Bode}, {Boer}, {Bogaert}, {Bohe}, {Bondu}, {Bonilla}, {Bonnand},
  {Boom}, {Bork}, {Boschi}, {Bose}, {Bossie}, {Bouffanais}, {Bozzi},
  {Bradaschia}, {Brady}, {Branchesi}, {Brau}, {Briant}, {Brillet}, {Brinkmann},
  {Brisson}, {Brockill}, {Broida}, {Brooks}, {Brown}, {Brown}, {Brunett},
  {Buchanan}, {Buikema}, {Bulik}, {Bulten}, {Buonanno}, {Buskulic}, {Buy},
  {Byer}, {Cabero}, {Cadonati}, {Cagnoli}, {Cahillane}, {Calder{\'o}n
  Bustillo}, {Callister}, {Calloni}, {Camp}, {Canepa}, {Canizares}, {Cannon},
  {Cao}, {Cao}, {Capano}, {Capocasa}, {Carbognani}, {Caride}, {Carney},
  {Casanueva Diaz}, {Casentini}, {Caudill}, {Cavagli{\`a}}, {Cavalier},
  {Cavalieri}, {Cella}, {Cepeda}, {Cerd{\'a}-Dur{\'a}n}, {Cerretani},
  {Cesarini}, {Chamberlin}, {Chan}, {Chao}, {Charlton}, {Chase},
  {Chassande-Mottin}, {Chatterjee}, {Chatziioannou}, {Cheeseboro}, {Chen},
  {Chen}, {Chen}, {Cheng}, {Chia}, {Chincarini}, {Chiummo}, {Chmiel}, {Cho},
  {Cho}, {Chow}, {Christensen}, {Chu}, {Chua}, {Chua}, {Chung}, {Chung},
  {Ciani}, {Ciolfi}, {Cirelli}, {Cirone}, {Clara}, {Clark}, {Clearwater},
  {Cleva}, {Cocchieri}, {Coccia}, {Cohadon}, {Cohen}, {Colla}, {Collette},
  {Cominsky}, {Constancio}, {Conti}, {Cooper}, {Corban}, {Corbitt},
  {Cordero-Carri{\'o}n}, {Corley}, {Cornish}, {Corsi}, {Cortese}, {Costa},
  {Coughlin}, {Coughlin}, {Coulon}, {Countryman}, {Couvares}, {Covas}, {Cowan},
  {Coward}, {Cowart}, {Coyne}, {Coyne}, {Creighton}, {Creighton}, {Cripe},
  {Crowder}, {Cullen}, {Cumming}, {Cunningham}, {Cuoco}, {Dal Canton},
  {D{\'a}lya}, {Danilishin}, {D'Antonio}, {Danzmann}, {Dasgupta}, {Da Silva
  Costa}, {Dattilo}, {Dave}, {Davier}, {Davis}, {Daw}, {Day}, {De}, {DeBra},
  {Degallaix}, {De Laurentis}, {Del{\'e}glise}, {Del Pozzo}, {Demos}, {Denker},
  {Dent}, {De Pietri}, {Dergachev}, {De Rosa}, {DeRosa}, {De Rossi}, {DeSalvo},
  {de Varona}, {Devenson}, {Dhurandhar}, {D{\'\i}az}, {Di Fiore}, {Di
  Giovanni}, {Di Girolamo}, {Di Lieto}, {Di Pace}, {Di Palma}, {Di Renzo},
  {Doctor}, {Dolique}, {Donovan}, {Dooley}, {Doravari}, {Dorrington},
  {Douglas}, {Dovale {\'A}lvarez}, {Downes}, {Drago}, {Dreissigacker},
  {Driggers}, {Du}, {Ducrot}, {Dupej}, {Dwyer}, {Edo}, {Edwards}, {Effler},
  {Eggenstein}, {Ehrens}, {Eichholz}, {Eikenberry}, {Eisenstein}, {Essick},
  {Estevez}, {Etienne}, {Etzel}, {Evans}, {Evans}, {Factourovich}, {Fafone},
  {Fair}, {Fairhurst}, {Fan}, {Farinon}, {Farr}, {Farr}, {Fauchon-Jones},
  {Favata}, {Fays}, {Fee}, {Fehrmann}, {Feicht}, {Fejer}, {Fernandez-Galiana},
  {Ferrante}, {Ferreira}, {Ferrini}, {Fidecaro}, {Finstad}, {Fiori},
  {Fiorucci}, {Fishbach}, {Fisher}, {Fitz-Axen}, {Flaminio}, {Fletcher},
  {Fong}, {Font}, {Forsyth}, {Forsyth}, {Fournier}, {Frasca}, {Frasconi},
  {Frei}, {Freise}, {Frey}, {Frey}, {Fries}, {Fritschel}, {Frolov}, {Fulda},
  {Fyffe}, {Gabbard}, {Gadre}, {Gaebel}, {Gair}, {Gammaitoni}, {Ganija},
  {Gaonkar}, {Garcia-Quiros}, {Garufi}, {Gateley}, {Gaudio}, {Gaur},
  {Gayathri}, {Gehrels}, {Gemme}, {Genin}, {Gennai}, {George}, {George},
  {Gergely}, {Germain}, {Ghonge}, {Ghosh}, {Ghosh}, {Ghosh}, {Giaime},
  {Giardina}, {Giazotto}, {Gill}, {Glover}, {Goetz}, {Goetz}, {Gomes},
  {Goncharov}, {Gonz{\'a}lez}, {Gonzalez Castro}, {Gopakumar}, {Gorodetsky},
  {Gossan}, {Gosselin}, {Gouaty}, {Grado}, {Graef}, {Granata}, {Grant}, {Gras},
  {Gray}, {Greco}, {Green}, {Gretarsson}, {Groot}, {Grote}, {Grunewald},
  {Gruning}, {Guidi}, {Guo}, {Gupta}, {Gupta}, {Gushwa}, {Gustafson},
  {Gustafson}, {Halim}, {Hall}, {Hall}, {Hamilton}, {Hammond}, {Haney},
  {Hanke}, {Hanks}, {Hanna}, {Hannam}, {Hannuksela}, {Hanson}, {Hardwick},
  {Harms}, {Harry}, {Harry}, {Hart}, {Haster}, {Haughian}, {Healy}, {Heidmann},
  {Heintze}, {Heitmann}, {Hello}, {Hemming}, {Hendry}, {Heng}, {Hennig},
  {Heptonstall}, {Heurs}, {Hild}, {Hinderer}, {Hoak}, {Hofman}, {Holt}, {Holz},
  {Hopkins}, {Horst}, {Hough}, {Houston}, {Howell}, {Hreibi}, {Hu}, {Huerta},
  {Huet}, {Hughey}, {Husa}, {Huttner}, {Huynh-Dinh}, {Indik}, {Inta}, {Intini},
  {Isa}, {Isac}, {Isi}, {Iyer}, {Izumi}, {Jacqmin}, {Jani}, {Jaranowski},
  {Jawahar}, {Jim{\'e}nez-Forteza}, {Johnson}, {Johnson-McDaniel}, {Jones},
  {Jones}, {Jonker}, {Ju}, {Junker}, {Kalaghatgi}, {Kalogera}, {Kamai},
  {Kandhasamy}, {Kang}, {Kanner}, {Kapadia}, {Karki}, {Karvinen}, {Kasprzack},
  {Kastaun}, {Katolik}, {Katsavounidis}, {Katzman}, {Kaufer}, {Kawabe},
  {K{\'e}f{\'e}lian}, {Keitel}, {Kemball}, {Kennedy}, {Kent}, {Key}, {Khalili},
  {Khan}, {Khan}, {Khan}, {Khazanov}, {Kijbunchoo}, {Kim}, {Kim}, {Kim}, {Kim},
  {Kim}, {Kim}, {Kimbrell}, {King}, {King}, {Kinley-Hanlon}, {Kirchhoff},
  {Kissel}, {Kleybolte}, {Klimenko}, {Knowles}, {Koch}, {Koehlenbeck}, {Koley},
  {Kondrashov}, {Kontos}, {Korobko}, {Korth}, {Kowalska}, {Kozak},
  {Kr{\"a}mer}, {Kringel}, {Krishnan}, {Kr{\'o}lak}, {Kuehn}, {Kumar}, {Kumar},
  {Kumar}, {Kuo}, {Kutynia}, {Kwang}, {Lackey}, {Lai}, {Landry}, {Lang},
  {Lange}, {Lantz}, {Lanza}, {Lartaux-Vollard}, {Lasky}, {Laxen}, {Lazzarini},
  {Lazzaro}, {Leaci}, {Leavey}, {Lee}, {Lee}, {Lee}, {Lee}, {Lee}, {Lehmann},
  {Lenon}, {Leonardi}, {Leroy}, {Letendre}, {Levin}, {Li}, {Linker},
  {Littenberg}, {Liu}, {Lo}, {Lockerbie}, {London}, {Lord}, {Lorenzini},
  {Loriette}, {Lormand}, {Losurdo}, {Lough}, {Lousto}, {Lovelace}, {L{\"u}ck},
  {Lumaca}, {Lundgren}, {Lynch}, {Ma}, {Macas}, {Macfoy}, {Machenschalk},
  {MacInnis}, {Macleod}, {Maga{\~n}a Hernandez}, {Maga{\~n}a-Sandoval},
  {Maga{\~n}a Zertuche}, {Magee}, {Majorana}, {Maksimovic}, {Man}, {Mandic},
  {Mangano}, {Mansell}, {Manske}, {Mantovani}, {Marchesoni}, {Marion},
  {M{\'a}rka}, {M{\'a}rka}, {Markakis}, {Markosyan}, {Markowitz}, {Maros},
  {Marquina}, {Martelli}, {Martellini}, {Martin}, {Martin}, {Martynov},
  {Mason}, {Massera}, {Masserot}, {Massinger}, {Masso-Reid}, {Mastrogiovanni},
  {Matas}, {Matichard}, {Matone}, {Mavalvala}, {Mazumder}, {McCarthy},
  {McClelland}, {McCormick}, {McCuller}, {McGuire}, {McIntyre}, {McIver},
  {McManus}, {McNeill}, {McRae}, {McWilliams}, {Meacher}, {Meadors}, {Mehmet},
  {Meidam}, {Mejuto-Villa}, {Melatos}, {Mendell}, {Mercer}, {Merilh},
  {Merzougui}, {Meshkov}, {Messenger}, {Messick}, {Metzdorff}, {Meyers},
  {Miao}, {Michel}, {Middleton}, {Mikhailov}, {Milano}, {Miller}, {Miller},
  {Miller}, {Millhouse}, {Milovich-Goff}, {Minazzoli}, {Minenkov}, {Ming},
  {Mishra}, {Mitra}, {Mitrofanov}, {Mitselmakher}, {Mittleman}, {Moffa},
  {Moggi}, {Mogushi}, {Mohan}, {Mohapatra}, {Montani}, {Moore}, {Moraru},
  {Moreno}, {Morriss}, {Mours}, {Mow-Lowry}, {Mueller}, {Muir}, {Mukherjee},
  {Mukherjee}, {Mukherjee}, {Mukund}, {Mullavey}, {Munch}, {Mu{\~n}iz},
  {Muratore}, {Murray}, {Napier}, {Nardecchia}, {Naticchioni}, {Nayak},
  {Neilson}, {Nelemans}, {Nelson}, {Nery}, {Neunzert}, {Nevin}, {Newport},
  {Newton}, {Ng}, {Nguyen}, {Nichols}, {Nielsen}, {Nissanke}, {Nitz}, {Noack},
  {Nocera}, {Nolting}, {North}, {Nuttall}, {Oberling}, {O'Dea}, {Ogin}, {Oh},
  {Oh}, {Ohme}, {Okada}, {Oliver}, {Oppermann}, {Oram}, {O'Reilly}, {Ormiston},
  {Ortega}, {O'Shaughnessy}, {Ossokine}, {Ottaway}, {Overmier}, {Owen}, {Pace},
  {Page}, {Page}, {Pai}, {Pai}, {Palamos}, {Palashov}, {Palomba}, {Pal-Singh},
  {Pan}, {Pan}, {Pang}, {Pang}, {Pankow}, {Pannarale}, {Pant}, {Paoletti},
  {Paoli}, {Papa}, {Parida}, {Parker}, {Pascucci}, {Pasqualetti},
  {Passaquieti}, {Passuello}, {Patil}, {Patricelli}, {Pearlstone}, {Pedraza},
  {Pedurand}, {Pekowsky}, {Pele}, {Penn}, {Perez}, {Perreca}, {Perri},
  {Pfeiffer}, {Phelps}, {Piccinni}, {Pichot}, {Piergiovanni}, {Pierro},
  {Pillant}, {Pinard}, {Pinto}, {Pirello}, {Pitkin}, {Poe}, {Poggiani},
  {Popolizio}, {Porter}, {Post}, {Powell}, {Prasad}, {Pratt}, {Pratten},
  {Predoi}, {Prestegard}, {Prijatelj}, {Principe}, {Privitera}, {Prodi},
  {Prokhorov}, {Puncken}, {Punturo}, {Puppo}, {P{\"u}rrer}, {Qi}, {Quetschke},
  {Quintero}, {Quitzow-James}, {Raab}, {Rabeling}, {Radkins}, {Raffai}, {Raja},
  {Rajan}, {Rajbhandari}, {Rakhmanov}, {Ramirez}, {Ramos-Buades}, {Rapagnani},
  {Raymond}, {Razzano}, {Read}, {Regimbau}, {Rei}, {Reid}, {Reitze}, {Ren},
  {Reyes}, {Ricci}, {Ricker}, {Rieger}, {Riles}, {Rizzo}, {Robertson}, {Robie},
  {Robinet}, {Rocchi}, {Rolland}, {Rollins}, {Roma}, {Romano}, {Romel},
  {Romie}, {Rosi{\'n}ska}, {Ross}, {Rowan}, {R{\"u}diger}, {Ruggi}, {Rutins},
  {Ryan}, {Sachdev}, {Sadecki}, {Sadeghian}, {Sakellariadou}, {Salconi},
  {Saleem}, {Salemi}, {Samajdar}, {Sammut}, {Sampson}, {Sanchez}, {Sanchez},
  {Sanchis-Gual}, {Sandberg}, {Sanders}, {Sassolas}, {Sathyaprakash},
  {Saulson}, {Sauter}, {Savage}, {Sawadsky}, {Schale}, {Scheel}, {Scheuer},
  {Schmidt}, {Schmidt}, {Schnabel}, {Schofield}, {Sch{\"o}nbeck}, {Schreiber},
  {Schuette}, {Schulte}, {Schutz}, {Schwalbe}, {Scott}, {Scott}, {Seidel},
  {Sellers}, {Sengupta}, {Sentenac}, {Sequino}, {Sergeev}, {Shaddock},
  {Shaffer}, {Shah}, {Shahriar}, {Shaner}, {Shao}, {Shapiro}, {Shawhan},
  {Sheperd}, {Shoemaker}, {Shoemaker}, {Siellez}, {Siemens}, {Sieniawska},
  {Sigg}, {Silva}, {Singer}, {Singh}, {Singhal}, {Sintes}, {Slagmolen},
  {Smith}, {Smith}, {Smith}, {Somala}, {Son}, {Sonnenberg}, {Sorazu},
  {Sorrentino}, {Souradeep}, {Spencer}, {Srivastava}, {Staats}, {Staley},
  {Steinke}, {Steinlechner}, {Steinlechner}, {Steinmeyer}, {Stevenson},
  {Stone}, {Stops}, {Strain}, {Stratta}, {Strigin}, {Strunk}, {Sturani},
  {Stuver}, {Summerscales}, {Sun}, {Sunil}, {Suresh}, {Sutton}, {Swinkels},
  {Szczepa{\'n}czyk}, {Tacca}, {Tait}, {Talbot}, {Talukder}, {Tanner},
  {T{\'a}pai}, {Taracchini}, {Tasson}, {Taylor}, {Taylor}, {Tewari}, {Theeg},
  {Thies}, {Thomas}, {Thomas}, {Thomas}, {Thorne}, {Thorne}, {Thrane},
  {Tiwari}, {Tiwari}, {Tokmakov}, {Toland}, {Tonelli}, {Tornasi},
  {Torres-Forn{\'e}}, {Torrie}, {T{\"o}yr{\"a}}, {Travasso}, {Traylor},
  {Trinastic}, {Tringali}, {Trozzo}, {Tsang}, {Tse}, {Tso}, {Tsukada}, {Tsuna},
  {Tuyenbayev}, {Ueno}, {Ugolini}, {Unnikrishnan}, {Urban}, {Usman},
  {Vahlbruch}, {Vajente}, {Valdes}, {van Bakel}, {van Beuzekom}, {van den
  Brand}, {Van Den Broeck}, {Vander-Hyde}, {van der Schaaf}, {van Heijningen},
  {van Veggel}, {Vardaro}, {Varma}, {Vass}, {Vas{\'u}th}, {Vecchio},
  {Vedovato}, {Veitch}, {Veitch}, {Venkateswara}, {Venugopalan}, {Verkindt},
  {Vetrano}, {Vicer{\'e}}, {Viets}, {Vinciguerra}, {Vine}, {Vinet}, {Vitale},
  {Vo}, {Vocca}, {Vorvick}, {Vyatchanin}, {Wade}, {Wade}, {Wade}, {Walet},
  {Walker}, {Wallace}, {Walsh}, {Wang}, {Wang}, {Wang}, {Wang}, {Wang}, {Ward},
  {Warner}, {Was}, {Watchi}, {Weaver}, {Wei}, {Weinert}, {Weinstein}, {Weiss},
  {Wen}, {Wessel}, {We{\ss}els}, {Westerweck}, {Westphal}, {Wette}, {Whelan},
  {Whitcomb}, {Whiting}, {Whittle}, {Wilken}, {Williams}, {Williams},
  {Williamson}, {Willis}, {Willke}, {Wimmer}, {Winkler}, {Wipf}, {Wittel},
  {Woan}, {Woehler}, {Wofford}, {Wong}, {Worden}, {Wright}, {Wu}, {Wysocki},
  {Xiao}, {Yamamoto}, {Yancey}, {Yang}, {Yap}, {Yazback}, {Yu}, {Yu}, {Yvert},
  {Zadro{\.z}ny}, {Zanolin}, {Zelenova}, {Zendri}, {Zevin}, {Zhang}, {Zhang},
  {Zhang}, {Zhang}, {Zhao}, {Zhou}, {Zhou}, {Zhu}, {Zhu}, {Zimmerman},
  {Zucker}, {Zweizig}, {(LIGO Scientific Collaboration}, {Virgo Collaboration},
  {Burns}, {Veres}, {Kocevski}, {Racusin}, {Goldstein}, {Connaughton},
  {Briggs}, {Blackburn}, {Hamburg}, {Hui}, {von Kienlin}, {McEnery}, {Preece},
  {Wilson-Hodge}, {Bissaldi}, {Cleveland}, {Gibby}, {Giles}, {Kippen},
  {McBreen}, {Meegan}, {Paciesas}, {Poolakkil}, {Roberts}, {Stanbro},
  {Gamma-ray Burst Monitor}, {Savchenko}, {Ferrigno}, {Kuulkers}, {Bazzano},
  {Bozzo}, {Brandt}, {Chenevez}, {Courvoisier}, {Diehl}, {Domingo}, {Hanlon},
  {Jourdain}, {Laurent}, {Lebrun}, {Lutovinov}, {Mereghetti}, {Natalucci},
  {Rodi}, {Roques}, {Sunyaev}, {Ubertini}, \&
  {(INTEGRAL}}]{AbbottAA_gamma_ray2017}
---. 2017{\natexlab{b}}, \apjl, 848, L13, \dodoi{10.3847/2041-8213/aa920c}

\bibitem[{{Abbott} {et~al.}(2019){Abbott}, {Abbott}, {Abbott}, {Abraham},
  {Acernese}, {Ackley}, {Adams}, {Adhikari}, {Adya}, {Affeldt}, {Agathos},
  {Agatsuma}, {Aggarwal}, {Aguiar}, {Aiello}, {Ain}, {Ajith}, {Allen},
  {Allocca}, {Aloy}, {Altin}, {Amato}, {Ananyeva}, {Anderson}, {Anderson},
  {Angelova}, {Antier}, {Appert}, {Arai}, {Araya}, {Areeda}, {Ar{\`e}ne},
  {Arnaud}, {Arun}, {Ascenzi}, {Ashton}, {Aston}, {Astone}, {Aubin}, {Aufmuth},
  {AultONeal}, {Austin}, {Avendano}, {Avila-Alvarez}, {Babak}, {Bacon},
  {Badaracco}, {Bader}, {Bae}, {Baker}, {Baldaccini}, {Ballardin}, {Ballmer},
  {Banagiri}, {Barayoga}, {Barclay}, {Barish}, {Barker}, {Barkett}, {Barnum},
  {Barone}, {Barr}, {Barsotti}, {Barsuglia}, {Barta}, {Bartlett}, {Bartos},
  {Bassiri}, {Basti}, {Bawaj}, {Bayley}, {Bazzan}, {B{\'e}csy}, {Bejger},
  {Belahcene}, {Bell}, {Beniwal}, {Berger}, {Bergmann}, {Bernuzzi}, {Bero},
  {Berry}, {Bersanetti}, {Bertolini}, {Betzwieser}, {Bhand are}, {Bidler},
  {Bilenko}, {Bilgili}, {Billingsley}, {Birch}, {Birney}, {Birnholtz},
  {Biscans}, {Biscoveanu}, {Bisht}, {Bitossi}, {Bizouard}, {Blackburn},
  {Blackman}, {Blair}, {Blair}, {Blair}, {Bloemen}, {Bode}, {Boer}, {Boetzel},
  {Bogaert}, {Bondu}, {Bonilla}, {Bonnand}, {Booker}, {Boom}, {Booth}, {Bork},
  {Boschi}, {Bose}, {Bossie}, {Bossilkov}, {Bosveld}, {Bouffanais}, {Bozzi},
  {Bradaschia}, {Brady}, {Bramley}, {Branchesi}, {Brau}, {Briant}, {Briggs},
  {Brighenti}, {Brillet}, {Brinkmann}, {Brisson}, {Brockill}, {Brooks},
  {Brown}, {Brunett}, {Buikema}, {Bulik}, {Bulten}, {Buonanno}, {Buskulic},
  {Bustamante Rosell}, {Buy}, {Byer}, {Cabero}, {Cadonati}, {Cagnoli},
  {Cahillane}, {Calder{\'o}n Bustillo}, {Callister}, {Calloni}, {Camp},
  {Campbell}, {Canepa}, {Cannon}, {Cao}, {Cao}, {Capocasa}, {Carbognani},
  {Caride}, {Carney}, {Carullo}, {Casanueva Diaz}, {Casentini}, {Caudill},
  {Cavagli{\`a}}, {Cavalier}, {Cavalieri}, {Cella}, {Cerd{\'a}-Dur{\'a}n},
  {Cerretani}, {Cesarini}, {Chaibi}, {Chakravarti}, {Chamberlin}, {Chan},
  {Chao}, {Charlton}, {Chase}, {Chassand e-Mottin}, {Chatterjee}, {Chaturvedi},
  {Chatziioannou}, {Cheeseboro}, {Chen}, {Chen}, {Chen}, {Cheng}, {Cheong},
  {Chia}, {Chincarini}, {Chiummo}, {Cho}, {Cho}, {Cho}, {Christensen}, {Chu},
  {Chua}, {Chung}, {Chung}, {Ciani}, {Ciobanu}, {Ciolfi}, {Cipriano}, {Cirone},
  {Clara}, {Clark}, {Clearwater}, {Cleva}, {Cocchieri}, {Coccia}, {Cohadon},
  {Cohen}, {Colgan}, {Colleoni}, {Collette}, {Collins}, {Cominsky},
  {Constancio}, {Conti}, {Cooper}, {Corban}, {Corbitt}, {Cordero-Carri{\'o}n},
  {Corley}, {Cornish}, {Corsi}, {Cortese}, {Costa}, {Cotesta}, {Coughlin},
  {Coughlin}, {Coulon}, {Countryman}, {Couvares}, {Covas}, {Cowan}, {Coward},
  {Cowart}, {Coyne}, {Coyne}, {Creighton}, {Creighton}, {Cripe}, {Croquette},
  {Crowder}, {Cullen}, {Cumming}, {Cunningham}, {Cuoco}, {Canton}, {D{\'a}lya},
  {Danilishin}, {D'Antonio}, {Danzmann}, {Dasgupta}, {Da Silva Costa},
  {Datrier}, {Dattilo}, {Dave}, {Davier}, {Davis}, {Daw}, {DeBra},
  {Deenadayalan}, {Degallaix}, {De Laurentis}, {Del{\'e}glise}, {Del Pozzo},
  {DeMarchi}, {Demos}, {Dent}, {De Pietri}, {Derby}, {De Rosa}, {De Rossi},
  {DeSalvo}, {de Varona}, {Dhurandhar}, {D{\'\i}az}, {Dietrich}, {Di Fiore},
  {Di Giovanni}, {Di Girolamo}, {Di Lieto}, {Ding}, {Di Pace}, {Di Palma}, {Di
  Renzo}, {Dmitriev}, {Doctor}, {Donovan}, {Dooley}, {Doravari}, {Dorrington},
  {Downes}, {Drago}, {Driggers}, {Du}, {Ducoin}, {Dupej}, {Dwyer}, {Easter},
  {Edo}, {Edwards}, {Effler}, {Ehrens}, {Eichholz}, {Eikenberry}, {Eisenmann},
  {Eisenstein}, {Essick}, {Estelles}, {Estevez}, {Etienne}, {Etzel}, {Evans},
  {Evans}, {Fafone}, {Fair}, {Fairhurst}, {Fan}, {Farinon}, {Farr}, {Farr},
  {Fauchon-Jones}, {Favata}, {Fays}, {Fazio}, {Fee}, {Feicht}, {Fejer}, {Feng},
  {Fernand ez-Galiana}, {Ferrante}, {Ferreira}, {Ferreira}, {Ferrini},
  {Fidecaro}, {Fiori}, {Fiorucci}, {Fishbach}, {Fisher}, {Fishner},
  {Fitz-Axen}, {Flaminio}, {Fletcher}, {Flynn}, {Fong}, {Font}, {Forsyth},
  {Fournier}, {Frasca}, {Frasconi}, {Frei}, {Freise}, {Frey}, {Frey},
  {Fritschel}, {Frolov}, {Fulda}, {Fyffe}, {Gabbard}, {Gadre}, {Gaebel},
  {Gair}, {Gammaitoni}, {Ganija}, {Gaonkar}, {Garcia},
  {Garc{\'\i}a-Quir{\'o}s}, {Garufi}, {Gateley}, {Gaudio}, {Gaur}, {Gayathri},
  {Gemme}, {Genin}, {Gennai}, {George}, {George}, {Gergely}, {Germain},
  {Ghonge}, {Ghosh}, {Ghosh}, {Ghosh}, {Giacomazzo}, {Giaime}, {Giardina},
  {Giazotto}, {Gill}, {Giordano}, {Glover}, {Godwin}, {Goetz}, {Goetz},
  {Goncharov}, {Gonz{\'a}lez}, {Gonzalez Castro}, {Gopakumar}, {Gorodetsky},
  {Gossan}, {Gosselin}, {Gouaty}, {Grado}, {Graef}, {Granata}, {Grant}, {Gras},
  {Grassia}, {Gray}, {Gray}, {Greco}, {Green}, {Green}, {Gretarsson}, {Groot},
  {Grote}, {Grunewald}, {Gruning}, {Guidi}, {Gulati}, {Guo}, {Gupta}, {Gupta},
  {Gustafson}, {Gustafson}, {Haegel}, {Halim}, {Hall}, {Hall}, {Hamilton},
  {Hammond}, {Haney}, {Hanke}, {Hanks}, {Hanna}, {Hannam}, {Hannuksela},
  {Hanson}, {Hardwick}, {Haris}, {Harms}, {Harry}, {Harry}, {Haster},
  {Haughian}, {Hayes}, {Healy}, {Heidmann}, {Heintze}, {Heitmann}, {Hello},
  {Hemming}, {Hendry}, {Heng}, {Hennig}, {Heptonstall}, {Hernandez Vivanco},
  {Heurs}, {Hild}, {Hinderer}, {Hoak}, {Hochheim}, {Hofman}, {Holgado},
  {Holland }, {Holt}, {Holz}, {Hopkins}, {Horst}, {Hough}, {Howell}, {Hoy},
  {Hreibi}, {Huang}, {Huerta}, {Huet}, {Hughey}, {Hulko}, {Husa}, {Huttner},
  {Huynh-Dinh}, {Idzkowski}, {Iess}, {Ingram}, {Inta}, {Intini}, {Irwin},
  {Isa}, {Isac}, {Isi}, {Iyer}, {Izumi}, {Jacqmin}, {Jadhav}, {Jani},
  {Janthalur}, {Jaranowski}, {Jenkins}, {Jiang}, {Johnson}, {Johnson-McDaniel},
  {Jones}, {Jones}, {Jones}, {Jonker}, {Ju}, {Junker}, {Kalaghatgi},
  {Kalogera}, {Kamai}, {Kand hasamy}, {Kang}, {Kanner}, {Kapadia}, {Karki},
  {Karvinen}, {Kashyap}, {Kasprzack}, {Katsanevas}, {Katsavounidis}, {Katzman},
  {Kaufer}, {Kawabe}, {Keerthana}, {K{\'e}f{\'e}lian}, {Keitel}, {Kennedy},
  {Key}, {Khalili}, {Khan}, {Khan}, {Khan}, {Khan}, {Khazanov}, {Khursheed},
  {Kijbunchoo}, {Kim}, {Kim}, {Kim}, {Kim}, {Kim}, {Kim}, {Kimball}, {King},
  {King}, {Kinley-Hanlon}, {Kirchhoff}, {Kissel}, {Kleybolte}, {Klika},
  {Klimenko}, {Knowles}, {Koch}, {Koehlenbeck}, {Koekoek}, {Koley},
  {Kondrashov}, {Kontos}, {Koper}, {Korobko}, {Korth}, {Kowalska}, {Kozak},
  {Kringel}, {Krishnendu}, {Kr{\'o}lak}, {Kuehn}, {Kumar}, {Kumar}, {Kumar},
  {Kumar}, {Kuo}, {Kutynia}, {Kwang}, {Lackey}, {Lai}, {Lam}, {Landry}, {Lane},
  {Lang}, {Lange}, {Lantz}, {Lanza}, {Lartaux-Vollard}, {Lasky}, {Laxen},
  {Lazzarini}, {Lazzaro}, {Leaci}, {Leavey}, {Lecoeuche}, {Lee}, {Lee}, {Lee},
  {Lee}, {Lee}, {Lee}, {Lehmann}, {Lenon}, {Leroy}, {Letendre}, {Levin}, {Li},
  {Li}, {Li}, {Li}, {Lin}, {Linde}, {Linker}, {Littenberg}, {Liu}, {Liu}, {Lo},
  {Lockerbie}, {London}, {Longo}, {Lorenzini}, {Loriette}, {Lormand},
  {Losurdo}, {Lough}, {Lousto}, {Lovelace}, {Lower}, {L{\"u}ck}, {Lumaca},
  {Lundgren}, {Lynch}, {Ma}, {Macas}, {Macfoy}, {MacInnis}, {Macleod},
  {Macquet}, {Maga{\~n}a-Sandoval}, {Maga{\~n}a Zertuche}, {Magee}, {Majorana},
  {Maksimovic}, {Malik}, {Man}, {Mandic}, {Mangano}, {Mansell}, {Manske},
  {Mantovani}, {Marchesoni}, {Marion}, {M{\'a}rka}, {M{\'a}rka}, {Markakis},
  {Markosyan}, {Markowitz}, {Maros}, {Marquina}, {Marsat}, {Martelli},
  {Martin}, {Martin}, {Martynov}, {Mason}, {Massera}, {Masserot}, {Massinger},
  {Masso-Reid}, {Mastrogiovanni}, {Matas}, {Matichard}, {Matone}, {Mavalvala},
  {Mazumder}, {McCann}, {McCarthy}, {McClelland }, {McCormick}, {McCuller},
  {McGuire}, {McIver}, {McManus}, {McRae}, {McWilliams}, {Meacher}, {Meadors},
  {Mehmet}, {Mehta}, {Meidam}, {Melatos}, {Mendell}, {Mercer}, {Mereni},
  {Merilh}, {Merzougui}, {Meshkov}, {Messenger}, {Messick}, {Metzdorff},
  {Meyers}, {Miao}, {Michel}, {Middleton}, {Mikhailov}, {Milano}, {Miller},
  {Miller}, {Millhouse}, {Mills}, {Milovich-Goff}, {Minazzoli}, {Minenkov},
  {Mishkin}, {Mishra}, {Mistry}, {Mitra}, {Mitrofanov}, {Mitselmakher},
  {Mittleman}, {Mo}, {Moffa}, {Mogushi}, {Mohapatra}, {Montani}, {Moore},
  {Moraru}, {Moreno}, {Morisaki}, {Mours}, {Mow-Lowry}, {Mukherjee},
  {Mukherjee}, {Mukherjee}, {Mukund}, {Mullavey}, {Munch}, {Mu{\~n}iz},
  {Muratore}, {Murray}, {Nagar}, {Nardecchia}, {Naticchioni}, {Nayak},
  {Neilson}, {Nelemans}, {Nelson}, {Nery}, {Neunzert}, {Ng}, {Ng}, {Nguyen},
  {Nichols}, {Nielsen}, {Nissanke}, {Nitz}, {Nocera}, {North}, {Nuttall},
  {Obergaulinger}, {Oberling}, {O'Brien}, {O'Dea}, {Ogin}, {Oh}, {Oh}, {Ohme},
  {Ohta}, {Okada}, {Oliver}, {Oppermann}, {Oram}, {O'Reilly}, {Ormiston},
  {Ortega}, {O'Shaughnessy}, {Ossokine}, {Ottaway}, {Overmier}, {Owen}, {Pace},
  {Pagano}, {Page}, {Pai}, {Pai}, {Palamos}, {Palashov}, {Palomba},
  {Pal-Singh}, {Pan}, {Pang}, {Pang}, {Pankow}, {Pannarale}, {Pant},
  {Paoletti}, {Paoli}, {Papa}, {Parida}, {Parker}, {Pascucci}, {Pasqualetti},
  {Passaquieti}, {Passuello}, {Patil}, {Patricelli}, {Pearlstone}, {Pedersen},
  {Pedraza}, {Pedurand}, {Pele}, {Penn}, {Perego}, {Perez}, {Perreca},
  {Pfeiffer}, {Phelps}, {Phukon}, {Piccinni}, {Pichot}, {Piergiovanni},
  {Pillant}, {Pinard}, {Pirello}, {Pitkin}, {Poggiani}, {Pong}, {Ponrathnam},
  {Popolizio}, {Porter}, {Powell}, {Prajapati}, {Prasad}, {Prasai}, {Prasanna},
  {Pratten}, {Prestegard}, {Privitera}, {Prodi}, {Prokhorov}, {Puncken},
  {Punturo}, {Puppo}, {P{\"u}rrer}, {Qi}, {Quetschke}, {Quinonez}, {Quintero},
  {Quitzow-James}, {Raab}, {Radkins}, {Radulescu}, {Raffai}, {Raja}, {Rajan},
  {Rajbhandari}, {Rakhmanov}, {Ramirez}, {Ramos-Buades}, {Rana}, {Rao},
  {Rapagnani}, {Raymond}, {Razzano}, {Read}, {Regimbau}, {Rei}, {Reid},
  {Reitze}, {Ren}, {Ricci}, {Richardson}, {Richardson}, {Ricker},
  {Riemenschneider}, {Riles}, {Rizzo}, {Robertson}, {Robie}, {Robinet},
  {Rocchi}, {Rolland}, {Rollins}, {Roma}, {Romanelli}, {Romano}, {Romel},
  {Romie}, {Rose}, {Rosi{\'n}ska}, {Rosofsky}, {Ross}, {Rowan}, {R{\"u}diger},
  {Ruggi}, {Rutins}, {Ryan}, {Sachdev}, {Sadecki}, {Sakellariadou}, {Salafia},
  {Salconi}, {Saleem}, {Salemi}, {Samajdar}, {Sammut}, {Sanchez}, {Sanchez},
  {Sanchis-Gual}, {Sandberg}, {Sand ers}, {Santiago}, {Sarin}, {Sassolas},
  {Sathyaprakash}, {Saulson}, {Sauter}, {Savage}, {Schale}, {Scheel},
  {Scheuer}, {Schmidt}, {Schnabel}, {Schofield}, {Sch{\"o}nbeck}, {Schreiber},
  {Schulte}, {Schutz}, {Schwalbe}, {Scott}, {Scott}, {Seidel}, {Sellers},
  {Sengupta}, {Sennett}, {Sentenac}, {Sequino}, {Sergeev}, {Setyawati},
  {Shaddock}, {Shaffer}, {Shahriar}, {Shaner}, {Shao}, {Sharma}, {Shawhan},
  {Shen}, {Shink}, {Shoemaker}, {Shoemaker}, {ShyamSundar}, {Siellez},
  {Sieniawska}, {Sigg}, {Silva}, {Singer}, {Singh}, {Singhal}, {Sintes},
  {Sitmukhambetov}, {Skliris}, {Slagmolen}, {Slaven-Blair}, {Smith}, {Smith},
  {Somala}, {Son}, {Sorazu}, {Sorrentino}, {Souradeep}, {Sowell}, {Spencer},
  {Srivastava}, {Srivastava}, {Staats}, {Stachie}, {Standke}, {Steer},
  {Steinke}, {Steinlechner}, {Steinlechner}, {Steinmeyer}, {Stevenson},
  {Stocks}, {Stone}, {Stops}, {Strain}, {Stratta}, {Strigin}, {Strunk},
  {Sturani}, {Stuver}, {Sudhir}, {Summerscales}, {Sun}, {Sunil}, {Suresh},
  {Sutton}, {Swinkels}, {Szczepa{\'n}czyk}, {Tacca}, {Tait}, {Talbot},
  {Talukder}, {Tanner}, {T{\'a}pai}, {Taracchini}, {Tasson}, {Taylor}, {Thies},
  {Thomas}, {Thomas}, {Thondapu}, {Thorne}, {Thrane}, {Tiwari}, {Tiwari},
  {Tiwari}, {Toland}, {Tonelli}, {Tornasi}, {Torres-Forn{\'e}}, {Torrie},
  {T{\"o}yr{\"a}}, {Travasso}, {Traylor}, {Tringali}, {Trovato}, {Trozzo},
  {Trudeau}, {Tsang}, {Tse}, {Tso}, {Tsukada}, {Tsuna}, {Tuyenbayev}, {Ueno},
  {Ugolini}, {Unnikrishnan}, {Urban}, {Usman}, {Vahlbruch}, {Vajente},
  {Valdes}, {van Bakel}, {van Beuzekom}, {van den Brand}, {Van Den Broeck},
  {Vander-Hyde}, {van Heijningen}, {van der Schaaf}, {van Veggel}, {Vardaro},
  {Varma}, {Vass}, {Vas{\'u}th}, {Vecchio}, {Vedovato}, {Veitch}, {Veitch},
  {Venkateswara}, {Venugopalan}, {Verkindt}, {Vetrano}, {Vicer{\'e}}, {Viets},
  {Vine}, {Vinet}, {Vitale}, {Vo}, {Vocca}, {Vorvick}, {Vyatchanin}, {Wade},
  {Wade}, {Wade}, {Walet}, {Walker}, {Wallace}, {Walsh}, {Wang}, {Wang},
  {Wang}, {Wang}, {Wang}, {Ward}, {Warden}, {Warner}, {Was}, {Watchi},
  {Weaver}, {Wei}, {Weinert}, {Weinstein}, {Weiss}, {Wellmann}, {Wen},
  {Wessel}, {We{\ss}els}, {Westhouse}, {Wette}, {Whelan}, {White}, {Whiting},
  {Whittle}, {Wilken}, {Williams}, {Williamson}, {Willis}, {Willke}, {Wimmer},
  {Winkler}, {Wipf}, {Wittel}, {Woan}, {Woehler}, {Wofford}, {Worden},
  {Wright}, {Wu}, {Wysocki}, {Xiao}, {Yamamoto}, {Yancey}, {Yang}, {Yap},
  {Yazback}, {Yeeles}, {Yu}, {Yu}, {Yuen}, {Yvert}, {Zadro{\.Z}ny}, {Zanolin},
  {Zappa}, {Zelenova}, {Zendri}, {Zevin}, {Zhang}, {Zhang}, {Zhang}, {Zhao},
  {Zhou}, {Zhou}, {Zhu}, {Zimmerman}, {Zlochower}, {Zucker}, {Zweizig}, {LIGO
  Scientific Collaboration}, \& {Virgo Collaboration}}]{AbbottAA2019}
---. 2019, Physical Review X, 9, 031040, \dodoi{10.1103/PhysRevX.9.031040}

\bibitem[{{Abbott} {et~al.}(2020){Abbott}, {Abbott}, {Abbott}, {Abraham},
  {Acernese}, {Ackley}, {Adams}, {Adya}, {Affeldt}, {Agathos}, \&
  et~al.}]{AbbottAA2020}
---. 2020, \prd, 101, 084002, \dodoi{10.1103/PhysRevD.101.084002}

\bibitem[{{Abbott} {et~al.}(2021){Abbott}, {Abbott}, {Abraham}, {Acernese},
  {Ackley}, {Adams}, {Adams}, {Adhikari}, {Adya}, {Affeldt}, {Agathos},
  {Agatsuma}, {Aggarwal}, {Aguiar}, {Aiello}, {Ain}, {Ajith}, {Akcay}, {Allen},
  {Allocca}, {Altin}, {Amato}, {Anand}, {Ananyeva}, {Anderson}, {Anderson},
  {Angelova}, {Ansoldi}, {Antelis}, {Antier}, {Appert}, {Arai}, {Araya},
  {Areeda}, {Ar{\`e}ne}, {Arnaud}, {Aronson}, {Arun}, {Asali}, {Ascenzi},
  {Ashton}, {Aston}, {Astone}, {Aubin}, {Aufmuth}, {AultONeal}, {Austin},
  {Avendano}, {Babak}, {Badaracco}, {Bader}, {Bae}, {Baer}, {Bagnasco},
  {Baird}, {Ball}, {Ballardin}, {Ballmer}, {Bals}, {Balsamo}, {Baltus},
  {Banagiri}, {Bankar}, {Bankar}, {Barayoga}, {Barbieri}, {Barish}, {Barker},
  {Barneo}, {Barnum}, {Barone}, {Barr}, {Barsotti}, {Barsuglia}, {Barta},
  {Bartlett}, {Bartos}, {Bassiri}, {Basti}, {Bawaj}, {Bayley}, {Bazzan},
  {Becher}, {B{\'e}csy}, {Bedakihale}, {Bejger}, {Belahcene}, {Beniwal},
  {Benjamin}, {Bennett}, {Bentley}, {Bergamin}, {Berger}, {Bergmann},
  {Bernuzzi}, {Berry}, {Bersanetti}, {Bertolini}, {Betzwieser}, {Bhandare},
  {Bhandari}, {Bhattacharjee}, {Bidler}, {Bilenko}, {Billingsley}, {Birney},
  {Birnholtz}, {Biscans}, {Bischi}, {Biscoveanu}, {Bisht}, {Bitossi},
  {Bizouard}, {Blackburn}, {Blackman}, {Blair}, {Blair}, {Blair}, {Blanch},
  {Bobba}, {Bode}, {Boer}, {Boetzel}, {Bogaert}, {Boldrini}, {Bondu},
  {Bonilla}, {Bonnand}, {Booker}, {Boom}, {Bork}, {Boschi}, {Bose},
  {Bossilkov}, {Boudart}, {Bouffanais}, {Bozzi}, {Bradaschia}, {Brady},
  {Bramley}, {Branchesi}, {Brau}, {Breschi}, {Briant}, {Briggs}, {Brighenti},
  {Brillet}, {Brinkmann}, {Brockill}, {Brooks}, {Brooks}, {Brown}, {Brunett},
  {Bruno}, {Bruntz}, {Buikema}, {Bulik}, {Bulten}, {Buonanno}, {Buscicchio},
  {Buskulic}, {Byer}, {Cabero}, {Cadonati}, {Caesar}, {Cagnoli}, {Cahillane},
  {Calder{\'o}n Bustillo}, {Callaghan}, {Callister}, {Calloni}, {Camp},
  {Canepa}, {Cannon}, {Cao}, {Cao}, {Carapella}, {Carbognani}, {Carney},
  {Carpinelli}, {Carullo}, {Carver}, {Casanueva Diaz}, {Casentini}, {Caudill},
  {Cavagli{\`a}}, {Cavalier}, {Cavalieri}, {Cella}, {Cerd{\'a}-Dur{\'a}n},
  {Cesarini}, {Chaibi}, {Chakravarti}, {Chan}, {Chan}, {Chandra}, {Chanial},
  {Chao}, {Charlton}, {Chase}, {Chassande-Mottin}, {Chatterjee},
  {Chattopadhyay}, {Chaturvedi}, {Chatziioannou}, {Chen}, {Chen}, {Chen},
  {Chen}, {Cheng}, {Cheong}, {Chia}, {Chiadini}, {Chierici}, {Chincarini},
  {Chiummo}, {Cho}, {Cho}, {Cho}, {Choate}, {Christensen}, {Chu}, {Chua},
  {Chung}, {Chung}, {Ciani}, {Ciecielag}, {Cie{\'s}lar}, {Cifaldi}, {Ciobanu},
  {Ciolfi}, {Cipriano}, {Cirone}, {Clara}, {Clark}, {Clark}, {Clarke},
  {Clearwater}, {Clesse}, {Cleva}, {Coccia}, {Cohadon}, {Cohen}, {Colleoni},
  {Collette}, {Collins}, {Colpi}, {Constancio}, {Conti}, {Cooper}, {Corban},
  {Corbitt}, {Cordero-Carri{\'o}n}, {Corezzi}, {Corley}, {Cornish}, {Corre},
  {Corsi}, {Cortese}, {Costa}, {Cotesta}, {Coughlin}, {Coughlin}, {Coulon},
  {Countryman}, {Cousins}, {Couvares}, {Covas}, {Coward}, {Cowart}, {Coyne},
  {Coyne}, {Creighton}, {Creighton}, {Croquette}, {Crowder}, {Cudell},
  {Cullen}, {Cumming}, {Cummings}, {Cunningham}, {Cuoco}, {Cury{\l}o},
  {Canton}, {D{\'a}lya}, {Dana}, {DaneshgaranBajastani}, {D'Angelo}, {Danila},
  {Danilishin}, {D'Antonio}, {Danzmann}, {Darsow-Fromm}, {Dasgupta}, {Datrier},
  {Dattilo}, {Dave}, {Davier}, {Davies}, {Davis}, {Daw}, {Dean}, {DeBra},
  {Deenadayalan}, {Degallaix}, {De Laurentis}, {Del{\'e}glise}, {Del Favero},
  {De Lillo}, {De Lillo}, {Del Pozzo}, {DeMarchi}, {De Matteis}, {D'Emilio},
  {Demos}, {Denker}, {Dent}, {Depasse}, {De Pietri}, {De Rosa}, {De Rossi},
  {DeSalvo}, {de Varona}, {Dhurandhar}, {D{\'\i}az}, {Diaz-Ortiz}, {Didio},
  {Dietrich}, {Di Fiore}, {DiFronzo}, {Di Giorgio}, {Di Giovanni}, {Di
  Giovanni}, {Di Girolamo}, {Di Lieto}, {Ding}, {Di Pace}, {Di Palma}, {Di
  Renzo}, {Divakarla}, {Dmitriev}, {Doctor}, {D'Onofrio}, {Donovan}, {Dooley},
  {Doravari}, {Dorrington}, {Downes}, {Drago}, {Driggers}, {Du}, {Ducoin},
  {Dupej}, {Durante}, {D'Urso}, {Duverne}, {Dwyer}, {Easter}, {Eddolls},
  {Edelman}, {Edo}, {Edy}, {Effler}, {Eichholz}, {Eikenberry}, {Eisenmann},
  {Eisenstein}, {Ejlli}, {Errico}, {Essick}, {Estell{\'e}s}, {Estevez},
  {Etienne}, {Etzel}, {Evans}, {Evans}, {Ewing}, {Fafone}, {Fair}, {Fairhurst},
  {Fan}, {Farah}, {Farinon}, {Farr}, {Farr}, {Fauchon-Jones}, {Favata}, {Fays},
  {Fazio}, {Feicht}, {Fejer}, {Feng}, {Fenyvesi}, {Ferguson},
  {Fernandez-Galiana}, {Ferrante}, {Ferreira}, {Fidecaro}, {Figura}, {Fiori},
  {Fiorucci}, {Fishbach}, {Fisher}, {Fishner}, {Fittipaldi}, {Fitz-Axen},
  {Fiumara}, {Flaminio}, {Floden}, {Flynn}, {Fong}, {Font}, {Forsyth},
  {Fournier}, {Frasca}, {Frasconi}, {Frei}, {Freise}, {Frey}, {Frey},
  {Fritschel}, {Frolov}, {Fronz{\'e}}, {Fulda}, {Fyffe}, {Gabbard}, {Gadre},
  {Gaebel}, {Gair}, {Gais}, {Galaudage}, {Gamba}, {Ganapathy}, {Ganguly},
  {Gaonkar}, {Garaventa}, {Garc{\'\i}a-Quir{\'o}s}, {Garufi}, {Gateley},
  {Gaudio}, {Gayathri}, {Gemme}, {Gennai}, {George}, {George}, {George},
  {Gergely}, {Ghonge}, {Ghosh}, {Ghosh}, {Ghosh}, {Giacomazzo}, {Giacoppo},
  {Giaime}, {Giardina}, {Gibson}, {Gier}, {Gill}, {Giri}, {Glanzer}, {Gleckl},
  {Godwin}, {Goetz}, {Goetz}, {Gohlke}, {Goncharov}, {Gonz{\'a}lez},
  {Gopakumar}, {Gossan}, {Gosselin}, {Gouaty}, {Grace}, {Grado}, {Granata},
  {Granata}, {Grant}, {Gras}, {Grassia}, {Gray}, {Gray}, {Greco}, {Green},
  {Green}, {Gretarsson}, {Griggs}, {Grignani}, {Grimaldi}, {Grimes}, {Grimm},
  {Grote}, {Grunewald}, {Gruning}, {Guerrero}, {Guidi}, {Guimaraes},
  {Guix{\'e}}, {Gulati}, {Guo}, {Gupta}, {Gupta}, {Gupta}, {Gustafson},
  {Gustafson}, {Guzman}, {Haegel}, {Halim}, {Hall}, {Hamilton}, {Hammond},
  {Haney}, {Hanke}, {Hanks}, {Hanna}, {Hannam}, {Hannuksela}, {Hannuksela},
  {Hansen}, {Hansen}, {Hanson}, {Harder}, {Hardwick}, {Haris}, {Harms},
  {Harry}, {Harry}, {Hartwig}, {Hasskew}, {Haster}, {Haughian}, {Hayes},
  {Healy}, {Heidmann}, {Heintze}, {Heinze}, {Heinzel}, {Heitmann}, {Hellman},
  {Hello}, {Helmling-Cornell}, {Hemming}, {Hendry}, {Heng}, {Hennes}, {Hennig},
  {Hennig}, {Hernandez Vivanco}, {Heurs}, {Hild}, {Hill}, {Hines}, {Hochheim},
  {Hofgard}, {Hofman}, {Hohmann}, {Holgado}, {Holland}, {Hollows}, {Holmes},
  {Holt}, {Holz}, {Hopkins}, {Horst}, {Hough}, {Howell}, {Hoy}, {Hoyland},
  {Huang}, {H{\"u}bner}, {Huddart}, {Huerta}, {Hughey}, {Hui}, {Husa},
  {Huttner}, {Hutzler}, {Huxford}, {Huynh-Dinh}, {Idzkowski}, {Iess},
  {Imperato}, {Inchauspe}, {Ingram}, {Intini}, {Isi}, {Iyer},
  {JaberianHamedan}, {Jacqmin}, {Jadhav}, {Jadhav}, {James}, {Jani},
  {Janssens}, {Janthalur}, {Jaranowski}, {Jariwala}, {Jaume}, {Jenkins},
  {Jeunon}, {Jiang}, {Johns}, {Johnson-McDaniel}, {Jones}, {Jones}, {Jones},
  {Jones}, {Jones}, {Jonker}, {Ju}, {Junker}, {Kalaghatgi}, {Kalogera},
  {Kamai}, {Kandhasamy}, {Kang}, {Kanner}, {Kapadia}, {Kapasi}, {Karathanasis},
  {Karki}, {Kashyap}, {Kasprzack}, {Kastaun}, {Katsanevas}, {Katsavounidis},
  {Katzman}, {Kawabe}, {K{\'e}f{\'e}lian}, {Keitel}, {Key}, {Khadka},
  {Khalili}, {Khan}, {Khan}, {Khazanov}, {Khetan}, {Khursheed}, {Kijbunchoo},
  {Kim}, {Kim}, {Kim}, {Kim}, {Kim}, {Kim}, {Kimball}, {King}, {Kinley-Hanlon},
  {Kirchhoff}, {Kissel}, {Kleybolte}, {Klimenko}, {Knowles}, {Knyazev}, {Koch},
  {Koehlenbeck}, {Koekoek}, {Koley}, {Kolstein}, {Komori}, {Kondrashov},
  {Kontos}, {Koper}, {Korobko}, {Korth}, {Kovalam}, {Kozak}, {Kr{\"a}mer},
  {Kringel}, {Krishnendu}, {Kr{\'o}lak}, {Kuehn}, {Kumar}, {Kumar}, {Kumar},
  {Kumar}, {Kuns}, {Kwang}, {Lackey}, {Laghi}, {Lalande}, {Lam}, {Lamberts},
  {Landry}, {Lane}, {Lang}, {Lange}, {Lantz}, {Lanza}, {La Rosa},
  {Lartaux-Vollard}, {Lasky}, {Laxen}, {Lazzarini}, {Lazzaro}, {Leaci},
  {Leavey}, {Lecoeuche}, {Lee}, {Lee}, {Lee}, {Lee}, {Lehmann}, {Leon},
  {Leroy}, {Letendre}, {Levin}, {Li}, {Li}, {Li}, {Li}, {Li}, {Linde},
  {Linker}, {Linley}, {Littenberg}, {Liu}, {Liu}, {Llorens-Monteagudo}, {Lo},
  {Lockwood}, {London}, {Longo}, {Lorenzini}, {Loriette}, {Lormand}, {Losurdo},
  {Lough}, {Lousto}, {Lovelace}, {L{\"u}ck}, {Lumaca}, {Lundgren}, {Ma},
  {Macas}, {MacInnis}, {Macleod}, {MacMillan}, {Macquet}, {Maga{\~n}a
  Hernandez}, {Maga{\~n}a-Sandoval}, {Magazz{\`u}}, {Magee}, {Majorana},
  {Maksimovic}, {Maliakal}, {Malik}, {Man}, {Mandic}, {Mangano}, {Mansell},
  {Manske}, {Mantovani}, {Mapelli}, {Marchesoni}, {Marion}, {M{\'a}rka},
  {M{\'a}rka}, {Markakis}, {Markosyan}, {Markowitz}, {Maros}, {Marquina},
  {Marsat}, {Martelli}, {Martin}, {Martin}, {Martinez}, {Martinez}, {Martynov},
  {Masalehdan}, {Mason}, {Massera}, {Masserot}, {Massinger}, {Masso-Reid},
  {Mastrogiovanni}, {Matas}, {Mateu-Lucena}, {Matichard}, {Matiushechkina},
  {Mavalvala}, {Maynard}, {McCann}, {McCarthy}, {McClelland}, {McCormick},
  {McCuller}, {McGuire}, {McIsaac}, {McIver}, {McManus}, {McRae}, {McWilliams},
  {Meacher}, {Meadors}, {Mehmet}, {Mehta}, {Melatos}, {Melchor}, {Mendell},
  {Menendez-Vazquez}, {Mercer}, {Mereni}, {Merfeld}, {Merilh}, {Merritt},
  {Merzougui}, {Meshkov}, {Messenger}, {Messick}, {Metzdorff}, {Meyers},
  {Meylahn}, {Mhaske}, {Miani}, {Miao}, {Michaloliakos}, {Michel}, {Middleton},
  {Milano}, {Miller}, {Millhouse}, {Mills}, {Milotti}, {Milovich-Goff},
  {Minazzoli}, {Minenkov}, {Mir}, {Mishkin}, {Mishra}, {Mistry}, {Mitra},
  {Mitrofanov}, {Mitselmakher}, {Mittleman}, {Mo}, {Mogushi}, {Mohapatra},
  {Mohite}, {Molina}, {Molina-Ruiz}, {Mondin}, {Montani}, {Moore}, {Moraru},
  {Morawski}, {Moreno}, {Morisaki}, {Mours}, {Mow-Lowry}, {Mozzon},
  {Muciaccia}, {Mukherjee}, {Mukherjee}, {Mukherjee}, {Mukherjee}, {Mukund},
  {Mullavey}, {Munch}, {Mu{\~n}iz}, {Murray}, {Nadji}, {Nagar}, {Nardecchia},
  {Naticchioni}, {Nayak}, {Neil}, {Neilson}, {Nelemans}, {Nelson}, {Nery},
  {Neunzert}, {Nitz}, {Ng}, {Ng}, {Nguyen}, {Nguyen}, {Nguyen}, {Nichols},
  {Nissanke}, {Nocera}, {Noh}, {North}, {Nothard}, {Nuttall}, {Oberling},
  {O'Brien}, {O'Dell}, {Oganesyan}, {Ogin}, {Oh}, {Oh}, {Ohme}, {Ohta},
  {Okada}, {Olivetto}, {Oppermann}, {Oram}, {O'Reilly}, {Ormiston}, {Ortega},
  {O'Shaughnessy}, {Ossokine}, {Osthelder}, {Ottaway}, {Overmier}, {Owen},
  {Pace}, {Pagano}, {Page}, {Pagliaroli}, {Pai}, {Pai}, {Palamos}, {Palashov},
  {Palomba}, {Pan}, {Panda}, {Pang}, {Pankow}, {Pannarale}, {Pant}, {Paoletti},
  {Paoli}, {Paolone}, {Parker}, {Pascucci}, {Pasqualetti}, {Passaquieti},
  {Passuello}, {Patel}, {Patricelli}, {Payne}, {Pechsiri}, {Pedraza},
  {Pegoraro}, {Pele}, {Penn}, {Perego}, {Perez}, {P{\'e}rigois}, {Perreca},
  {Perri{\`e}s}, {Petermann}, {Petterson}, {Pfeiffer}, {Pham}, {Phukon},
  {Piccinni}, {Pichot}, {Piendibene}, {Piergiovanni}, {Pierini}, {Pierro},
  {Pillant}, {Pilo}, {Pinard}, {Pinto}, {Piotrzkowski}, {Pirello}, {Pitkin},
  {Placidi}, {Plastino}, {Pluchar}, {Poggiani}, {Polini}, {Pong}, {Ponrathnam},
  {Popolizio}, {Porter}, {Poverman}, {Powell}, {Pracchia}, {Prajapati},
  {Prasai}, {Prasanna}, {Pratten}, {Prestegard}, {Principe}, {Prodi},
  {Prokhorov}, {Prosposito}, {Prudenzi}, {Puecher}, {Punturo}, {Puosi},
  {Puppo}, {P{\"u}rrer}, {Qi}, {Quetschke}, {Quinonez}, {Quitzow-James},
  {Raab}, {Raaijmakers}, {Radkins}, {Radulesco}, {Raffai}, {Rafferty}, {Rail},
  {Raja}, {Rajan}, {Rajbhandari}, {Rakhmanov}, {Ramirez}, {Ramirez},
  {Ramos-Buades}, {Rana}, {Rao}, {Rapagnani}, {Rapol}, {Ratto}, {Raymond},
  {Razzano}, {Read}, {Regimbau}, {Rei}, {Reid}, {Reitze}, {Rettegno}, {Ricci},
  {Richardson}, {Richardson}, {Richardson}, {Ricker}, {Riemenschneider},
  {Riles}, {Rizzo}, {Robertson}, {Robinet}, {Rocchi}, {Rocha}, {Rodriguez},
  {Rodriguez-Soto}, {Rolland}, {Rollins}, {Roma}, {Romanelli}, {Romano},
  {Romel}, {Romero}, {Romero-Shaw}, {Romie}, {Ronchini}, {Rose}, {Rose},
  {Rose}, {Rosell}, {Rosi{\'n}ska}, {Rosofsky}, {Ross}, {Rowan}, {Rowlinson},
  {Roy}, {Roy}, {Ruggi}, {Ryan}, {Sachdev}, {Sadecki}, {Sadiq},
  {Sakellariadou}, {Salafia}, {Salconi}, {Saleem}, {Samajdar}, {Sanchez},
  {Sanchez}, {Sanchez}, {Sanchis-Gual}, {Sanders}, {Sandles}, {Santiago},
  {Santos}, {Saravanan}, {Sarin}, {Sassolas}, {Sathyaprakash}, {Sauter},
  {Savage}, {Savant}, {Sawant}, {Sayah}, {Schaetzl}, {Schale}, {Scheel},
  {Scheuer}, {Schindler-Tyka}, {Schmidt}, {Schnabel}, {Schofield},
  {Sch{\"o}nbeck}, {Schreiber}, {Schulte}, {Schutz}, {Schwarm}, {Schwartz},
  {Scott}, {Scott}, {Seglar-Arroyo}, {Seidel}, {Sellers}, {Sengupta},
  {Sennett}, {Sentenac}, {Sequino}, {Sergeev}, {Setyawati}, {Shaffer},
  {Shahriar}, {Sharifi}, {Sharma}, {Sharma}, {Shawhan}, {Shen}, {Shikauchi},
  {Shink}, {Shoemaker}, {Shoemaker}, {Shukla}, {ShyamSundar}, {Sieniawska},
  {Sigg}, {Singer}, {Singh}, {Singh}, {Singha}, {Singhal}, {Sintes}, {Sipala},
  {Skliris}, {Slagmolen}, {Slaven-Blair}, {Smetana}, {Smith}, {Smith},
  {Somala}, {Son}, {Soni}, {Soni}, {Sorazu}, {Sordini}, {Sorrentino},
  {Sorrentino}, {Soulard}, {Souradeep}, {Sowell}, {Spencer}, {Spera},
  {Srivastava}, {Srivastava}, {Staats}, {Stachie}, {Steer}, {Steinhoff},
  {Steinke}, {Steinlechner}, {Steinlechner}, {Steinmeyer}, {Stevenson},
  {Stolle-McAllister}, {Stops}, {Stover}, {Strain}, {Stratta}, {Strunk},
  {Sturani}, {Stuver}, {S{\"u}dbeck}, {Sudhagar}, {Sudhir}, {Suh},
  {Summerscales}, {Sun}, {Sun}, {Sunil}, {Sur}, {Suresh}, {Sutton}, {Swinkels},
  {Szczepa{\'n}czyk}, {Tacca}, {Tait}, {Talbot}, {Tanasijczuk}, {Tanner},
  {Tao}, {Tapia}, {Tapia San Martin}, {Tasson}, {Taylor}, {Tenorio},
  {Terkowski}, {Thirugnanasambandam}, {Thomas}, {Thomas}, {Thomas}, {Thompson},
  {Thondapu}, {Thorne}, {Thrane}, {Tiwari}, {Tiwari}, {Tiwari}, {Toland},
  {Tolley}, {Tonelli}, {Tornasi}, {Torres-Forn{\'e}}, {Torrie}, {e Melo},
  {T{\"o}yr{\"a}}, {Tran}, {Trapananti}, {Travasso}, {Traylor}, {Tringali},
  {Tripathee}, {Trovato}, {Trudeau}, {Tsai}, {Tsang}, {Tse}, {Tso}, {Tsukada},
  {Tsuna}, {Tsutsui}, {Turconi}, {Ubhi}, {Udall}, {Ueno}, {Ugolini},
  {Unnikrishnan}, {Urban}, {Usman}, {Utina}, {Vahlbruch}, {Vajente}, {Vajpeyi},
  {Valdes}, {Valentini}, {Valsan}, {van Bakel}, {van Beuzekom}, {van den
  Brand}, {Van Den Broeck}, {Vander-Hyde}, {van der Schaaf}, {van Heijningen},
  {Vardaro}, {Vargas}, {Varma}, {Vass}, {Vas{\'u}th}, {Vecchio}, {Vedovato},
  {Veitch}, {Veitch}, {Venkateswara}, {Venneberg}, {Venugopalan}, {Verkindt},
  {Verma}, {Veske}, {Vetrano}, {Vicer{\'e}}, {Viets}, {Vijaykumar},
  {Villa-Ortega}, {Vinet}, {Vitale}, {Vo}, {Vocca}, {Vorvick}, {Vyatchanin},
  {Wade}, {Wade}, {Wade}, {Walet}, {Walker}, {Wallace}, {Wallace}, {Walsh},
  {Wang}, {Wang}, {Wang}, {Wang}, {Ward}, {Warner}, {Was}, {Washington},
  {Watchi}, {Weaver}, {Wei}, {Weinert}, {Weinstein}, {Weiss}, {Wellmann},
  {Wen}, {We{\ss}els}, {Westhouse}, {Wette}, {Whelan}, {White}, {White},
  {Whiting}, {Whittle}, {Wilken}, {Williams}, {Williams}, {Williamson},
  {Willis}, {Willke}, {Wilson}, {Wimmer}, {Winkler}, {Wipf}, {Woan}, {Woehler},
  {Wofford}, {Wong}, {Wrangel}, {Wright}, {Wu}, {Wysocki}, {Xiao}, {Yamamoto},
  {Yang}, {Yang}, {Yang}, {Yap}, {Yeeles}, {Yoon}, {Yu}, {Yu}, {Yuen},
  {Zadro{\.Z}ny}, {Zanolin}, {Zelenova}, {Zendri}, {Zevin}, {Zhang}, {Zhang},
  {Zhang}, {Zhang}, {Zhao}, {Zhao}, {Zheng}, {Zhou}, {Zhou}, {Zhu},
  {Zimmerman}, {Zlochower}, {Zucker}, {Zweizig}, {LIGO Scientific
  Collaboration}, \& {Virgo Collaboration}}]{GWTC2-2021}
{Abbott}, R., {Abbott}, T.~D., {Abraham}, S., {et~al.} 2021, Physical Review X,
  11, 021053, \dodoi{10.1103/PhysRevX.11.021053}

\bibitem[{{Akhshi} {et~al.}(2021){Akhshi}, {Alimohammadi}, {Baghram}, {Rahvar},
  {Tabar}, \& {Arfaei}}]{AkhshiAB2021}
{Akhshi}, A., {Alimohammadi}, H., {Baghram}, S., {et~al.} 2021, Scientific
  Reports, 11, 20507, \dodoi{10.1038/s41598-021-98821-z}

\bibitem[{{Andresen} {et~al.}(2017){Andresen}, {M{\"u}ller}, {M{\"u}ller}, \&
  {Janka}}]{AndresenMM2017}
{Andresen}, H., {M{\"u}ller}, B., {M{\"u}ller}, E., \& {Janka}, H.~T. 2017,
  \mnras, 468, 2032, \dodoi{10.1093/mnras/stx618}

\bibitem[{Bedrosian(1962)}]{Bedrosian1962}
Bedrosian, E. 1962, A Product Theorem for Hilbert Transforms (Santa Monica, CA:
  RAND Corporation)

\bibitem[{{Blondin} {et~al.}(2003){Blondin}, {Mezzacappa}, \&
  {DeMarino}}]{BlondinMD2003}
{Blondin}, J.~M., {Mezzacappa}, A., \& {DeMarino}, C. 2003, \apj, 584, 971,
  \dodoi{10.1086/345812}

\bibitem[{{Boh{\'e}} {et~al.}(2017){Boh{\'e}}, {Shao}, {Taracchini},
  {Buonanno}, {Babak}, {Harry}, {Hinder}, {Ossokine}, {P{\"u}rrer}, {Raymond},
  {Chu}, {Fong}, {Kumar}, {Pfeiffer}, {Boyle}, {Hemberger}, {Kidder},
  {Lovelace}, {Scheel}, \& {Szil{\'a}gyi}}]{BoheST2017}
{Boh{\'e}}, A., {Shao}, L., {Taracchini}, A., {et~al.} 2017, \prd, 95, 044028,
  \dodoi{10.1103/PhysRevD.95.044028}

\bibitem[{{Burrows} \& {Vartanyan}(2021)}]{BurrowsV2021}
{Burrows}, A., \& {Vartanyan}, D. 2021, \nat, 589, 29,
  \dodoi{10.1038/s41586-020-03059-w}

\bibitem[{{Camp} {et~al.}(2007){Camp}, {Cannizzo}, \& {Numata}}]{Camp2007}
{Camp}, J.~B., {Cannizzo}, J.~K., \& {Numata}, K. 2007, \prd, 75, 061101,
  \dodoi{10.1103/PhysRevD.75.061101}

\bibitem[{Chatterji {et~al.}(2004)Chatterji, Blackburn, Martin, \&
  Katsavounidis}]{Chatterji2004}
Chatterji, S., Blackburn, L., Martin, G., \& Katsavounidis, E. 2004, Classical
  and Quantum Gravity, 21

\bibitem[{{Foster}(1996)}]{Foster1996}
{Foster}, G. 1996, \aj, 112, 1709, \dodoi{10.1086/118137}

\bibitem[{{Fryxell} {et~al.}(2000){Fryxell}, {Olson}, {Ricker}, {Timmes},
  {Zingale}, {Lamb}, {MacNeice}, {Rosner}, {Truran}, \& {Tufo}}]{FryxellOR2000}
{Fryxell}, B., {Olson}, K., {Ricker}, P., {et~al.} 2000, \apjs, 131, 273,
  \dodoi{10.1086/317361}

\bibitem[{{Hu} {et~al.}(2014){Hu}, {Chou}, {Yang}, \& {Su}}]{Hu2014}
{Hu}, C.-P., {Chou}, Y., {Yang}, T.-C., \& {Su}, Y.-H. 2014, \apj, 788, 31,
  \dodoi{10.1088/0004-637X/788/1/31}

\bibitem[{{Hu} {et~al.}(2019){Hu}, {Mihara}, {Sugizaki}, {Ueda}, \&
  {Enoto}}]{HuMS2019}
{Hu}, C.-P., {Mihara}, T., {Sugizaki}, M., {Ueda}, Y., \& {Enoto}, T. 2019,
  \apj, 885, 123, \dodoi{10.3847/1538-4357/ab48e4}

\bibitem[{Huang {et~al.}(2021)Huang, Zhou, Liu, Zhu, \& Shao}]{HuangZL2021}
Huang, L., Zhou, Y., Liu, L., Zhu, F., \& Shao, L. 2021, in 2021 IEEE/CVF
  Conference on Computer Vision and Pattern Recognition (CVPR), 9507--9516

\bibitem[{Huang \& Attoh-Okine(2005)}]{HuangA2005}
Huang, N., \& Attoh-Okine, N. 2005, The Hilbert-Huang Transform in Engineering
  (CRC Press).
\newblock \url{https://books.google.com.tw/books?id=bO\_KBQAAQBAJ}

\bibitem[{Huang \& Shen(2014)}]{HuangS2014}
Huang, N., \& Shen, S. 2014, Hilbert-Huang Transform and Its Applications,
  EBSCO ebook academic collection (World Scientific).
\newblock \url{https://books.google.com.tw/books?id=aJ26CgAAQBAJ}

\bibitem[{{Huang} {et~al.}(1999){Huang}, {Shen}, \& {Long}}]{HuangSL1999}
{Huang}, N.~E., {Shen}, Z., \& {Long}, S.~R. 1999, Annual Review of Fluid
  Mechanics, 31, 417, \dodoi{10.1146/annurev.fluid.31.1.417}

\bibitem[{Huang {et~al.}(2009)Huang, Wu, Long, Arnold, Chen, \&
  Blank}]{Huang2009}
Huang, N.~E., Wu, Z., Long, S.~R., {et~al.} 2009, Adv. Data Sci. Adapt. Anal.,
  1, 177, \dodoi{10.1142/S1793536909000096}

\bibitem[{{Huang} {et~al.}(1998){Huang}, {Shen}, {Long}, {Wu}, {Shih}, {Zheng},
  {Yen}, {Tung}, \& {Liu}}]{Huang1998}
{Huang}, N.~E., {Shen}, Z., {Long}, S.~R., {et~al.} 1998, Royal Society of
  London Proceedings Series A, 454, 903, \dodoi{10.1098/rspa.1998.0193}

\bibitem[{{Janka} {et~al.}(2016){Janka}, {Melson}, \& {Summa}}]{JankaMS2016}
{Janka}, H.-T., {Melson}, T., \& {Summa}, A. 2016, Annual Review of Nuclear and
  Particle Science, 66, 341, \dodoi{10.1146/annurev-nucl-102115-044747}

\bibitem[{{Kaneyama} {et~al.}(2016){Kaneyama}, {Oohara}, {Takahashi},
  {Sekiguchi}, {Tagoshi}, \& {Shibata}}]{KaneyamaOT2016}
{Kaneyama}, M., {Oohara}, K.-i., {Takahashi}, H., {et~al.} 2016, \prd, 93,
  123010, \dodoi{10.1103/PhysRevD.93.123010}

\bibitem[{{Kuroda} {et~al.}(2016){Kuroda}, {Kotake}, \&
  {Takiwaki}}]{KurodaKT2016}
{Kuroda}, T., {Kotake}, K., \& {Takiwaki}, T. 2016, \apjl, 829, L14,
  \dodoi{10.3847/2041-8205/829/1/L14}

\bibitem[{{Lattimer} \& {Swesty}(1991)}]{LattimerS1991}
{Lattimer}, J.~M., \& {Swesty}, D.~F. 1991, \nphysa, 535, 331,
  \dodoi{10.1016/0375-9474(91)90452-C}

\bibitem[{{Liebend{\"o}rfer} {et~al.}(2009){Liebend{\"o}rfer}, {Whitehouse}, \&
  {Fischer}}]{LiebendoerferWF2009}
{Liebend{\"o}rfer}, M., {Whitehouse}, S.~C., \& {Fischer}, T. 2009, \apj, 698,
  1174, \dodoi{10.1088/0004-637X/698/2/1174}

\bibitem[{{LIGO Scientific Collaboration}(2018)}]{lalsuite}
{LIGO Scientific Collaboration}. 2018, {LIGO} {A}lgorithm {L}ibrary -
  {LALS}uite, free software (GPL), \dodoi{10.7935/GT1W-FZ16}

\bibitem[{{Macleod} {et~al.}(2020){Macleod}, {Urban}, {Coughlin}, {Massinger},
  {Pitkin}, {Paulaltin}, {Areeda}, {Quintero}, {Badger}, {Singer}, \&
  {Leinweber}}]{MacleodUC2020}
{Macleod}, D., {Urban}, A.~L., {Coughlin}, S., {et~al.} 2020, {gwpy/gwpy:
  1.0.1}, v1.0.1,  Zenodo, \dodoi{10.5281/zenodo.3598469}

\bibitem[{{Marek} {et~al.}(2006){Marek}, {Dimmelmeier}, {Janka}, {M{\"u}ller},
  \& {Buras}}]{MarekDJ2006}
{Marek}, A., {Dimmelmeier}, H., {Janka}, H.~T., {M{\"u}ller}, E., \& {Buras},
  R. 2006, \aap, 445, 273, \dodoi{10.1051/0004-6361:20052840}

\bibitem[{MATLAB(2018)}]{matlab2018b}
MATLAB. 2018, 9.7.0 (R2018b) (Natick, Massachusetts: The MathWorks Inc.)

\bibitem[{{Mezzacappa} {et~al.}(2020){Mezzacappa}, {Marronetti}, {Landfield},
  {Lentz}, {Yakunin}, {Bruenn}, {Hix}, {Messer}, {Endeve}, {Blondin}, \&
  {Harris}}]{MezzacappaML2020}
{Mezzacappa}, A., {Marronetti}, P., {Landfield}, R.~E., {et~al.} 2020, \prd,
  102, 023027, \dodoi{10.1103/PhysRevD.102.023027}

\bibitem[{{M{\"u}ller} {et~al.}(2013){M{\"u}ller}, {Janka}, \&
  {Marek}}]{MullerJM2013}
{M{\"u}ller}, B., {Janka}, H.-T., \& {Marek}, A. 2013, \apj, 766, 43,
  \dodoi{10.1088/0004-637X/766/1/43}

\bibitem[{{Murphy} {et~al.}(2009){Murphy}, {Ott}, \& {Burrows}}]{MurphyOB2009}
{Murphy}, J.~W., {Ott}, C.~D., \& {Burrows}, A. 2009, \apj, 707, 1173,
  \dodoi{10.1088/0004-637X/707/2/1173}

\bibitem[{{Pan} {et~al.}(2021){Pan}, {Liebend{\"o}rfer}, {Couch}, \&
  {Thielemann}}]{PanLC2021}
{Pan}, K.-C., {Liebend{\"o}rfer}, M., {Couch}, S.~M., \& {Thielemann}, F.-K.
  2021, \apj, 914, 140, \dodoi{10.3847/1538-4357/abfb05}

\bibitem[{{P{\"u}rrer}(2016)}]{Purrer2016}
{P{\"u}rrer}, M. 2016, \prd, 93, 064041, \dodoi{10.1103/PhysRevD.93.064041}

\bibitem[{{Sakai} {et~al.}(2017){Sakai}, {Oohara}, {Kaneyama}, \&
  {Takahashi}}]{SakaiOM2017}
{Sakai}, K., {Oohara}, K.-i., {Kaneyama}, M., \& {Takahashi}, H. 2017, ICIC
  Express Letters, 11, 45, \dodoi{10.24507/icicel.11.01.45}

\bibitem[{{Scargle} {et~al.}(2013){Scargle}, {Norris}, {Jackson}, \&
  {Chiang}}]{ScargleNJ2013}
{Scargle}, J.~D., {Norris}, J.~P., {Jackson}, B., \& {Chiang}, J. 2013, \apj,
  764, 167, \dodoi{10.1088/0004-637X/764/2/167}

\bibitem[{{Scheidegger} {et~al.}(2008){Scheidegger}, {Fischer}, {Whitehouse},
  \& {Liebend{\"o}rfer}}]{ScheideggerFW2008}
{Scheidegger}, S., {Fischer}, T., {Whitehouse}, S.~C., \& {Liebend{\"o}rfer},
  M. 2008, \aap, 490, 231, \dodoi{10.1051/0004-6361:20078577}

\bibitem[{{Sekhar} \& {Sreenivas}(2004)}]{SekharS2004}
{Sekhar}, S.~C., \& {Sreenivas}, T. 2004, EURASIP Journal on Applied Signal
  Processing, 2004, 249858, \dodoi{10.1155/S111086570440417X}

\bibitem[{{Stroeer} {et~al.}(2009){Stroeer}, {Cannizzo}, {Camp}, \&
  {Gagarin}}]{StroeerCC2009}
{Stroeer}, A., {Cannizzo}, J.~K., {Camp}, J.~B., \& {Gagarin}, N. 2009, \prd,
  79, 124022, \dodoi{10.1103/PhysRevD.79.124022}

\bibitem[{Su(2017)}]{Su2017_HHT}
Su, Y.-H. 2017, HHTpywrapper v0.1.dev, v0.1.dev,  Zenodo,
  \dodoi{10.5281/zenodo.584082}.
\newblock \url{https://doi.org/10.5281/zenodo.584082}

\bibitem[{{Su} {et~al.}(2015){Su}, {Chou}, {Hu}, \& {Yang}}]{SuCH2015}
{Su}, Y.-H., {Chou}, Y., {Hu}, C.-P., \& {Yang}, T.-C. 2015, \apj, 815, 74,
  \dodoi{10.1088/0004-637X/815/1/74}

\bibitem[{{Takeda} {et~al.}(2021){Takeda}, {Hiranuma}, {Kanda}, {Kotake},
  {Kuroda}, {Negishi}, {Oohara}, {Sakai}, {Sakai}, {Sawada}, {Takahashi},
  {Tsuchida}, {Watanabe}, \& {Yokozawa}}]{TakedaHK2021}
{Takeda}, M., {Hiranuma}, Y., {Kanda}, N., {et~al.} 2021, \prd, 104, 084063,
  \dodoi{10.1103/PhysRevD.104.084063}

\bibitem[{{Taracchini} {et~al.}(2014){Taracchini}, {Buonanno}, {Pan},
  {Hinderer}, {Boyle}, {Hemberger}, {Kidder}, {Lovelace}, {Mrou{\'e}},
  {Pfeiffer}, {Scheel}, {Szil{\'a}gyi}, {Taylor}, \&
  {Zenginoglu}}]{TaracchiniBP2014}
{Taracchini}, A., {Buonanno}, A., {Pan}, Y., {et~al.} 2014, \prd, 89, 061502,
  \dodoi{10.1103/PhysRevD.89.061502}

\bibitem[{{The LIGO Scientific Collaboration} {et~al.}(2021){The LIGO
  Scientific Collaboration}, {the Virgo Collaboration}, {the KAGRA
  Collaboration}, {Abbott}, {Abbott}, {Acernese}, {Ackley}, {Adams},
  {Adhikari}, {Adhikari}, {Adya}, {Affeldt}, {Agarwal}, {Agathos}, {Agatsuma},
  {Aggarwal}, {Aguiar}, {Aiello}, {Ain}, {Ajith}, {Akcay}, {Akutsu},
  {Albanesi}, {Allocca}, {Altin}, {Amato}, {Anand}, {Anand}, {Ananyeva},
  {Anderson}, {Anderson}, {Ando}, {Andrade}, {Andres}, {Andri{\'c}},
  {Angelova}, {Ansoldi}, {Antelis}, {Antier}, {Appert}, {Arai}, {Arai}, {Arai},
  {Araki}, {Araya}, {Araya}, {Areeda}, {Ar{\`e}ne}, {Aritomi}, {Arnaud},
  {Arogeti}, {Aronson}, {Arun}, {Asada}, {Asali}, {Ashton}, {Aso}, {Assiduo},
  {Aston}, {Astone}, {Aubin}, {Austin}, {Babak}, {Badaracco}, {Bader},
  {Badger}, {Bae}, {Bae}, {Baer}, {Bagnasco}, {Bai}, {Baiotti}, {Baird},
  {Bajpai}, {Ball}, {Ballardin}, {Ballmer}, {Balsamo}, {Baltus}, {Banagiri},
  {Bankar}, {Barayoga}, {Barbieri}, {Barish}, {Barker}, {Barneo}, {Barone},
  {Barr}, {Barsotti}, {Barsuglia}, {Barta}, {Bartlett}, {Barton}, {Bartos},
  {Bassiri}, {Basti}, {Bawaj}, {Bayley}, {Baylor}, {Bazzan}, {B{\'e}csy},
  {Bedakihale}, {Bejger}, {Belahcene}, {Benedetto}, {Beniwal}, {Bennett},
  {Bentley}, {BenYaala}, {Bergamin}, {Berger}, {Bernuzzi}, {Berry},
  {Bersanetti}, {Bertolini}, {Betzwieser}, {Beveridge}, {Bhandare}, {Bhardwaj},
  {Bhattacharjee}, {Bhaumik}, {Bilenko}, {Billingsley}, {Bini}, {Birney},
  {Birnholtz}, {Biscans}, {Bischi}, {Biscoveanu}, {Bisht}, {Biswas}, {Bitossi},
  {Bizouard}, {Blackburn}, {Blair}, {Blair}, {Blair}, {Bobba}, {Bode}, {Boer},
  {Bogaert}, {Boldrini}, {Bonavena}, {Bondu}, {Bonilla}, {Bonnand}, {Booker},
  {Boom}, {Bork}, {Boschi}, {Bose}, {Bose}, {Bossilkov}, {Boudart},
  {Bouffanais}, {Bozzi}, {Bradaschia}, {Brady}, {Bramley}, {Branch},
  {Branchesi}, {Brandt}, {Brau}, {Breschi}, {Briant}, {Briggs}, {Brillet},
  {Brinkmann}, {Brockill}, {Brooks}, {Brooks}, {Brown}, {Brunett}, {Bruno},
  {Bruntz}, {Bryant}, {Bulik}, {Bulten}, {Buonanno}, {Buscicchio}, {Buskulic},
  {Buy}, {Byer}, {Cabourn Davies}, {Cadonati}, {Cagnoli}, {Cahillane},
  {Calder{\'o}n Bustillo}, {Callaghan}, {Callister}, {Calloni}, {Cameron},
  {Camp}, {Canepa}, {Canevarolo}, {Cannavacciuolo}, {Cannon}, {Cao}, {Cao},
  {Capocasa}, {Capote}, {Carapella}, {Carbognani}, {Carlin}, {Carney},
  {Carpinelli}, {Carrillo}, {Carullo}, {Carver}, {Casanueva Diaz}, {Casentini},
  {Castaldi}, {Caudill}, {Cavagli{\`a}}, {Cavalier}, {Cavalieri}, {Ceasar},
  {Cella}, {Cerd{\'a}-Dur{\'a}n}, {Cesarini}, {Chaibi}, {Chakravarti},
  {Chalathadka Subrahmanya}, {Champion}, {Chan}, {Chan}, {Chan}, {Chan},
  {Chan}, {Chandra}, {Chanial}, {Chao}, {Chapman-Bird}, {Charlton}, {Chase},
  {Chassande-Mottin}, {Chatterjee}, {Chatterjee}, {Chatterjee}, {Chaturvedi},
  {Chaty}, {Chatziioannou}, {Chen}, {Chen}, {Chen}, {Chen}, {Chen}, {Chen},
  {Chen}, {Chen}, {Cheng}, {Cheong}, {Cheung}, {Chia}, {Chiadini}, {Chiang},
  {Chiarini}, {Chierici}, {Chincarini}, {Chiofalo}, {Chiummo}, {Cho}, {Cho},
  {Choudhary}, {Choudhary}, {Christensen}, {Chu}, {Chu}, {Chu}, {Chua},
  {Chung}, {Ciani}, {Ciecielag}, {Cie{\'s}lar}, {Cifaldi}, {Ciobanu}, {Ciolfi},
  {Cipriano}, {Cirone}, {Clara}, {Clark}, {Clark}, {Clarke}, {Clearwater},
  {Clesse}, {Cleva}, {Coccia}, {Codazzo}, {Cohadon}, {Cohen}, {Cohen},
  {Colleoni}, {Collette}, {Colombo}, {Colpi}, {Compton}, {Constancio}, {Conti},
  {Cooper}, {Corban}, {Corbitt}, {Cordero-Carri{\'o}n}, {Corezzi}, {Corley},
  {Cornish}, {Corre}, {Corsi}, {Cortese}, {Costa}, {Cotesta}, {Coughlin},
  {Coulon}, {Countryman}, {Cousins}, {Couvares}, {Coward}, {Cowart}, {Coyne},
  {Coyne}, {Creighton}, {Creighton}, {Criswell}, {Croquette}, {Crowder},
  {Cudell}, {Cullen}, {Cumming}, {Cummings}, {Cunningham}, {Cuoco},
  {Cury{\l}o}, {Dabadie}, {Dal Canton}, {Dall'Osso}, {D{\'a}lya}, {Dana},
  {DaneshgaranBajastani}, {D'Angelo}, {Danila}, {Danilishin}, {D'Antonio},
  {Danzmann}, {Darsow-Fromm}, {Dasgupta}, {Datrier}, {Datta}, {Dattilo},
  {Dave}, {Davier}, {Davis}, {Davis}, {Daw}, {de Alarc{\'o}n}, {Dean}, {DeBra},
  {Deenadayalan}, {Degallaix}, {De Laurentis}, {Del{\'e}glise}, {Del Favero},
  {De Lillo}, {De Lillo}, {Del Pozzo}, {DeMarchi}, {De Matteis}, {D'Emilio},
  {Demos}, {Dent}, {Depasse}, {De Pietri}, {De Rosa}, {De Rossi}, {DeSalvo},
  {De Simone}, {Dhurandhar}, {D{\'\i}az}, {Diaz-Ortiz}, {Didio}, {Dietrich},
  {Di Fiore}, {Di Fronzo}, {Di Giorgio}, {Di Giovanni}, {Di Giovanni}, {Di
  Girolamo}, {Di Lieto}, {Ding}, {Di Pace}, {Di Palma}, {Di Renzo},
  {Divakarla}, {Dmitriev}, {Doctor}, {D'Onofrio}, {Donovan}, {Dooley},
  {Doravari}, {Dorrington}, {Drago}, {Driggers}, {Drori}, {Ducoin}, {Dupej},
  {Durante}, {D'Urso}, {Duverne}, {Dwyer}, {Eassa}, {Easter}, {Ebersold},
  {Eckhardt}, {Eddolls}, {Edelman}, {Edo}, {Edy}, {Effler}, {Eguchi},
  {Eichholz}, {Eikenberry}, {Eisenmann}, {Eisenstein}, {Ejlli}, {Engelby},
  {Enomoto}, {Errico}, {Essick}, {Estell{\'e}s}, {Estevez}, {Etienne}, {Etzel},
  {Evans}, {Evans}, {Ewing}, {Fafone}, {Fair}, {Fairhurst}, {Farah}, {Farinon},
  {Farr}, {Farr}, {Farrow}, {Fauchon-Jones}, {Favaro}, {Favata}, {Fays},
  {Fazio}, {Feicht}, {Fejer}, {Fenyvesi}, {Ferguson}, {Fernandez-Galiana},
  {Ferrante}, {Ferreira}, {Fidecaro}, {Figura}, {Fiori}, {Fishbach}, {Fisher},
  {Fittipaldi}, {Fiumara}, {Flaminio}, {Floden}, {Fong}, {Font}, {Fornal},
  {Forsyth}, {Franke}, {Frasca}, {Frasconi}, {Frederick}, {Freed}, {Frei},
  {Freise}, {Frey}, {Fritschel}, {Frolov}, {Fronz{\'e}}, {Fujii}, {Fujikawa},
  {Fukunaga}, {Fukushima}, {Fulda}, {Fyffe}, {Gabbard}, {Gabella}, {Gadre},
  {Gair}, {Gais}, {Galaudage}, {Gamba}, {Ganapathy}, {Ganguly}, {Gao},
  {Gaonkar}, {Garaventa}, {Garc{\'\i}a}, {Garc{\'\i}a-N{\'u}{\~n}ez},
  {Garc{\'\i}a-Quir{\'o}s}, {Garufi}, {Gateley}, {Gaudio}, {Gayathri}, {Ge},
  {Gemme}, {Gennai}, {George}, {George}, {Gerberding}, {Gergely}, {Gewecke},
  {Ghonge}, {Ghosh}, {Ghosh}, {Ghosh}, {Ghosh}, {Giacomazzo}, {Giacoppo},
  {Giaime}, {Giardina}, {Gibson}, {Gier}, {Giesler}, {Giri}, {Gissi},
  {Glanzer}, {Gleckl}, {Godwin}, {Goetz}, {Goetz}, {Gohlke}, {Golomb},
  {Goncharov}, {Gonz{\'a}lez}, {Gopakumar}, {Gosselin}, {Gouaty}, {Gould},
  {Grace}, {Grado}, {Granata}, {Granata}, {Grant}, {Gras}, {Grassia}, {Gray},
  {Gray}, {Greco}, {Green}, {Green}, {Gretarsson}, {Gretarsson}, {Griffith},
  {Griffiths}, {Griggs}, {Grignani}, {Grimaldi}, {Grimm}, {Grote}, {Grunewald},
  {Gruning}, {Guerra}, {Guidi}, {Guimaraes}, {Guix{\'e}}, {Gulati}, {Guo},
  {Guo}, {Gupta}, {Gupta}, {Gupta}, {Gustafson}, {Gustafson}, {Guzman}, {Ha},
  {Haegel}, {Hagiwara}, {Haino}, {Halim}, {Hall}, {Hamilton}, {Hammond}, {Han},
  {Haney}, {Hanks}, {Hanna}, {Hannam}, {Hannuksela}, {Hansen}, {Hansen},
  {Hanson}, {Harder}, {Hardwick}, {Haris}, {Harms}, {Harry}, {Harry},
  {Hartwig}, {Hasegawa}, {Haskell}, {Hasskew}, {Haster}, {Hattori}, {Haughian},
  {Hayakawa}, {Hayama}, {Hayes}, {Healy}, {Heidmann}, {Heidt}, {Heintze},
  {Heinze}, {Heinzel}, {Heitmann}, {Hellman}, {Hello}, {Helmling-Cornell},
  {Hemming}, {Hendry}, {Heng}, {Hennes}, {Hennig}, {Hennig}, {Hernandez},
  {Hernandez Vivanco}, {Heurs}, {Hild}, {Hill}, {Himemoto}, {Hines},
  {Hiranuma}, {Hirata}, {Hirose}, {Hochheim}, {Hofman}, {Hohmann}, {Holcomb},
  {Holland}, {Holley-Bockelmann}, {Hollows}, {Holmes}, {Holt}, {Holz}, {Hong},
  {Hopkins}, {Hough}, {Hourihane}, {Howell}, {Hoy}, {Hoyland}, {Hreibi},
  {Hsieh}, {Hsu}, {Huang}, {Huang}, {Huang}, {Huang}, {Huang}, {Huang},
  {H{\"u}bner}, {Huddart}, {Hughey}, {Hui}, {Hui}, {Husa}, {Huttner},
  {Huxford}, {Huynh-Dinh}, {Ide}, {Idzkowski}, {Iess}, {Ikenoue}, {Imam},
  {Inayoshi}, {Ingram}, {Inoue}, {Ioka}, {Isi}, {Isleif}, {Ito}, {Itoh},
  {Iyer}, {Izumi}, {JaberianHamedan}, {Jacqmin}, {Jadhav}, {Jadhav}, {James},
  {Jan}, {Jani}, {Janquart}, {Janssens}, {Janthalur}, {Jaranowski}, {Jariwala},
  {Jaume}, {Jenkins}, {Jenner}, {Jeon}, {Jeunon}, {Jia}, {Jin}, {Johns},
  {Johnson-McDaniel}, {Jones}, {Jones}, {Jones}, {Jones}, {Jones}, {Jonker},
  {Ju}, {Jung}, {Jung}, {Junker}, {Juste}, {Kaihotsu}, {Kajita}, {Kakizaki},
  {Kalaghatgi}, {Kalogera}, {Kamai}, {Kamiizumi}, {Kanda}, {Kandhasamy},
  {Kang}, {Kanner}, {Kao}, {Kapadia}, {Kapasi}, {Karat}, {Karathanasis},
  {Karki}, {Kashyap}, {Kasprzack}, {Kastaun}, {Katsanevas}, {Katsavounidis},
  {Katzman}, {Kaur}, {Kawabe}, {Kawaguchi}, {Kawai}, {Kawasaki},
  {K{\'e}f{\'e}lian}, {Keitel}, {Key}, {Khadka}, {Khalili}, {Khan}, {Khazanov},
  {Khetan}, {Khursheed}, {Kijbunchoo}, {Kim}, {Kim}, {Kim}, {Kim}, {Kim},
  {Kim}, {Kimball}, {Kimura}, {Kinley-Hanlon}, {Kirchhoff}, {Kissel}, {Kita},
  {Kitazawa}, {Kleybolte}, {Klimenko}, {Knee}, {Knowles}, {Knyazev}, {Koch},
  {Koekoek}, {Kojima}, {Kokeyama}, {Koley}, {Kolitsidou}, {Kolstein}, {Komori},
  {Kondrashov}, {Kong}, {Kontos}, {Koper}, {Korobko}, {Kotake}, {Kovalam},
  {Kozak}, {Kozakai}, {Kozu}, {Kringel}, {Krishnendu}, {Kr{\'o}lak}, {Kuehn},
  {Kuei}, {Kuijer}, {Kulkarni}, {Kumar}, {Kumar}, {Kumar}, {Kumar}, {Kume},
  {Kuns}, {Kuo}, {Kuo}, {Kuromiya}, {Kuroyanagi}, {Kusayanagi}, {Kuwahara},
  {Kwak}, {Lagabbe}, {Laghi}, {Lalande}, {Lam}, {Lamberts}, {Landry}, {Lane},
  {Lang}, {Lange}, {Lantz}, {La Rosa}, {Lartaux-Vollard}, {Lasky}, {Laxen},
  {Lazzarini}, {Lazzaro}, {Leaci}, {Leavey}, {Lecoeuche}, {Lee}, {Lee}, {Lee},
  {Lee}, {Lee}, {Lee}, {Lehmann}, {Lema{\^\i}tre}, {Leonardi}, {Leroy},
  {Letendre}, {Levesque}, {Levin}, {Leviton}, {Leyde}, {Li}, {Li}, {Li}, {Li},
  {Li}, {Li}, {Lin}, {Lin}, {Lin}, {Lin}, {Lin}, {Linde}, {Linker}, {Linley},
  {Littenberg}, {Liu}, {Liu}, {Liu}, {Liu}, {Llamas}, {Llorens-Monteagudo},
  {Lo}, {Lockwood}, {Loh}, {London}, {Longo}, {Lopez}, {Lopez Portilla},
  {Lorenzini}, {Loriette}, {Lormand}, {Losurdo}, {Lott}, {Lough}, {Lousto},
  {Lovelace}, {Lucaccioni}, {L{\"u}ck}, {Lumaca}, {Lundgren}, {Luo}, {Lynam},
  {Macas}, {MacInnis}, {Macleod}, {MacMillan}, {Macquet}, {Maga{\~n}a
  Hernandez}, {Magazz{\`u}}, {Magee}, {Maggiore}, {Magnozzi}, {Mahesh},
  {Majorana}, {Makarem}, {Maksimovic}, {Maliakal}, {Malik}, {Man}, {Mandic},
  {Mangano}, {Mango}, {Mansell}, {Manske}, {Mantovani}, {Mapelli},
  {Marchesoni}, {Marchio}, {Marion}, {Mark}, {M{\'a}rka}, {M{\'a}rka},
  {Markakis}, {Markosyan}, {Markowitz}, {Maros}, {Marquina}, {Marsat},
  {Martelli}, {Martin}, {Martin}, {Martinez}, {Martinez}, {Martinez},
  {Martinovic}, {Martynov}, {Marx}, {Masalehdan}, {Mason}, {Massera},
  {Masserot}, {Massinger}, {Masso-Reid}, {Mastrogiovanni}, {Matas},
  {Mateu-Lucena}, {Matichard}, {Matiushechkina}, {Mavalvala}, {McCann},
  {McCarthy}, {McClelland}, {McClincy}, {McCormick}, {McCuller}, {McGhee},
  {McGuire}, {McIsaac}, {McIver}, {McRae}, {McWilliams}, {Meacher}, {Mehmet},
  {Mehta}, {Meijer}, {Melatos}, {Melchor}, {Mendell}, {Menendez-Vazquez},
  {Menoni}, {Mercer}, {Mereni}, {Merfeld}, {Merilh}, {Merritt}, {Merzougui},
  {Meshkov}, {Messenger}, {Messick}, {Meyers}, {Meylahn}, {Mhaske}, {Miani},
  {Miao}, {Michaloliakos}, {Michel}, {Michimura}, {Middleton}, {Milano},
  {Miller}, {Miller}, {Miller}, {Millhouse}, {Mills}, {Milotti}, {Minazzoli},
  {Minenkov}, {Mio}, {Mir}, {Miravet-Ten{\'e}s}, {Mishra}, {Mishra}, {Mistry},
  {Mitra}, {Mitrofanov}, {Mitselmakher}, {Mittleman}, {Miyakawa}, {Miyamoto},
  {Miyazaki}, {Miyo}, {Miyoki}, {Mo}, {Modafferi}, {Moguel}, {Mogushi},
  {Mohapatra}, {Mohite}, {Molina}, {Molina-Ruiz}, {Mondin}, {Montani}, {Moore},
  {Moraru}, {Morawski}, {More}, {Moreno}, {Moreno}, {Mori}, {Morisaki},
  {Moriwaki}, {Morr{\'a}s}, {Mours}, {Mow-Lowry}, {Mozzon}, {Muciaccia},
  {Mukherjee}, {Mukherjee}, {Mukherjee}, {Mukherjee}, {Mukherjee}, {Mukund},
  {Mullavey}, {Munch}, {Mu{\~n}iz}, {Murray}, {Musenich}, {Muusse}, {Nadji},
  {Nagano}, {Nagano}, {Nagar}, {Nakamura}, {Nakano}, {Nakano}, {Nakashima},
  {Nakayama}, {Napolano}, {Nardecchia}, {Narikawa}, {Naticchioni}, {Nayak},
  {Nayak}, {Negishi}, {Neil}, {Neilson}, {Nelemans}, {Nelson}, {Nery},
  {Neubauer}, {Neunzert}, {Ng}, {Ng}, {Nguyen}, {Nguyen}, {Nguyen}, {Nguyen
  Quynh}, {Ni}, {Nichols}, {Nishizawa}, {Nissanke}, {Nitoglia}, {Nocera},
  {Norman}, {North}, {Nozaki}, {Nu{\~n}o Siles}, {Nuttall}, {Oberling},
  {O'Brien}, {Obuchi}, {O'Dell}, {Oelker}, {Ogaki}, {Oganesyan}, {Oh}, {Oh},
  {Oh}, {Ohashi}, {Ohishi}, {Ohkawa}, {Ohme}, {Ohta}, {Okada}, {Okutani},
  {Okutomi}, {Olivetto}, {Oohara}, {Ooi}, {Oram}, {O'Reilly}, {Ormiston},
  {Ormsby}, {Ortega}, {O'Shaughnessy}, {O'Shea}, {Oshino}, {Ossokine},
  {Osthelder}, {Otabe}, {Ottaway}, {Overmier}, {Pace}, {Pagano}, {Page},
  {Pagliaroli}, {Pai}, {Pai}, {Palamos}, {Palashov}, {Palomba}, {Pan}, {Pan},
  {Panda}, {Pang}, {Pang}, {Pankow}, {Pannarale}, {Pant}, {Panther},
  {Paoletti}, {Paoli}, {Paolone}, {Parisi}, {Park}, {Park}, {Parker},
  {Pascucci}, {Pasqualetti}, {Passaquieti}, {Passuello}, {Patel}, {Pathak},
  {Patricelli}, {Patron}, {Paul}, {Payne}, {Pedraza}, {Pegoraro}, {Pele},
  {Pe{\~n}a Arellano}, {Penn}, {Perego}, {Pereira}, {Pereira}, {Perez},
  {P{\'e}rigois}, {Perkins}, {Perreca}, {Perri{\`e}s}, {Petermann},
  {Petterson}, {Pfeiffer}, {Pham}, {Phukon}, {Piccinni}, {Pichot},
  {Piendibene}, {Piergiovanni}, {Pierini}, {Pierro}, {Pillant}, {Pillas},
  {Pilo}, {Pinard}, {Pinto}, {Pinto}, {Piotrzkowski}, {Piotrzkowski},
  {Pirello}, {Pitkin}, {Placidi}, {Planas}, {Plastino}, {Pluchar}, {Poggiani},
  {Polini}, {Pong}, {Ponrathnam}, {Popolizio}, {Porter}, {Poulton}, {Powell},
  {Pracchia}, {Pradier}, {Prajapati}, {Prasai}, {Prasanna}, {Pratten},
  {Principe}, {Prodi}, {Prokhorov}, {Prosposito}, {Prudenzi}, {Puecher},
  {Punturo}, {Puosi}, {Puppo}, {P{\"u}rrer}, {Qi}, {Quetschke},
  {Quitzow-James}, {Qutob}, {Raab}, {Raaijmakers}, {Radkins}, {Radulesco},
  {Raffai}, {Rail}, {Raja}, {Rajan}, {Ramirez}, {Ramirez}, {Ramos-Buades},
  {Rana}, {Rapagnani}, {Rapol}, {Ray}, {Raymond}, {Raza}, {Razzano}, {Read},
  {Rees}, {Regimbau}, {Rei}, {Reid}, {Reid}, {Reitze}, {Relton}, {Renzini},
  {Rettegno}, {Reza}, {Rezac}, {Ricci}, {Richards}, {Richardson}, {Richardson},
  {Riemenschneider}, {Riles}, {Rinaldi}, {Rink}, {Rizzo}, {Robertson}, {Robie},
  {Robinet}, {Rocchi}, {Rodriguez}, {Rolland}, {Rollins}, {Romanelli},
  {Romano}, {Romel}, {Romero-Rodr{\'\i}guez}, {Romero-Shaw}, {Romie},
  {Ronchini}, {Rosa}, {Rose}, {Rosi{\'n}ska}, {Ross}, {Rowan}, {Rowlinson},
  {Roy}, {Roy}, {Roy}, {Rozza}, {Ruggi}, {Ruiz-Rocha}, {Ryan}, {Sachdev},
  {Sadecki}, {Sadiq}, {Sago}, {Saito}, {Saito}, {Sakai}, {Sakai},
  {Sakellariadou}, {Sakuno}, {Salafia}, {Salconi}, {Saleem}, {Salemi},
  {Samajdar}, {Sanchez}, {Sanchez}, {Sanchez}, {Sanchis-Gual}, {Sanders},
  {Sanuy}, {Saravanan}, {Sarin}, {Sassolas}, {Satari}, {Sathyaprakash}, {Sato},
  {Sato}, {Sauter}, {Savage}, {Sawada}, {Sawant}, {Sawant}, {Sayah},
  {Schaetzl}, {Scheel}, {Scheuer}, {Schiworski}, {Schmidt}, {Schmidt},
  {Schnabel}, {Schneewind}, {Schofield}, {Sch{\"o}nbeck}, {Schulte}, {Schutz},
  {Schwartz}, {Scott}, {Scott}, {Seglar-Arroyo}, {Sekiguchi}, {Sekiguchi},
  {Sellers}, {Sengupta}, {Sentenac}, {Seo}, {Sequino}, {Sergeev}, {Setyawati},
  {Shaffer}, {Shahriar}, {Shams}, {Shao}, {Sharma}, {Sharma}, {Shawhan},
  {Shcheblanov}, {Shibagaki}, {Shikauchi}, {Shimizu}, {Shimoda}, {Shimode},
  {Shinkai}, {Shishido}, {Shoda}, {Shoemaker}, {Shoemaker}, {ShyamSundar},
  {Sieniawska}, {Sigg}, {Singer}, {Singh}, {Singh}, {Singha}, {Sintes},
  {Sipala}, {Skliris}, {Slagmolen}, {Slaven-Blair}, {Smetana}, {Smith},
  {Smith}, {Soldateschi}, {Somala}, {Somiya}, {Son}, {Soni}, {Soni}, {Sordini},
  {Sorrentino}, {Sorrentino}, {Sotani}, {Soulard}, {Souradeep}, {Sowell},
  {Spagnuolo}, {Spencer}, {Spera}, {Srinivasan}, {Srivastava}, {Srivastava},
  {Staats}, {Stachie}, {Steer}, {Steinhoff}, {Steinlechner}, {Steinlechner},
  {Stevenson}, {Stops}, {Stover}, {Strain}, {Strang}, {Stratta}, {Strunk},
  {Sturani}, {Stuver}, {Sudhagar}, {Sudhir}, {Sugimoto}, {Suh}, {Sullivan},
  {Sullivan}, {Summerscales}, {Sun}, {Sun}, {Sunil}, {Sur}, {Suresh}, {Sutton},
  {Suzuki}, {Suzuki}, {Swinkels}, {Szczepa{\'n}czyk}, {Szewczyk}, {Tacca},
  {Tagoshi}, {Tait}, {Takahashi}, {Takahashi}, {Takamori}, {Takano}, {Takeda},
  {Takeda}, {Talbot}, {Talbot}, {Tanaka}, {Tanaka}, {Tanaka}, {Tanaka},
  {Tanaka}, {Tanasijczuk}, {Tanioka}, {Tanner}, {Tao}, {Tao}, {Tapia San
  Mart{\'\i}n}, {Taranto}, {Tasson}, {Telada}, {Tenorio}, {Terhune},
  {Terkowski}, {Thirugnanasambandam}, {Thomas}, {Thomas}, {Thomas}, {Thompson},
  {Thondapu}, {Thorne}, {Thrane}, {Tiwari}, {Tiwari}, {Tiwari}, {Toivonen},
  {Toland}, {Tolley}, {Tomaru}, {Tomigami}, {Tomura}, {Tonelli},
  {Torres-Forn{\'e}}, {Torrie}, {Tosta e Melo}, {T{\"o}yr{\"a}}, {Trapananti},
  {Travasso}, {Traylor}, {Trevor}, {Tringali}, {Tripathee}, {Troiano},
  {Trovato}, {Trozzo}, {Trudeau}, {Tsai}, {Tsai}, {Tsang}, {Tsang}, {Tsao},
  {Tse}, {Tso}, {Tsubono}, {Tsuchida}, {Tsukada}, {Tsuna}, {Tsutsui},
  {Tsuzuki}, {Turbang}, {Turconi}, {Tuyenbayev}, {Ubhi}, {Uchikata},
  {Uchiyama}, {Udall}, {Ueda}, {Uehara}, {Ueno}, {Ueshima}, {Unnikrishnan},
  {Uraguchi}, {Urban}, {Ushiba}, {Utina}, {Vahlbruch}, {Vajente}, {Vajpeyi},
  {Valdes}, {Valentini}, {Valsan}, {van Bakel}, {van Beuzekom}, {van den
  Brand}, {Van Den Broeck}, {Vander-Hyde}, {van der Schaaf}, {van Heijningen},
  {Vanosky}, {van Putten}, {van Remortel}, {Vardaro}, {Vargas}, {Varma},
  {Vas{\'u}th}, {Vecchio}, {Vedovato}, {Veitch}, {Veitch}, {Venneberg},
  {Venugopalan}, {Verkindt}, {Verma}, {Verma}, {Veske}, {Vetrano},
  {Vicer{\'e}}, {Vidyant}, {Viets}, {Vijaykumar}, {Villa-Ortega}, {Vinet},
  {Virtuoso}, {Vitale}, {Vo}, {Vocca}, {von Reis}, {von Wrangel}, {Vorvick},
  {Vyatchanin}, {Wade}, {Wade}, {Wagner}, {Walet}, {Walker}, {Wallace},
  {Wallace}, {Walsh}, {Wang}, {Wang}, {Wang}, {Ward}, {Warner}, {Was},
  {Washimi}, {Washington}, {Watchi}, {Weaver}, {Webster}, {Weinert},
  {Weinstein}, {Weiss}, {Weller}, {Weller}, {Wellmann}, {Wen}, {We{\ss}els},
  {Wette}, {Whelan}, {White}, {Whiting}, {Whittle}, {Wilken}, {Williams},
  {Williams}, {Williams}, {Williamson}, {Willis}, {Willke}, {Wilson},
  {Winkler}, {Wipf}, {Wlodarczyk}, {Woan}, {Woehler}, {Wofford}, {Wong}, {Wu},
  {Wu}, {Wu}, {Wu}, {Wysocki}, {Xiao}, {Xu}, {Yamada}, {Yamamoto}, {Yamamoto},
  {Yamamoto}, {Yamamoto}, {Yamashita}, {Yamazaki}, {Yang}, {Yang}, {Yang},
  {Yang}, {Yang}, {Yap}, {Yeeles}, {Yelikar}, {Ying}, {Yokogawa}, {Yokoyama},
  {Yokozawa}, {Yoo}, {Yoshioka}, {Yu}, {Yu}, {Yuzurihara}, {Zadro{\.z}ny},
  {Zanolin}, {Zeidler}, {Zelenova}, {Zendri}, {Zevin}, {Zhan}, {Zhang},
  {Zhang}, {Zhang}, {Zhang}, {Zhang}, {Zhao}, {Zhao}, {Zhao}, {Zhao}, {Zheng},
  {Zhou}, {Zhou}, {Zhu}, {Zhu}, {Zimmerman}, {Zlochower}, {Zucker}, \&
  {Zweizig}}]{GWTC3-2021}
{The LIGO Scientific Collaboration}, {the Virgo Collaboration}, {the KAGRA
  Collaboration}, {et~al.} 2021, arXiv e-prints, arXiv:2111.03606.
\newblock \doarXiv{2111.03606}

\bibitem[{Torres {et~al.}(2011)Torres, Colominas, Schlotthauer, \&
  Flandrin}]{TorresCS2011}
Torres, M.~E., Colominas, M.~A., Schlotthauer, G., \& Flandrin, P. 2011, in
  2011 IEEE International Conference on Acoustics, Speech and Signal Processing
  (ICASSP), 4144--4147

\bibitem[{{Tsukada} {et~al.}(2018){Tsukada}, {Cannon}, {Hanna}, {Keppel},
  {Meacher}, \& {Messick}}]{TsukadaCH2018}
{Tsukada}, L., {Cannon}, K., {Hanna}, C., {et~al.} 2018, \prd, 97, 103009,
  \dodoi{10.1103/PhysRevD.97.103009}

\bibitem[{{Usman} {et~al.}(2016){Usman}, {Nitz}, {Harry}, {Biwer}, {Brown},
  {Cabero}, {Capano}, {Dal Canton}, {Dent}, {Fairhurst}, {Kehl}, {Keppel},
  {Krishnan}, {Lenon}, {Lundgren}, {Nielsen}, {Pekowsky}, {Pfeiffer},
  {Saulson}, {West}, \& {Willis}}]{UsmanNH2016}
{Usman}, S.~A., {Nitz}, A.~H., {Harry}, I.~W., {et~al.} 2016, Classical and
  Quantum Gravity, 33, 215004, \dodoi{10.1088/0264-9381/33/21/215004}

\bibitem[{{Vallisneri} {et~al.}(2015){Vallisneri}, {Kanner}, {Williams},
  {Weinstein}, \& {Stephens}}]{VallisneriKW2015}
{Vallisneri}, M., {Kanner}, J., {Williams}, R., {Weinstein}, A., \& {Stephens},
  B. 2015, in Journal of Physics Conference Series, Vol. 610, Journal of
  Physics Conference Series, 012021

\bibitem[{Van~Rossum \& Drake(2009)}]{python3}
Van~Rossum, G., \& Drake, F.~L. 2009, Python 3 Reference Manual (Scotts Valley,
  CA: CreateSpace)

\bibitem[{{Wang} {et~al.}(2014){Wang}, {Yeh}, {Young}, {Hu}, \&
  {Lo}}]{WangYY2014}
{Wang}, Y.-H., {Yeh}, C.-H., {Young}, H.-W.~V., {Hu}, K., \& {Lo}, M.-T. 2014,
  Physica A Statistical Mechanics and its Applications, 400, 159,
  \dodoi{10.1016/j.physa.2014.01.020}

\bibitem[{{Woosley} \& {Heger}(2007)}]{WoosleyH2007}
{Woosley}, S.~E., \& {Heger}, A. 2007, \physrep, 442, 269,
  \dodoi{10.1016/j.physrep.2007.02.009}

\bibitem[{{Wu} \& {Huang}(2004)}]{Wu2004}
{Wu}, Z., \& {Huang}, N.~E. 2004, Royal Society of London Proceedings Series A,
  460, 1597, \dodoi{10.1098/rspa.2003.1221}

\bibitem[{Wu \& Huang(2009)}]{Wu2009}
Wu, Z., \& Huang, N.~E. 2009, AADA, 1, 1, \dodoi{10.1142/S1793536909000047}

\bibitem[{Xu \& Yan(2006)}]{XuY2006}
Xu, Y., \& Yan, D. 2006, Proceedings of the American Mathematical Society, 134,
  2719, \dodoi{10.2307/4098122}

\bibitem[{{Yeh} {et~al.}(2010){Yeh}, {Shieh}, \& {Huang}}]{YehSH2010}
{Yeh}, J.-R., {Shieh}, J.-S., \& {Huang}, N.~E. 2010, Advances in Adaptive Data
  Analysis, 02, 135, \dodoi{10.1142/S1793536910000422}

\end{thebibliography}
\end{document}